\documentclass[conference]{IEEEtran}
\usepackage{geometry}                		% See geometry.pdf to learn the layout options. There are lots.
\usepackage[parfill]{parskip}    		        % Activate to begin paragraphs with an empty line rather than an indent
\usepackage{graphicx}				% Use pdf, png, jpg, or eps§ with pdflatex; use eps in DVI mode
\usepackage{amssymb}				% !TEX TS-program = latex   pdflatexmk
\usepackage{amsmath}
\usepackage{textcomp}
\usepackage{bbm}
\usepackage{algorithmic}
\usepackage[linesnumbered,ruled,vlined]{algorithm2e}
\usepackage{tabularx}
\usepackage{latexsym}
\usepackage{subcaption}
\usepackage{booktabs} 
\usepackage{eucal}
\usepackage{float}
\usepackage{parskip}
\usepackage{textcomp}
\usepackage{geometry}                		% See geometry.pdf to learn the layout options. There are lots.
\usepackage{tabularx}
\usepackage{mathrsfs}
\usepackage{calrsfs}
\usepackage{mathtools} 
\usepackage{stmaryrd}
\usepackage{wasysym} 
\usepackage{microtype}
\usepackage{yfonts}
\usepackage{listings} 
\usepackage{url}
\usepackage{multirow}
\usepackage{xcolor}
\usepackage{pdfpages}
\usepackage{graphicx} 
\usepackage{xurl}
\usepackage[utf8]{inputenc}
\usepackage [english]{babel}
\usepackage [autostyle, english = american]{csquotes}
\usepackage[style=numeric,natbib=true]{biblatex} 
\graphicspath{{./Images/}} 
\usepackage{mathtools}
\usepackage{amsmath}
\usepackage{fancyhdr} 
\usepackage{titlesec}

\usepackage[backref=true,colorlinks=true,linkcolor=blue, urlcolor=blue, citecolor=blue]{hyperref}
\hypersetup{
    pagebackref=true,
    colorlinks=true,
    citecolor=blue,
    linkcolor=blue,
    filecolor=magenta,      
    urlcolor=blue,
}
\definecolor{dkgreen}{rgb}{0,0.6,0}
\definecolor{gray}{rgb}{0.5,0.5,0.5}
\definecolor{mauve}{rgb}{0.58,0,0.82}
\definecolor{lightsilver}{rgb}{0.96,0.96,0.98} 
\definecolor{icterine}{rgb}{0.99, 0.97, 0.37}
\definecolor{lemonchiffon}{rgb}{1.0, 0.98, 0.8}
\lstdefinestyle{R} {
backgroundcolor=\color{lemonchiffon},
language=Python,
basicstyle=\tiny,
} 
\titleformat{\chapter}[display]
{\normalfont\Huge\bfseries\raggedright}{\chaptertitlename\  7}
{15pt}{\Huge} 
\title{Control Systems Analysis of a 3-Axis Photovoltatic Solar Tracker for Water Pumping}
\author{Justin London}
\begin{document}

\author{\IEEEauthorblockN{Justin London \\}
\IEEEauthorblockA{\textit{Department of Electrical Engineering and Computer Science} \\
\textit{University of North Dakota}\\
Grand Forks, North Dakota USA \\
justin.london@und.edu}
}

\maketitle

\begin{abstract}
    We propose 3-axis solar tracker water pumping system.  The solar tracker can rotate and tilt using stepper/DC motors and can rise and lower on a tripod using a linear actuator.  The charge generated from solar energy absorbed by photovoltaic (PV) cells in the solar panel is stored in a 12V battery that in turn powers two water diaphragm pumps using a solar charge controller. The PV uses four light photocell resistors/sensors to measure light intensity.  A solar tracking algorithm  determines the optimal angle for PV positioning. 
    Using an ultrasonic sensor to measure the water level in a reservoir water tank, water is pumped from one water tank to the reservoir.  Based on soil moisture sensor levels, a second water pump supplies water from the reservoir to the plant.  The system is analyzed from a control systems perspective. The transfer functions, root loci, and Bode plots are generated and simulated and experimental results are provided as well as stability and steady-state error analysis.  
\end{abstract}

\section{Introduction}
    \ \ \ In agricultural and rural areas, power grids may not exist or be too costly to build to pump water for irrigation. Renewable and abundant energy sources like solar and wind have become popular alternatives especially in remote areas \cite{Habib:2023}.  Low-maintenance, no pollution, easy installation, reliability are key advantages.  However, solar power systems have a high initial cost and variable water production.  
    
    Solar water pumping systems use the photovoltaic (PV) effect to convert sunlight into electrical energy that can be used to power motors to pump water for agriculture.    A solar PV system consists of five major components: a PV array, a power conditioning unit, pump-motor load, water tank storage, and pipe distribution system \cite{Khatib:2016}.   
    
    PV pumping systems can be battery-coupled or directly coupled \cite{Khatib:2021}.   A battery-coupled system has PV panels, charge regulator, batteries, pump controller, tank, and DC pump.  Alternatively, a directly coupled system has no batteries and water has to be stored in a the tank so as to be used at night or on cloudy days.   Most PV arrays are connected to DC links via a DC/DC boost or buck converter as a means to enable maximum power point tracking (MPPT) control, while using three-phase voltage source inverters to maintain DC-link voltages \cite{Sharma:2019}.  

    There are two main categories of water pumps used in standalone PV systems: rotating and positive displacement.  Centrifugal, rotating vane, and screw drive are rotating types that move water continuously when power is provided to the pump.   The output of the pumps depend on head, solar radiation (current produced), and operating voltage.  

    The performance of a PV pump system (PVPS) depends on numerous meteorological variables (e.g., solar radiation, ambient temperature, wind speed, humidity, and shadow effects), PV module specifications (e.g. conversion efficiency and tilt angle), and motor pump-hydraulic system characteristics.   

    To illustrate the performance and efficiency of PV water pump systems, we introduce a novel 3-axis solar tracker water pumping system that is applied for plant watering and can be used for irrigation.  The solar tracker uses solar energy to power a 12 V DC lead acid battery that in turn powers two water diaphragm pumps.  Water in one tank is pumped by one water pump to a second tank.  A second water pump pushes water from the second tank through a hose that can water plants via a spray nozzle.  
    
    System components such as motor drivers, relay switches, and sensors are controlled using an Arduino Uno microcontroller that provides a built-in analog-to-digital converter (ADC) with pulse width modulation (PWM).  PWM can be used to regulate motor speed and its duty cycle.   

\section{Photovoltaic Array Model}

    A solar cell \enquote{can be represented as a double-diode model \cite{Khatib:2021}.}  The output current of a solar cell can be expressed as:
%\begin{multline}
%    I_{c} = I_{Ph} - I_{o_{1}} \bigg [ \text{exp} \bigg (\frac{V_{c}+I_{c}R_{s}}{V_{t_1}} \bigg )  -1 \bigg] \\ - I_{o_{2}} \bigg [\text{exp} \bigg (\frac{V_{c}+I_{c}R_{s}}{V_{t_{2}}} \bigg ) -1 \bigg] \bigg - \frac{V_{c} + I_{c}R_{s}}{R_{p}}
%\end{multline}
\begin{multline}
    I_{c} = I_{Ph} - I_{o_{1}} \bigg [ \text{exp} \bigg (\frac{V_{c}+I_{c}R_{s}}{V_{t_{1}}} \bigg )  -1 \bigg] \\ - I_{o_{2}} \bigg [\text{exp} \bigg (\frac{V_{c}+I_{c}R_{s}}{V_{t_{2}}} \bigg ) -1 \bigg]  - \frac{V_{c} + I_{c}R_{s}}{R_{p}}
\end{multline}
where $I_{c}$ and $V_{c}$ are the output current (A) and voltage (V) of the solar cell, respectively; $I_{ph}$ is the photocurrent; $I_{o_{1}}$ and $I_{o_{2}}$ are the diode saturation currents of the first and second diodes, respectively in ampere (A);  $R_{s}$ is the series resistance ($\Omega$); $R_{p}$ is the resistance ($\Omega$); and $V_{t_{1}}$ and $V_{t_{2}}$ are the diode thermal voltages given by:
\begin{equation} 
    V_{t_{1}} = \frac{a_{1}k_{B}T_{c}}{q}
\end{equation}
\begin{equation}
     V_{t_{2}} = \frac{a_{2}k_{B}T_{c}}{q}
\end{equation}
where $q$ is the electron charge (1.602 $\times 10^{-19}$ Colombs), $k_{B}$ is Boltzmann's constant (1.381 $\times 10^{-23}$ J/K), $T_{C}$ is the cell temperature (Kelvin), and $a_{1}$ and $a_{2}$ are the diode ideality factors that represent the components of the diffusion and recombination currents, respectively \cite{Khatib:2021}.

The PV array consists of $N_{s}$ and $N_{p}$ cell modules, which are connected in series and parallel, respectively, to constitute the power demand by the load.  The output current of a PV array can be expressed as:
    \begin{multline}
        I_{a} = N_{p}I_{Ph} - \\ N_{p}I_{o_{1}} \bigg [\text{exp} \bigg (\frac{1}{V_{t_{1}}} \bigg (\frac{V_{a}}{N_{s}} + \frac{I_{a}R_{s}}{N_{p}} \bigg ) \bigg )-1 \bigg] \\
        - N_{p}I_{o_{2}} \bigg [\text{exp} \bigg (\frac{1}{V_{t_{2}}} \bigg (\frac{V_{a}}{N_{s}} + \frac{I_{a}R_{s}}{N_{p}} \bigg ) \bigg ) -1 \bigg] \\ - \frac{N_{p}}{R_{p}} \bigg (\frac{V_{a}}{N_{s}} + \frac{I_{a}R_{s}}{N_{p}} \bigg )
    \end{multline}
    where $V_{a}$ and $I_{a}$ are the output voltage (V) and current (A) of the PV array, respectively.  The PV cell is essentially a p-n junction that is fabricated on a thin film semiconductor like silicon.  As the cell is exposed to light energy (photons), the photon that hits the cells will be absorbed by semiconducting material as the PV panel is polycrystalline silicon.  
    
\begin{figure}[H] 
	\centering
	\includegraphics[width=0.85\columnwidth]{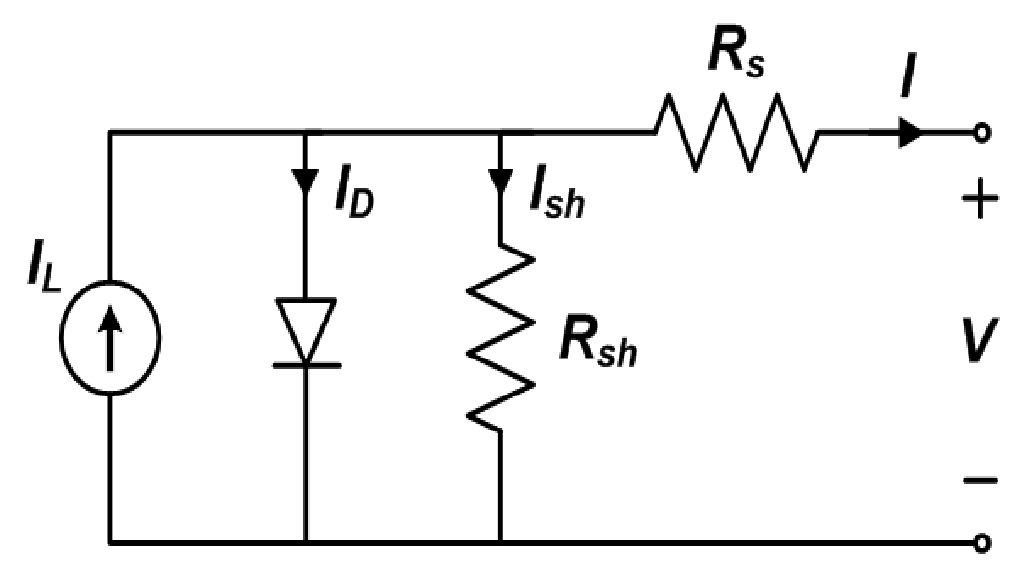} 
	\caption{PV cell as single diode circuit. Source: \cite{Sharma:2019}}
	\label{fig:cell} 
\end{figure}

    In an ideal cell $R_{s}$ is 0 and $R_sh$ is infinite
    \ \ The efficiency of the PV array can be computed as
    \begin{equation}
        \zeta_{PV}  =. \frac{V_{a}I_{a}}{AG_{T}}
    \end{equation}
    where $V_{a}$ and $I_{a}$ are the output voltage (V) and current (A) of the PV array, respectively.  Figure \ref{fig:cell} shows the equivalent circuit of a PV cell modeled as a single diode circuit.
    %The panels have a 
\section{Solar Tracking}
\ \ \ The model in this paper builds upon a two-axis solar tracker with a feedback control system using a stepper motor and linear actuator established though an electronic circuit using photodiodes \cite{Moron:2017}. 

Based on the incident angle of light hitting the panel, the solar panel is designed to drive the PV panel between 0 and 180 degrees at a low speed.  Pulse width modulation (PWM), a continuous square wave with a period of 20ms, is used to control the motor. The the PWM signal, the output shaft of the servo motor changes. 

The position of the sun during any time of day is calculated by knowing the azimuth and the elevation angles.  A dual axis solar tracker has two axes of rotation: the azimuth axis and altitude axis.  An azimuth-altitude dual axis tracker (AADAT) may \enquote{used to track the sun using a servo motor for the azimuth axis (horizontal) and a linear actuator for the zenith axis (vertical).  Light dependent resistors (LDRs), placed at 45 are used measure the light intensity and angle.}
If the sun is at a 90 degree angle to the LDR sensors, the voltages on the relevant tracking cells (north-south and east-west) are equal, and the solar panel stays in its position.   

\begin{figure}[H] 
	\centering     
  \includegraphics[width=0.9\columnwidth]{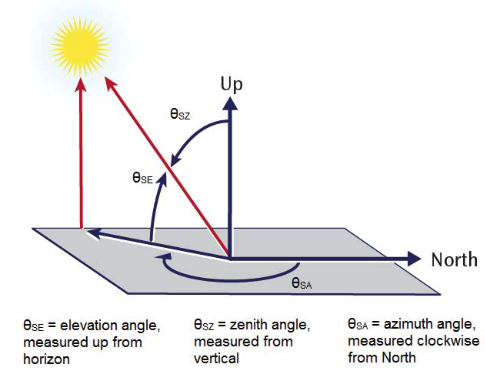} 
	\caption{Illustration of elevation, zenith, and azimuth angles. Source: \cite{Riley:2014}}
	\label{fig:sun2} 
\end{figure}

The solar elevation angle $\theta_{SE}$ is the angle between the observer's horizon and the sun, usually expressed in degrees, and is defined on the interval [$-90^{\circ},90^{\circ}$] (negative angles occur when the sun falls below the observer's horizon.)
\ \ At sunrise and sunset, $\theta_{SE} = 0^{\circ}$, and it reaches a maximum at solar noon. The solar zenith angle $\theta_{SZ}$ is the complement of the solar elevation angle and is defined as \enquote{the angle between the sun and a vector pointed directly overhead} the observer \cite{Riley:2014}.    

\ \ The solar azimuth angle, $\theta_{SA}$ as shown in Figure \ref{fig:sun2} describes the direction of the sun as a bearing on the Earth's surface.  As a bearing, the azimuth angle is defined as \enquote{the number of degrees clockwise from true north and ranges over the interval [$0^{\circ},360^{\circ}]$.  When the sun is due north of the observer, the azimuth if $0^{\circ}$, when the sun is due east of the observer the azimuth is $90^{\circ}$ (south = $180^{\circ}$, west = $270^{\circ}$)} \cite{Riley:2014}. 
    Solar irradiance can be computed by measuring the power of the light source of the luminous flux \cite{Musa:2023}.  The tilt angle $\theta_{tilt}$ is the angle formed by the solar tracking system and the horizontal axis and the angle of incidence is a variety of tile angles.  The declination angle $\delta$ is \enquote{the angle formed between the equator and a line drawn from the center of the sun to the center of the Earth} and can be expressed as:
    \begin{equation}
        \delta = -23.45 \times \text{cos} \bigg ( \frac{360}{365} \times (n+10) \bigg )
    \end{equation}
    The elevation angle is the angle between the sun's center and the horizon, while the zenith angle is the angle formed by the center of the sun and the vertical.  The elevation and zenith angles are defined as
   \begin{equation}
        \theta_{e} = 90^{\circ} - \theta_{z} \nonumber \\
    \end{equation}
    \begin{multline}
        \theta_{z} = \text{cos}^{-1} \big (\text{sin}(L_{st})\text{sin}(\delta) \\ + \text{cos}(L_{st})\text{cos}(\delta)\text{cos}(\text{ST}) \big ) \nonumber 
    \end{multline}
    where $L_{st}$ is the standard longitude that is positive for the east region and negative for the west. ST is the standard time \cite{Musa:2023}.  
    
\begin{figure}[H] 
	\centering
	\includegraphics[width=0.85\columnwidth]{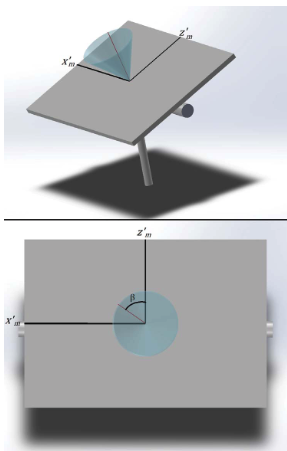} 
	\caption{Solar Tracker Views. Source: \cite{Riley:2014}}
	\label{fig:tracker} 
\end{figure}

The tilt angle of a tracker face $\theta_{tilt}$ is related to the tracker elevation angle $\theta_{TE}$ as $\theta_{tilt} = \lvert 90 - \theta_{TE} \rvert$
The solar angle of incidence (AOI), $\alpha \in [0,90)$, is the angle between the PV panel's normal vector and the vector pointing to the middle of the sun \cite{Riley:2014}.  The angle of incidence is defined as
\begin{multline}
    \alpha = \text{cos}^{-1} 
    \big (\text{sin}(\theta_{SE})\text{sin}(\theta_{TE}) + \\ 
    \text{cos}(\theta_{SE})\text{cos}(\theta_{TE})\text{cos}(\theta_{SA} - \theta_{TA}) \big ) 
\end{multline}
By projecting the sun unit vector
\begin{equation}
\textbf{s} = \begin{bmatrix} \text{sin}(\theta_{SA})\text{cos}(\theta_{SE}) & \\
                \text{cos}(\theta_{SA})\text{cos}(\theta_{SE}) & \\
                \text{sin}(\theta_{SE}) 
\end{bmatrix}
\end{equation}
in a topocentric right-handed, Cartesian coordinate system with unit vector $\mathbf{\hat{x}} = [1 \ 0 \ 0]'$, $\mathbf{\hat{y}} = [0 \ 1 \ 0]'$, and $\mathbf{\hat{z}} = [0 \  0  \ 1]'$  onto the rotated tracker surface as shown in Figure \ref{fig:tracker}, it can be shown that for a dual axis solar tracker, the angle of incidence direction is
\begin{equation}
    \beta = \text{tan}^{-1} \bigg ( \frac{\mathbf{s \cdot x'_{m}}}{\mathbf{s \cdot z'_{m}}} \bigg ) 
\end{equation}
where
\begin{multline}
        (\mathbf{s \cdot x'_{m}}) = \text{cos}(\theta_{SE})\text{cos}(\theta_{TA})\text{sin}(\theta_{SA}) \\ - \text{cos}(\theta_{SE})\text{cos}(\theta_{SA}\text{sin}(\theta_{TA}) \nonumber 
\end{multline}
\begin{multline}
    (\mathbf{s \cdot z'_{m}}) = \text{cos}(\theta_{TE})\text{sin}(\theta_{SE}) 
    \\- \text{sin}(\theta_{TE})\text{cos}(\theta_{SE})\text{cos}(\theta_{SA} - \theta_{TA}) \nonumber 
\end{multline}
    The top of Figure \ref{fig:tracker} shows \enquote{an isometric view of a solar tracker with reference axes.  The red line indicates the direction of the sun, the cone indicates all possible sun positions with the same AOI ($\alpha$) relative the tracker. \cite{Riley:2014}}  The bottom of Figure \ref{fig:tracker} shows a direct view of the face of the tracker with incident angle $\beta$. 
where 
\begin{equation}
  \mathbf{x'_{m}} = 
\begin{bmatrix}
            \text{cos}(\theta_{TA}) & \\
             -\text{sin}(\theta_{TA}) & \\
             0 
\end{bmatrix}
\end{equation}
\begin{equation}
\textbf{y}'_{m} =
\begin{bmatrix}
            \text{sin}(\theta_{TA})\text{cos}(\theta_{TE}) & \\
             -\text{cos}(\theta_{TA})\text{cos}(\theta_{TE}) & \\
             \text{sin}(\theta_{TE}) 
\end{bmatrix}
\end{equation}
\begin{equation}
\textbf{z}'_{m} = 
\begin{bmatrix}
            -\text{sin}(\theta_{TE})\text{sin}(\theta_{TA}) & \\
             -\text{cos}(\theta_{TE})\text{cos}(\theta_{T}) & \\
             \text{cos}(\theta_{TE}) 
\end{bmatrix}
\end{equation}
    The optimal elevation and tracking angles are derived in \cite{Riley:2014}
\begin{align}
    \theta_{TE} = \text{tan}^{-1} \big(F,R) \\
    \theta_{TA} = \theta_{SA} - \text{tan}^{-1}(G,H)
\end{align}
where 
\tiny
\begin{align}
F &= \frac{D^{2}A^{2} + D^{2}N^{2} + A^{2} + N^{2} - R^{2}(D^{2}A^{2} + 1)}{2AN(D^2+1)} \nonumber \\
G &= \frac{D^{3}N^{2} + D^{3}A^{2} - DA^{2} + DN^{2} + R^{2}(D^{3}A^{2}+D)}{(D^{2}+1)CNR} \nonumber \\
H &= \frac{A^{2} + D^2A^2 - N^2 - D^2N^2 + R^2(D^2A^2+1)}{(D^2 +1)CAR} \nonumber
\end{align}
\normalsize
and
\begin{align}
C &= \text{cos}(\theta_{SE}) \\
N &= \text{sin}(\theta_{SE}) \\
D &= \text{tan}(\beta) \\
A &= \text{cos}(\alpha) 
\end{align}
and $R$ are the roots of the fourth order polynomial:
\begin{equation}
   aw^4 + bw^2 + c = 0 
\end{equation}
where
\small
\begin{align}
    a &= (D^2A^2 + 1)^2 \nonumber \\
    b &= -2(D^2 + 1)(A^4D^2 - A^2D^2N^2 - 2A^2N^2 \nonumber \\ 
    & \ \ \ \ \ \ \ \ \ \ \ \ \ \ \ \ \ \ \ \ + A^2 + N^2) \nonumber \\
    c &= (D^2 + 1)^2(A^2 - N^2)^{2} \nonumber
\end{align}
\normalsize
%\begin{equation}
%\text{cos}(\theta) = \text{cos} \ \text{cos}\theta_{Z} + \text{sin}\beta \ \text{sin}\theta_{Z}\text{cos}(\gamma_{S}-\gamma)
%\label{eq:cos}
%\end{equation}
%where $\theta_{Z}$ and $\gamma_{S}$ are the zenith and solar azimuth angles and can be determined from the time and location using various algorithms \cite{Michalsky: }. To introduce the angle $\alpha$ for optimal tracking into equation \ref{fig:cos} and to remove $\beta$ and $\gamma$, substitutions are made using trigonometric identities yielding the resulting expression \cite{}:
%\begin{multline}
%    \text{cos}(\theta) = \text{cos}(\alpha) [ \text{sin}(\theta_{Z}) \text{cos}(\gamma_{S} - \gamma_{a})\text{sin}(\beta_{a}) \\ + \ \text{cos}(\theta_{Z}\text{cos}(\beta_{a}) + \text{sin}(\alpha)\text{sin}(\theta_{Z})\text{sin}(\gamma_{S} - \gamma_{a})  
%\end{multline}
%Figure \ref{fig:solartracking} is a flowchart of the 
Algorithm \ref{alg:mmpt} provides the solar tracking algorithm that positions the PV panel based on the magnitude differences between the LPR sensors.
%\begin{figure}[H] 
%	\centering  \includegraphics[width=0.9\columnwidth]{Images/solartracking.png} 
%	\caption{Solar Tracking Algorithm. Source: \cite{Elhammoumi:2020}}
%	\label{fig:mmpt} 
%\end{figure} 
\begin{algorithm}
\caption{{Solar Tracking Algorithm}}
\label{alg:mmpt}
\textbf{Set solar tracker initial position}\\
\While{Power On}{
	Read the analog value from each LDR sensor: topright, topleft, bottomright, bottomleft \\
	Calculate the average values for each LDR sensor: Eqs. (15)-(18)\\
	Calculate the differences (azimuth and elevation) : Eqs. (19)-(20)\\	
%	\eIf{$avgsum < 8$}{Rotate the Servo Motors to the initial position}{
%		\eIf{$ azimuth\_diff \leq 10 $}{Stop the Up-down Servo Motor}{
%			\eIf{$ diffazi > 0 $}{Left-Right Servo Motor move PV panel right}{Left-Right Servo Motor move PV panel left	}}
%		}
		\If{$avgsum < 8$}{Rotate the Servo Motors to the initial position\\ \textbf{break}}
		\eIf{$ azimuth\_diff \leq 10 $}{Stop the Up-down Servo Motor }{
						\eIf{$ diffazi > 0 $}{Left-Right Servo Motor move PV panel right}{Left-Right Servo Motor move PV panel left	}
                        \textbf{end}
                        }
				\If{$\lvert diffelev \rvert \leq 10$}{Stop the Up-down Servo Motor\\ \textbf{break}}
				\If{$ diffelev  > 0$}{	Up-Down Servo Motor move PV panel up\\ \textbf{break}}
		Up-Down Servo Motor move PV panel down \\
%		\If{$ azimuth\_diff \leq 10 $}{Stop the Up-down Servo Motor\\ }
%		\If{$ diffazi > 0 $}{Stop the Up-down Servo Motor\\ \textbf{Break}}
	}
\textbf{end}
\end{algorithm}
\subsection{MPPT}
Maximum power point tracking (MPPT) is an electronic DC to DC converter that optimizes the power match between the solar array (PV panels) and the battery (or battery bank). In essence, they convert and adjust a higher voltage DC output from solar panels down to the lower voltage needed to charge batteries.  All MPPTs are microprocessor controlled.   

A step-down (step-up) chopper or boost/DC-DC converter can be applied to MPPT systems in which the output voltage needs to be less (greater) than the input voltage. The relation between the output and input voltages ($V_{out}$ and $V_{in}$, respectively) is as follows
\begin{equation}
    \frac{V_{out}}{V_{in}} = \frac{1}{1-D}
\end{equation}
where $D$, $0 \leq D \leq 1$, is the duty cycle \cite{Amar:2022}.

Digital MPPT controllers know when to adjust the power being sent to the battery.  There are two common local optimizing hill climbing algorithms: (1) Perturb and Observe (P\&O) and (2) incremental conductance (IC).  
\subsection{Perturb and Observe}
\ \ P\&O determines on which side of the maximum power point (MPP), the hill climb optimization of the PV curve movies (left or right) by observing the sign of the previous change in voltage ($d$V) and change in power ($d$P). 

Depending on that information, the voltage increases (V++) or decreases (V--) the voltage.  If it is on the left side of the MPP, it moves right (increases) and if it is on the right of the MPP, the voltage moves left.  In particular, if the $d$V and $d$P are both positive or both negative, the voltage increases.  Otherwise, if dV and dP differ, the voltage decreases.  

    The P\&O method uses periodic perturbation via incremental increases or decreases of voltage (or current) generated by a PV source, along with a trial assessment of the power output of the present cycle to the power output of the preceding perturbation cycle \cite{Sharma:2019}.  The P\&O algorithm continuously searches the MPP in subsequent perturbation cycles.   
    %Figure \ref{fig:mmpt} shows a flow diagram of the MMPT P\&O algorithm.
    Algorithm \ref{alg:mmpt2} shows the MPPT P\&O algorithm.
\begin{algorithm}
\caption{{Perturb and Observe Algorithm }}
\label{alg:mmpt2}
\textbf{Initialization:} $\tau = 0$,  $\mathcal{T}_{max}$, $\mathcal{I}(\tau)$, $\mathcal{V}(\tau)$\\
\If{$\tau \leq \mathcal{T}_{max}$}{
    $\mathcal{P}(\tau) = \mathcal{I}(\tau) \times \mathcal{V}(\tau)$ \\
    \eIf{$d\mathcal{P} > 0$} {
	   \eIf{$d\mathcal{V} > 0$} {
        {$\mathcal{V}_{ref} = \mathcal{V}_{ref} - d\mathcal{V}$} \\
      }{
        {$\mathcal{V}_{ref} = \mathcal{V}_{ref} + \mathcal{V}$} \\
      }
      \textbf{end} \\
      }
      {
        \eIf{$d\mathcal{V} < 0$} {
        {$\mathcal{V}_{ref} = \mathcal{V}_{ref} + d\mathcal{V}$} \\
      }{
        {$\mathcal{V}_{ref} = \mathcal{V}_{ref} - \mathcal{V}$} \\
      }
      \textbf{end} \\
      }
     \textbf{end} \\
%\textbf{end}
$\mathcal{I}(\tau)$ = $\mathcal{I}(\tau-1)$ \\
$\mathcal{V}(\tau)$ = $\mathcal{V}(\tau-1)$ \\
$\mathcal{P}(\tau)$ = $\mathcal{P}(\tau-1)$ \\
$\tau = \tau +1$ \\
}
\textbf{end}
\end{algorithm}
%\begin{figure}[H] 
%	\centering  \includegraphics[width=0.9\columnwidth]{Images/%MMPT2.png} 
%	\caption{MMPT P\&O Flow Chart. Source: \cite{Sharma:2019}}
%	\label{fig:mmpt} 
%\end{figure} 

\subsection{Incremental Conductance}
    P\&O has the drawback that it oscillates around the MPP and may initially move in the incorrect direction when environmental conditions change.  Incremental conductance (IC) overcomes this issue is by representing the derivative of the power as:
    \begin{equation}
        \frac{dP}{dV} = \frac{d(I \cdot V)}{dV} \rightarrow \frac{1}{V}\frac{dP}{dV} = \frac{dI}{dV} + \frac{I}{V}
    \end{equation}
where $\frac{dI}{dV}$ is the incremental conductance.  The sign of the expression on the right side is always the same as the sign of the power change.
%\subsection{Q-Learning}
    %Figure \ref{fig:ic} 
    Algorithm \ref{alg:alg3} shows a flow diagram for the incremental conductance algorithm.
%\begin{figure}[H] 
%	\centering  \includegraphics[width=1\columnwidth]{Images/IC.png} 
%	\caption{MMPT IC Flow Chart. Source: \cite{Selvan:2013}} 
%	\label{fig:ic} 
%\end{figure} 

\begin{algorithm}
	\caption{{MMPT IC Algorithm}}
    \label{alg:alg3}
    \textbf{Initial Sensor Data:} $\tau = 0$, $\mathcal{T}_{max}$, 
    $\Delta \mathcal{V} = \mathcal{V}(\tau) - \mathcal{V}(\tau-1)$,
    $\Delta \mathcal{I} = \mathcal{I}(\tau) - \mathcal{I}(\tau-1)$ \\
    \If{$\Delta \mathcal{V} = 0$} { 
        \If{$\Delta \mathcal{I} = 0$} {
             $\mathcal{V}_{ref} = \mathcal{V}_{ref}$ \\
        }
        \ElseIf{$\Delta \mathcal{I} > 0$} {
            $\mathcal{V}_{ref} = \mathcal{V}_{ref} + d\mathcal{V}$ \\
        }
        \Else{$\mathcal{V}_{ref} = \mathcal{V}_{ref} - d\mathcal{V}$} 
        \textbf{end} 
    } 
    \ElseIf{$d\mathcal{I}/d\mathcal{V} = -\mathcal{I}/\mathcal{V}$} { 
        $\mathcal{V}_{ref} = \mathcal{V}_{ref}$ 
    }
    \ElseIf{$d\mathcal{I}/d\mathcal{V} > -\mathcal{I}/\mathcal{V}$} {
        $\mathcal{V}_{ref} = \mathcal{V}_{ref} + d\mathcal{V}$ \\
    }
    \Else{$\mathcal{V}_{ref} = \mathcal{V}_{ref} - d\mathcal{V}$}
    \textbf{end} \\
    $\mathcal{V}(\tau) = \mathcal{V}(\tau-1)$ \\
    $\mathcal{I}(\tau) = \mathcal{I}(\tau-1)$ \\
    $\tau = \tau +1$ \\
    %\If{$\tau \leq \mathcal{T}_{max}$}{
    %    \If{$ \Delta \mathcal{V} == 0$} {
    %        \If{$ \Delta I{V} == 0$}
                       
    %    \eIf{$ \Delta \mathcal{I} > 0$}
    %}
    %\textbf{end}
\end{algorithm}

\begin{figure}[H] 
	\centering  \includegraphics[width=0.9\columnwidth]{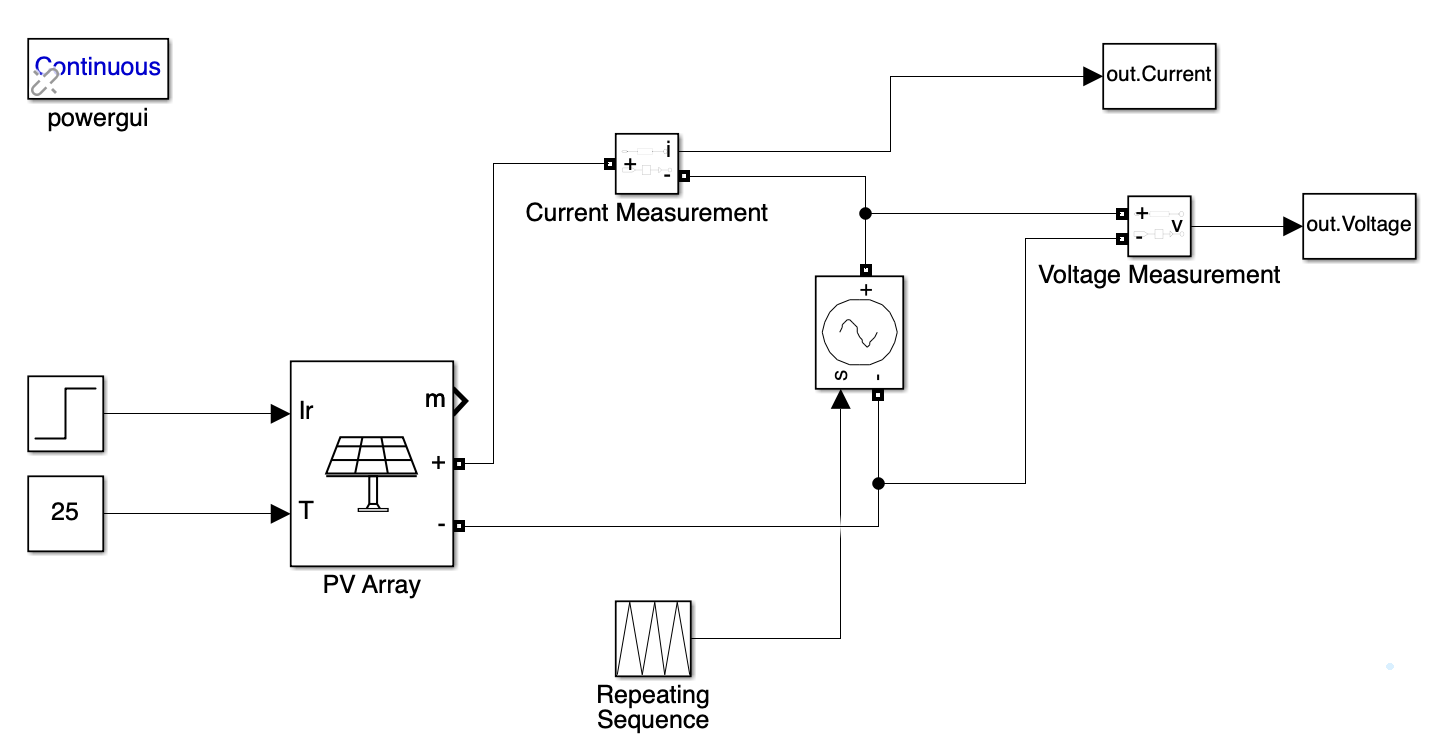} 
	\caption{Simulink Model to Generate IV-PV}
	\label{fig:model3} 
\end{figure}

\begin{figure}[H] 
	\centering  \includegraphics[width=0.9\columnwidth]{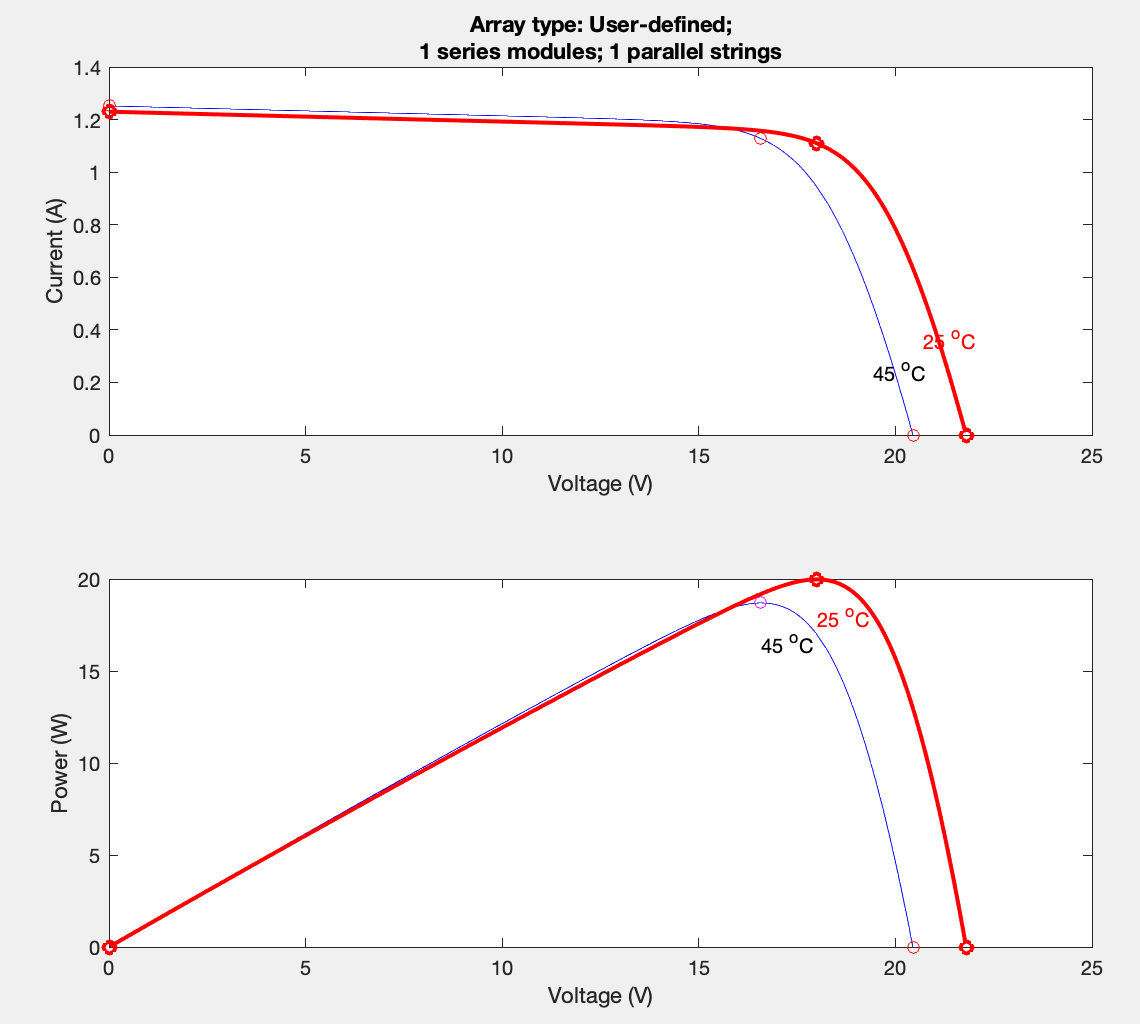} 
	\caption{IV and PV Performance of PV array @1000 $\text{W/m}^{2}$}
	\label{fig:model4} 
\end{figure}

\begin{figure}[H] 
	\centering  \includegraphics[width=0.9\columnwidth]{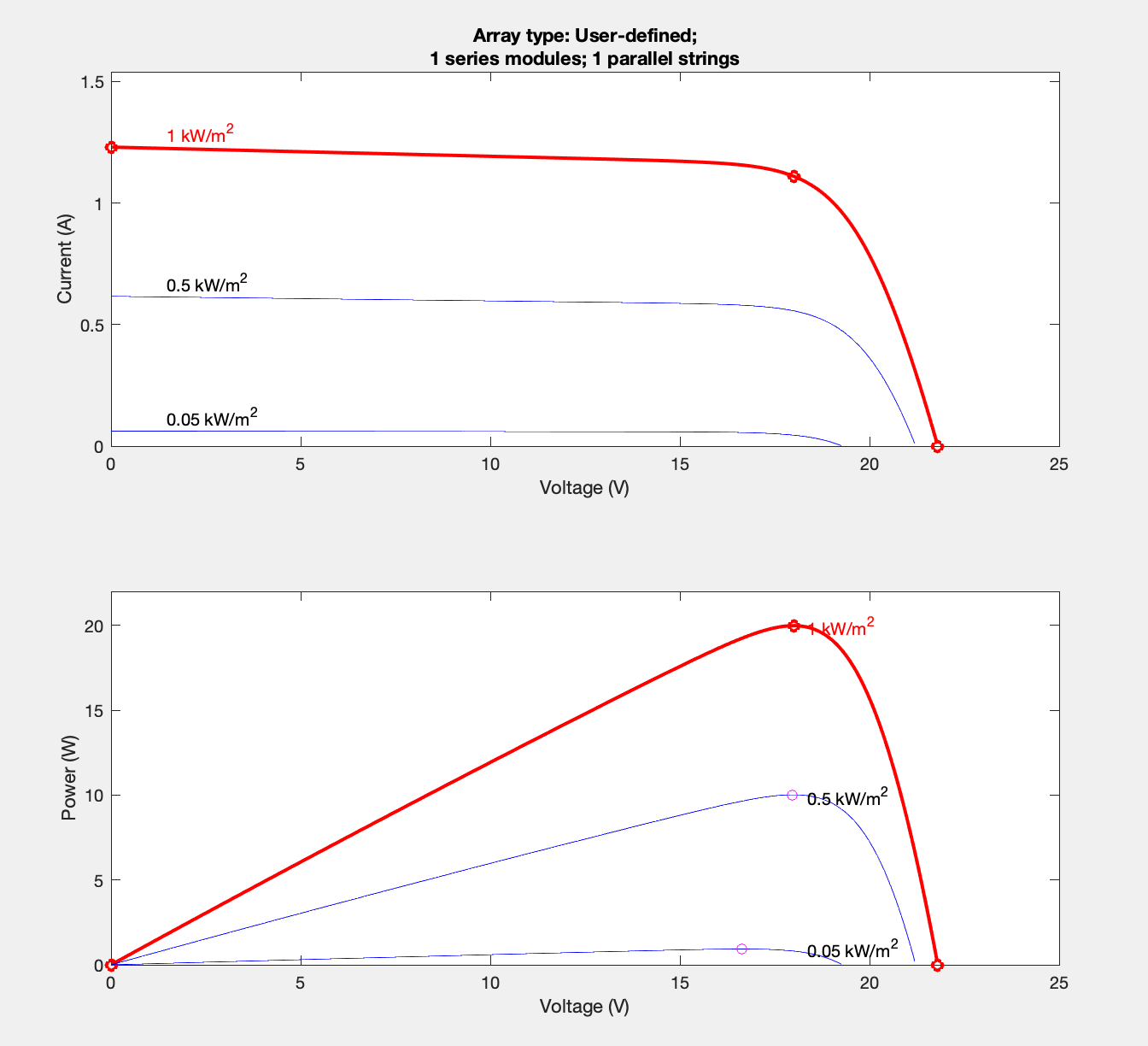} 
	\caption{IV and PV Performance @ $25^{\circ}$ C}
	\label{fig:model8} 
\end{figure}

    Figure \ref{fig:model3} shows the Simulink diagram to generate (current-voltage) IV and PV (power-voltage) characteristics of a PV array @ $1000 W/m^{2}$.  As shown, the higher the temperature, the lower the power and current at voltages greater than 30.
    Figure \ref{fig:model4} shows a plot of the IV and PV characteristics of the PV array @ $1000 W/m^{2}$. generated from the Simulink model MPPT algorithm assuming the maximum voltage of 200W.  As shown, the higher the temperature, the lower the power and current at voltages greater than 30.
     Figure \ref{fig:model8} shows plots of the PV array @ $25^{\circ}$ C and specified irradiances.  
    As the irradiance increases as measured in $\text{kW/m}^{2}$, the higher the maximum power point (MPP).
    %Figure \ref{fig:model4} shows a plot of the IV and PV curves generated from the Simulink model MPPT algorithm assuming the maximum voltage of 20W. 
%\begin{figure}[H] 
%	\centering  \includegraphics[width=0.9\columnwidth]{Images/curve1.png} 
%	\caption{IV and PV Performance of PV array @1000 $\text{W/m}^{2}$}
%	\label{fig:model4} 
%\end{figure} 
%     Figure \ref{fig:model8} shows plots of the PV array @ $25^{\circ}$ C and specified irradiances.  
%\begin{figure}[H] 
%	\centering  \includegraphics[width=0.9\columnwidth]{Images/curve2.png} 
%	\caption{IV and PV Performance @ $25^{\circ}$ C}
%	\label{fig:model8} 
%\end{figure} 
%    As the irradiance increases as measured in $\text{kW/m}^{2}$, the higher the maximum power point (MPP). 
\section{Motor Model}
\ \ \ A DC-DC boost converter increases the voltage from the PV generator side to the load side.  In general, the DC motor is proportional to the armature current and the strength of the magnetic field.   

\begin{figure}[H] 
	\centering     \includegraphics[width=1\columnwidth]{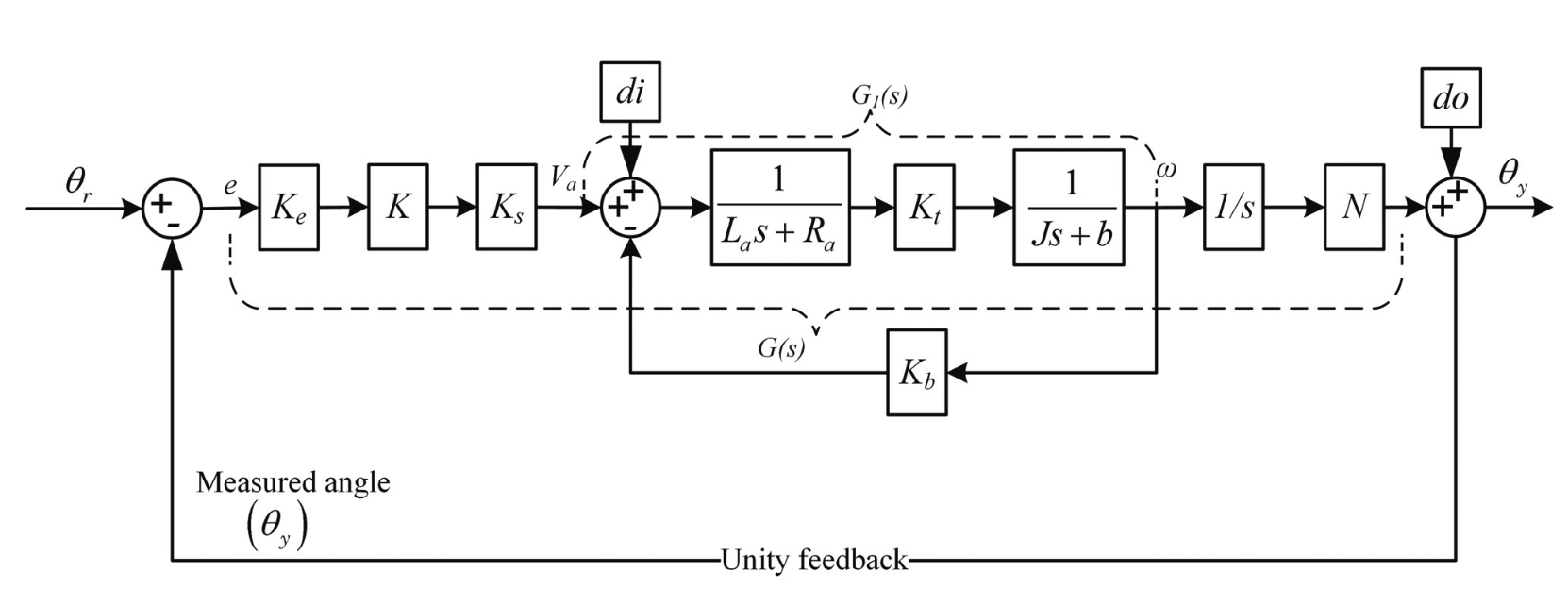} 
	\caption{Closed Feedback Loop for Solar Tracking System. Source: \cite{Hanwate:2018}}
	\label{fig:TF} 
\end{figure}

We use an armature-controlled motor with the following parameters:
\begin{itemize}
    \item Moment of inertia (J) = 0.01 $kg.m^{2}/s^{2}$ 
    \item Damping ratio ($\zeta$) = 0.1 Nm 
    \item Electromotive force constant ($K_{t}$) = 0.0125 Nm/Amp 
    \item Back emf (motor torque) constant ($K_{e}$) = 0.0125 s/rad 
    \item Amplifier gain ($K_{a}$) = 10,000
    \item Electric resistance ($R$) = 1 Ohm 
    \item Electric inductance ($L$) = 0.5H 
    \item Input voltage ($V$) 
    \item Output angle ($\theta$) 
    \item Input angle ($\theta_{r}$)
    \item Friction coefficient ($b$) = $10^{-6}$ Nm
    \item Gear ratio ($N$)
\end{itemize}
    The damping ratio $\zeta$ is also the motor viscous friction constant.  The following motor equations are used to model the DC motor: %with the following parameters.
    The back electromotive force (emf) $e$ is proportional to the angular velocity of the shaft by a constant factor $K_{e}$:
\begin{equation}
    e = K_{e}\dot{\theta}
\end{equation}    
and the torque $\tau$ is:
\begin{equation}
    \tau = K_{t}i
\end{equation}
    In SI units, $K_{t} = K_{e}$.  Thus, $K$ can be used to represent them both.
    The following equations are derived based on Newton's second law and Kirchhoff's voltage law:
\begin{align}
    J \ddot{\theta} + b\dot{\theta} &= Ki \\
    L\frac{di}{dt} + Ri &= V - K\dot{\theta}
\end{align}
    Taking the Laplace transform, we rewrite these equations as:
\begin{align}
    s(Js + b)\theta(s) &= KI(s) \\
    (Ls + R)I(s) &= V(s) - Ks\theta(s) 
\end{align}
yielding the following transfer function:
\begin{equation}
     \frac{\theta(s)}{V(s)} = \frac{K}{(Js + b)(Ls + R) + K^{2}} 
     \label{eq:tf4}
\end{equation} 
    Integrating eq. \ref{eq:tf4}, the position control transfer function may be written as \cite{Chin:2011}:
\begin{equation}
    \frac{\theta(s)}{V(s)} = \frac{1}{s} \bigg [ \frac{K_{t}}{(Js + K_{e})(Ls + R) + K_{t}K_{e}} \bigg ].  \nonumber
%\begin{align}
    \label{eq:tf1}
\end{equation}
    yielding
\begin{equation}
         \frac{\theta(s)}{V(s)} = \frac{K_{t}}{LJs^{3}+(L+RJ)s^{2} + (R + K_{t}K_{b})s}  
\end{equation}
    The transfer function units are in $\frac{\text{rad/sec}}{V}$.  
       The solar tracker interfaces with some additional components such as error discriminator ($K_{d}$), amplifier gain ($K_{a}$), servo amplifier ($K_{s}$) and gear ratio (N). If all these parameters are considered, then open loop transfer function for this sun
    tracker system becomes
    \small
    \begin{equation}
        \frac{\theta(s)}{\theta_{r}(s)} = \frac{K_{s}K_{a}K_{d}K_{t}N}{LJs^{3} + (bL + RJ)s^{2} + (Rb + K_{t}K_{e})s} 
    \end{equation}
    Figure \ref{fig:TF} shows the closed feedback loop model of the solar tracking system.
    Using the given parameter values, the transfer function becomes:
\begin{equation}
    G(s) = \frac{0.0001563}{1.2e^{-8}s^3 + 7.51e^{-6}s^2 + 0.0001625s}
    \label{eq:motortf}
\end{equation}
\normalsize
    We can also write the motor equations as:
\begin{align}
   \frac{di}{dt} &= \frac{1}{L} \big (V - K_{e}K_{f}\frac{d\theta}{dt} - Ri \big ) \\
    \tau &= K_{t}K_{e}i
\end{align}
%where $T$ is the torque.  
    The PV panel equation is:
\begin{equation}
    \frac{d^2{\theta}}{dt^2} = \frac{1}{J}(\tau - Kd )\frac{d\theta}{dt}
\end{equation}
where $d$ is the motor displacement.
    The DC motor model can be modeled using a PID controller and look-up tables.  The look-up table uses the PWM as an input to rotate the motor to a pre-determined angle. For instance, if the pulse width changes from 1.25 ms to 1.75 ms, the panel angle changes from $0^{\circ}$ to $180^{\circ}$ in a linear manner \cite{Chin:2011}. The second look-up table on the feedback path provides the actual pulse width results.  The actual and desired pulse-width are compared to obtain an error signal for a Proportional-Integral-Derivative (PID) controller to drive the motor to the desired angle \cite{Chin:2011}.
    
    Stepper motors are typically operated in open-loop systems due to their microstepping with discrete steps, but they can also used in closed-loop systems such as with encoder feedback.  The benefits of using feedback with stepper motors include not losing position when overloaded, handling higher torque loads as the feedback keeps the magnetic force optimally aligned, and the motor runs cooler sicne the feedback adjusts control current depending on the load.  The disadvantage is that they closed-loop stepper motors are more expensive than their open-loop counterparts.

\subsection{Transfer Function}

\begin{figure}[H] 
	\centering     
     \includegraphics[width=0.9\columnwidth]{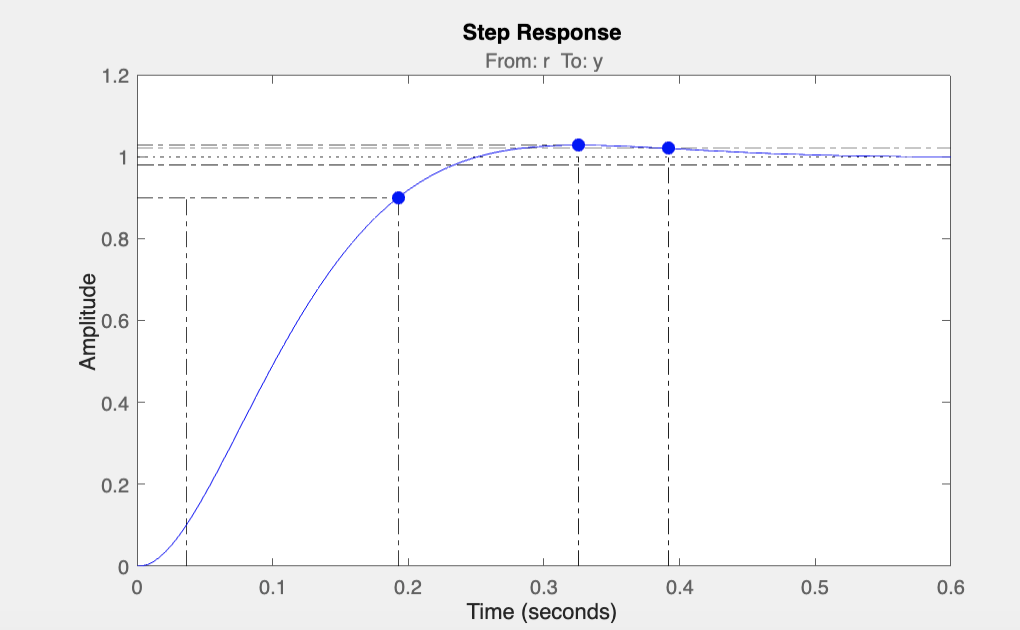} 
	\caption{Step Response for $K=1e^5$}
	\label{fig:step} 
\end{figure}

\begin{figure}
    \centering
    \includegraphics[width=0.8\columnwidth]{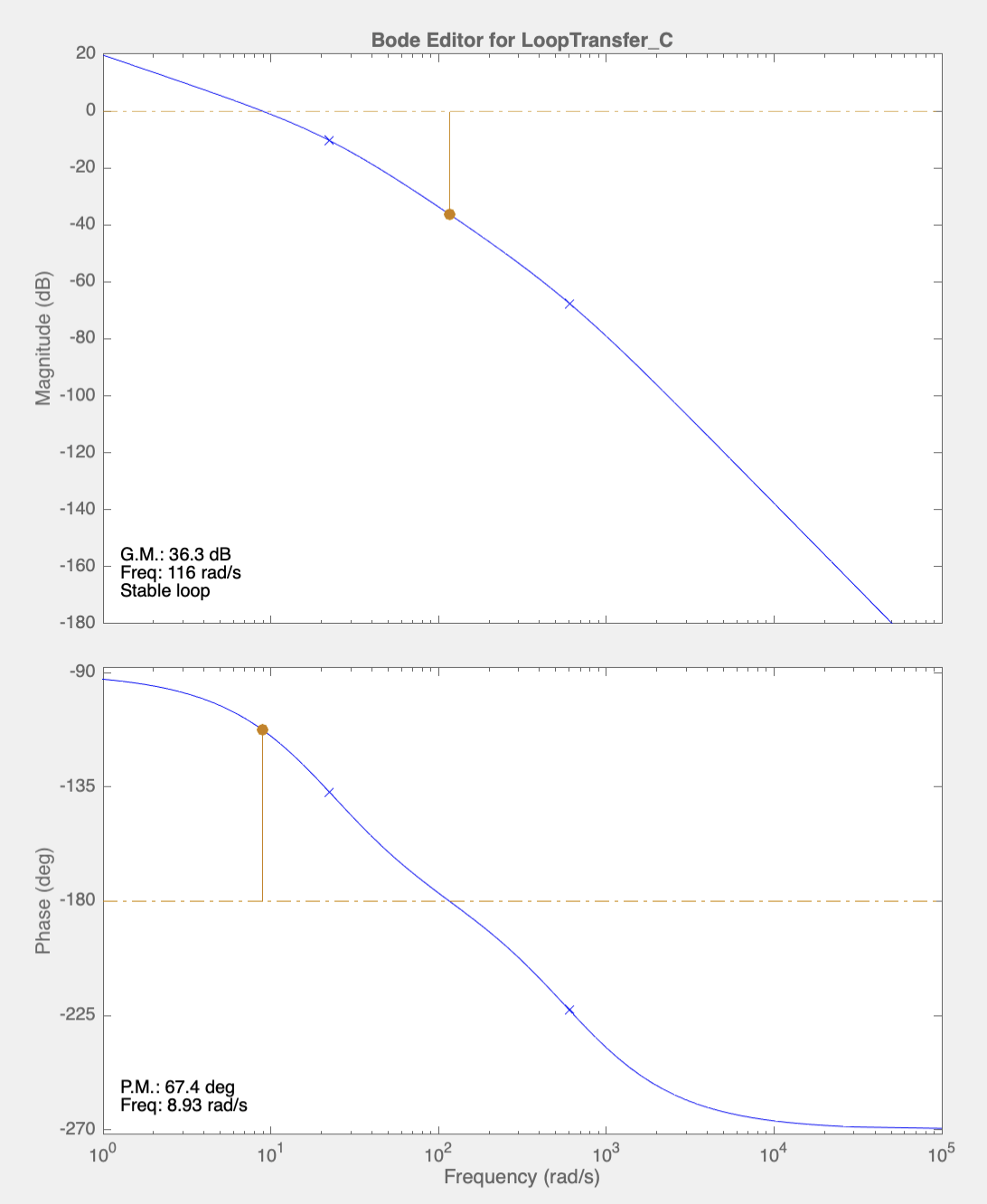} 
    \caption{Bode Plot}
    \label{fig:bode2}
\end{figure}

Figure \ref{fig:step} shows the step response of the closed loop transfer function in eq. \ref{eq:motortf} if we use a gain of $K = 1e^5$. 
    The rise time of the step response in Figure \ref{fig:step} is 0.156 seconds, the overshoot is 2.79\% at 0.325 seconds, and the steady state is 1.  The DC gain of the open-loop transfer function is 0.962 while it is 0.000156 if the transfer function has closed-loop feedback.
    Figure \ref{fig:bode2} shows a Bode plot of the transfer function.
    The gain magnitude is -36.3 dB a frequency of 116 radians/sec.  The phase magnitude is $67.4^{\circ}$ at 8.93 rad/sec.
    
\begin{figure}[H] 
	\centering     
    \includegraphics[width=0.9\columnwidth]{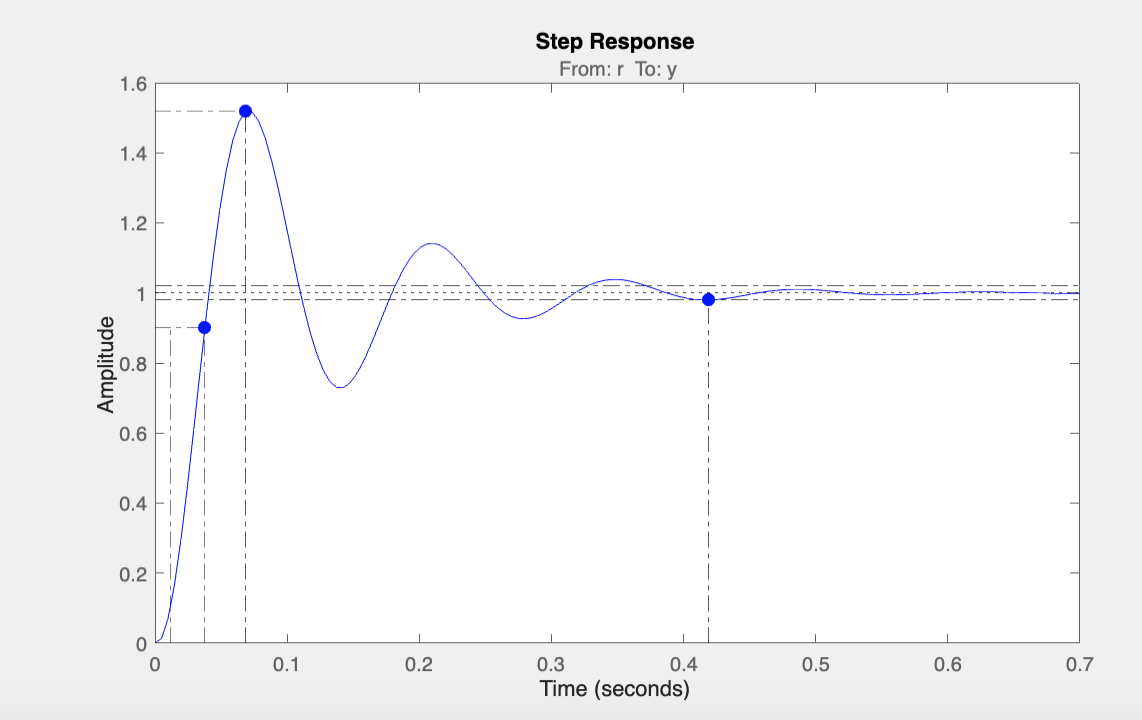} 
	\caption{Step Response for $K=1e^6$} 
	\label{fig:step2} 
\end{figure}

    Figure \ref{fig:step2} shows the step response for a gain parameter of $K = 1e^6$.   
    The rise time of the step response in Figure \ref{fig:step2} is 0.0264 seconds, the overshoot is 51.7\% at 0.0687 seconds, the peak is 1.52, the settling time is 0.419 sec, and the steady state value is 1.  The gain margin is 16.3 dB @ 116 rad/s and the phase margin is 23 deg @ 43.8 rad/s.  Thus, increasing the magnitude of the gain reduces the rise time, but substantially increases overshooting.

\begin{figure}[H] 
	\centering     
    \includegraphics[width=0.9\columnwidth]{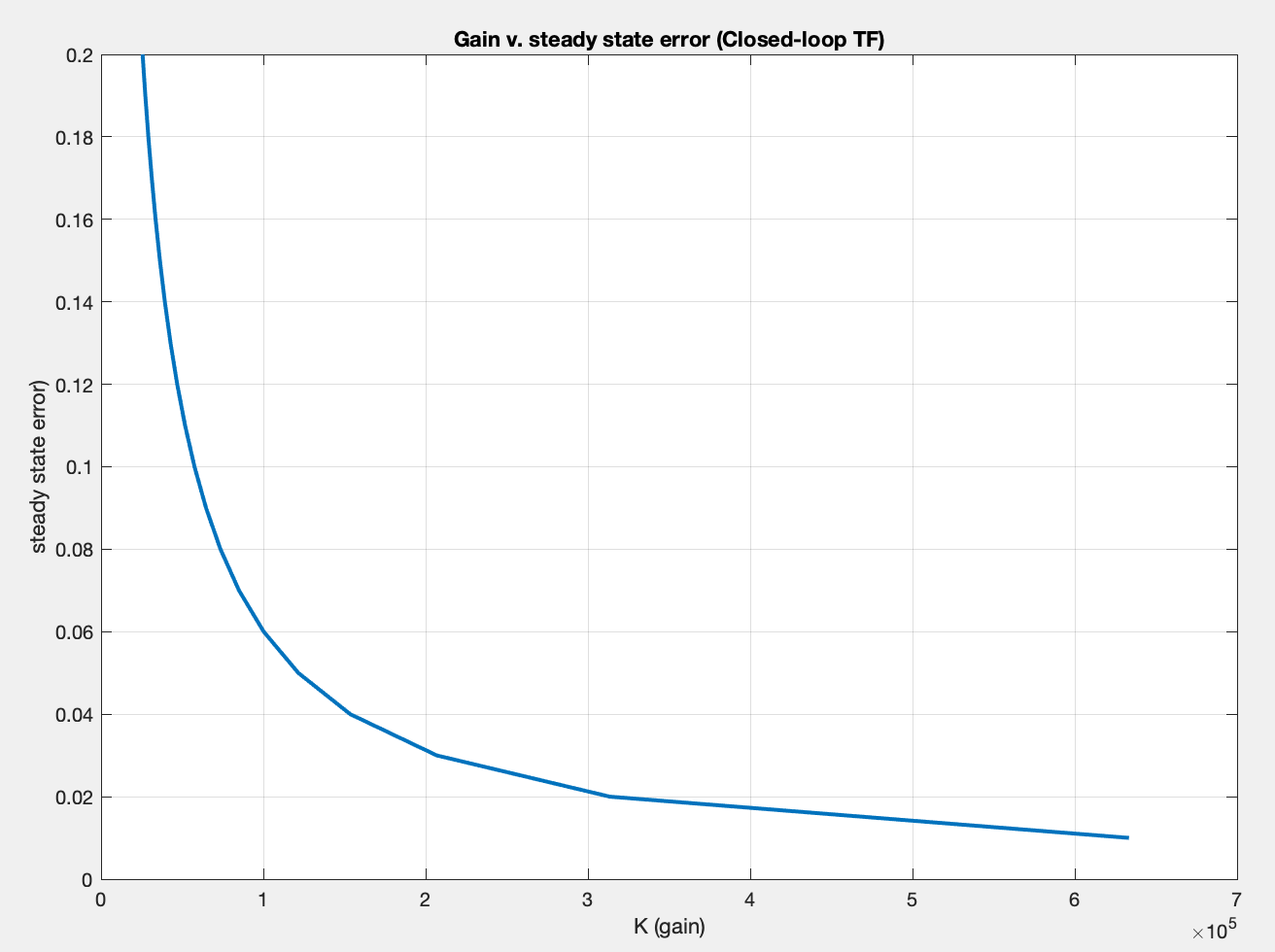} 
	\caption{K v. Steady-state error} 
	\label{fig:closed} 
\end{figure}

    Figure \ref{fig:closed} shows the tradeoff of gain ($K$) and steady-state error $e$.  
    As expected, there is an inverse relationship between $K$ and steady-state error.   For a desired steady-state error of 0.1 of a closed-loop transfer function with a step input, $K=57,582$.   However, for the open-loop transfer function, $K = 9.36$.  If a steady-state error $e(\infty)$ = 0.01 is desired, $K$ is required to be 102.9 for the open loop system and $K$ is required to be 633,400 for the closed loop system.  
    
\subsection{PID Controller}

\ \ \ We can use a PID controller for the DC stepper motor to provide fast and stable responses, reduce the steady-state error, and minimize overshooting and oscillations. The PID controller can be expressed as
\begin{equation}
    C(s) = \frac{K_{d}s^{2} + K_{p}s + K_{i}}{s} 
\end{equation}
where $K_{p} = 9.51202$, $K_{d} = 0.00022$, and $K_{i} = 5.6443$ \cite{Hanwate:2018}.   The closed-loop characteristic equation for the PID controller $C(s)$ and plant $G(s)$ is 
\begin{equation}
    \Delta(s) = 1 + G(s)C(s)
    \label{eq:line}
\end{equation}
or after simplifying eq. \ref{eq:line} and equating to zero \cite{Hanwate:2018},
\begin{multline}
    s^4 + \frac{b_{2}}{b_{1}}s^3 + \bigg ( \frac{b_{3} + a_{4}K_{d}}{b_{1}} \bigg )s^{2} \\ + \bigg (\frac{b_{4} + a_{4}K_{p}}{b_{1}} \bigg )s + \frac{a_{4}}{b_{1}K_{i}} = 0
\end{multline}
\normalsize
where $a_{4} = K_{s}K_{a}K_{e}K_{t}N$, $b_{1} = LJ$, $b_{2} = (bL+RJ)$, $b_{3} = (Rb + K_{t}K_{b})$, and $b_{4} = 0$.  

\begin{figure}[H] 
	\centering     
     \includegraphics[width=0.9\columnwidth]{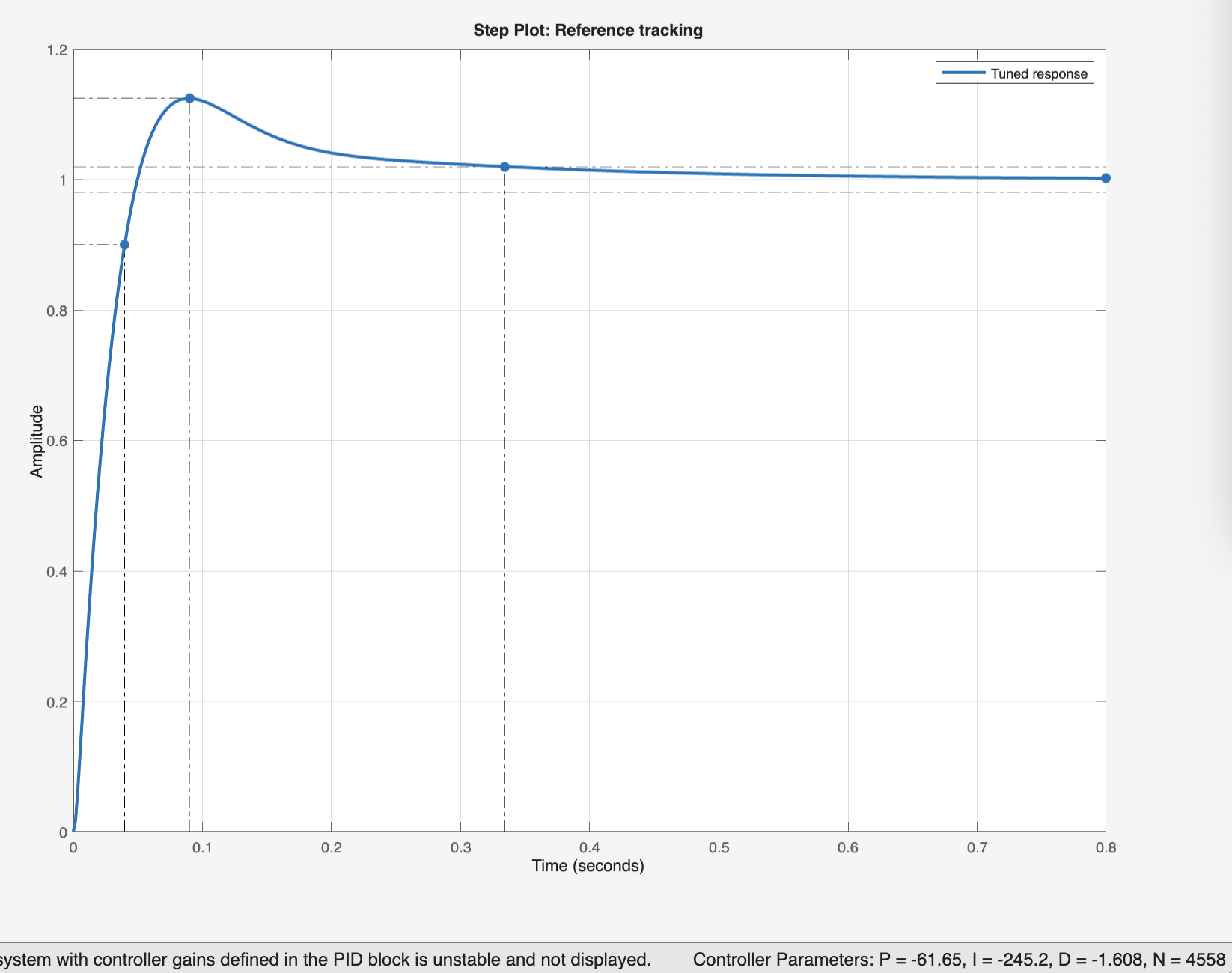} 
	\caption{PID-tuned step response}
	\label{fig:bodemotor1} 
\end{figure}

Plugging in the given values, we get the characteristic equation
\small
\begin{equation}
s^4 + 625.8s^3 + 1.382e4s^2 + 1.239e7s + 7.349e6 = 0 \nonumber 
\end{equation}
\normalsize
    Using the Routh-Horwitz Table \ref{table:table2}, we find that the system is stable.  
\begin{table}
\centering
\begin{tabular}{c | c  c }
\hline
    coeff. & 1.0e+7*   & \\
\hline
   $a_{3}$ & 0.0001  & 1.238 \\
   $a_{2}$ & 0.0014 & 0.7349 \\
   $a_{1}$ & 1.2052 &  0 \\
   $a_{0}$ & 0.7349 &  0  \\
   \hline 
\end{tabular}
\caption{Routh-Hurwitz table} 
\label{table:table2}
\end{table}
    %The mathematical 
\normalsize
    Figure \ref{fig:bodemotor1} shows a PID-tuned plot of the step response.  The tuned control parameters are $K_{p} = -69.94$, $K_{i} = -729.46$, $K_{d} = -1.651$, and $N = 4558.36$.  
    The performance and robustness of the PID controller step response generated by Simulink PID Tuner is provided in Table \ref{tab:table1}.
%\begin{figure}[H] 
%	\centering     
%     \includegraphics[width=0.9\columnwidth]{Images/nums.png} 
%	\label{fig:nums} 
%\end{figure}
\begin{table}[H]
\begin{flushleft}
\centering
\begin{tabular}{ | c | c | }
\hline
   \textbf{Metric}  & \textbf{Tuned} \\
\hline 
    Rise Time & 0.0303 sec \\
    \hline
    Settling Time & 0.179 sec \\
    \hline
    Overshoot & 23.2\% \\
    \hline
    Peak & 1.23 \\
    \hline
    Gain margin & 42.8 dB @ 1.63e+03 rad/s \\
    \hline 
    Phase margin & 60 deg @ 39.9 rad/s \\
\hline
\end{tabular}
\end{flushleft}
\caption{Performance and Robustness}
\label{tab:table1}
\end{table}

\subsection{Root Locus and Bode Plots}
\begin{figure}[H] 
	\centering     
     \includegraphics[width=0.9\columnwidth]{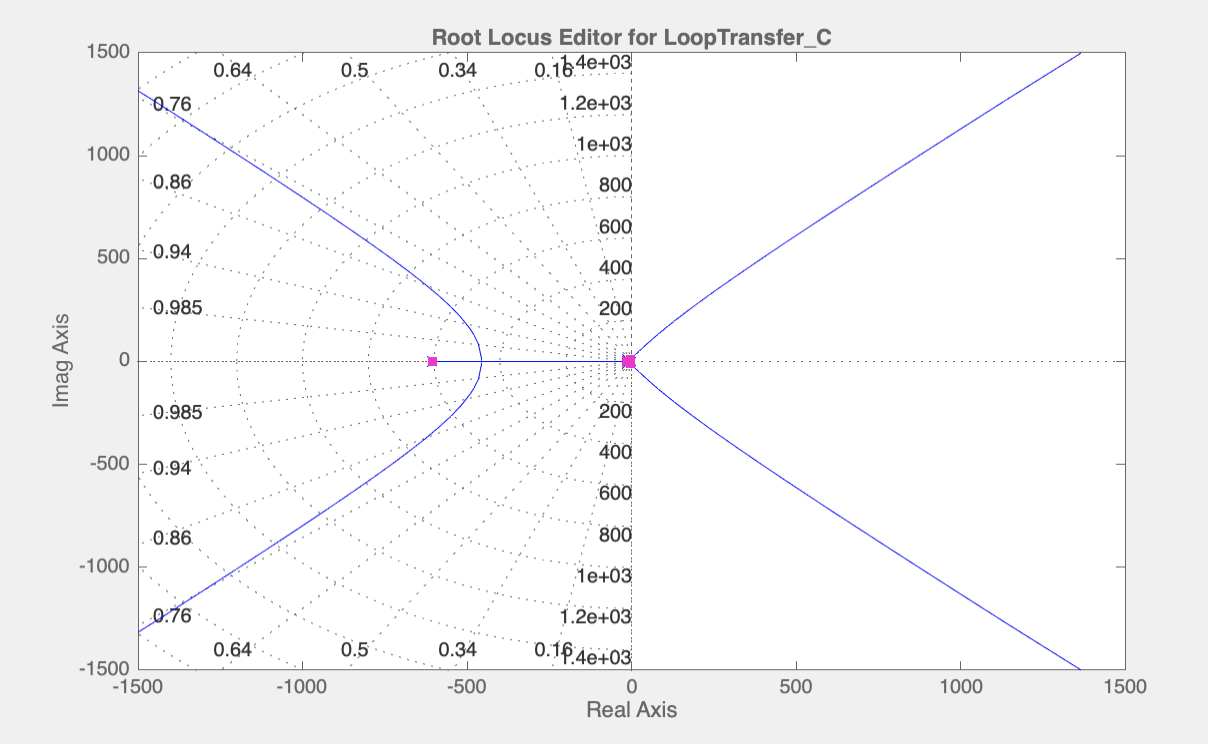} 
	\caption{Root Locus}
	\label{fig:rlocus1} 
\end{figure}

\begin{figure}[H] 
	\centering     
     \includegraphics[width=0.9\columnwidth]{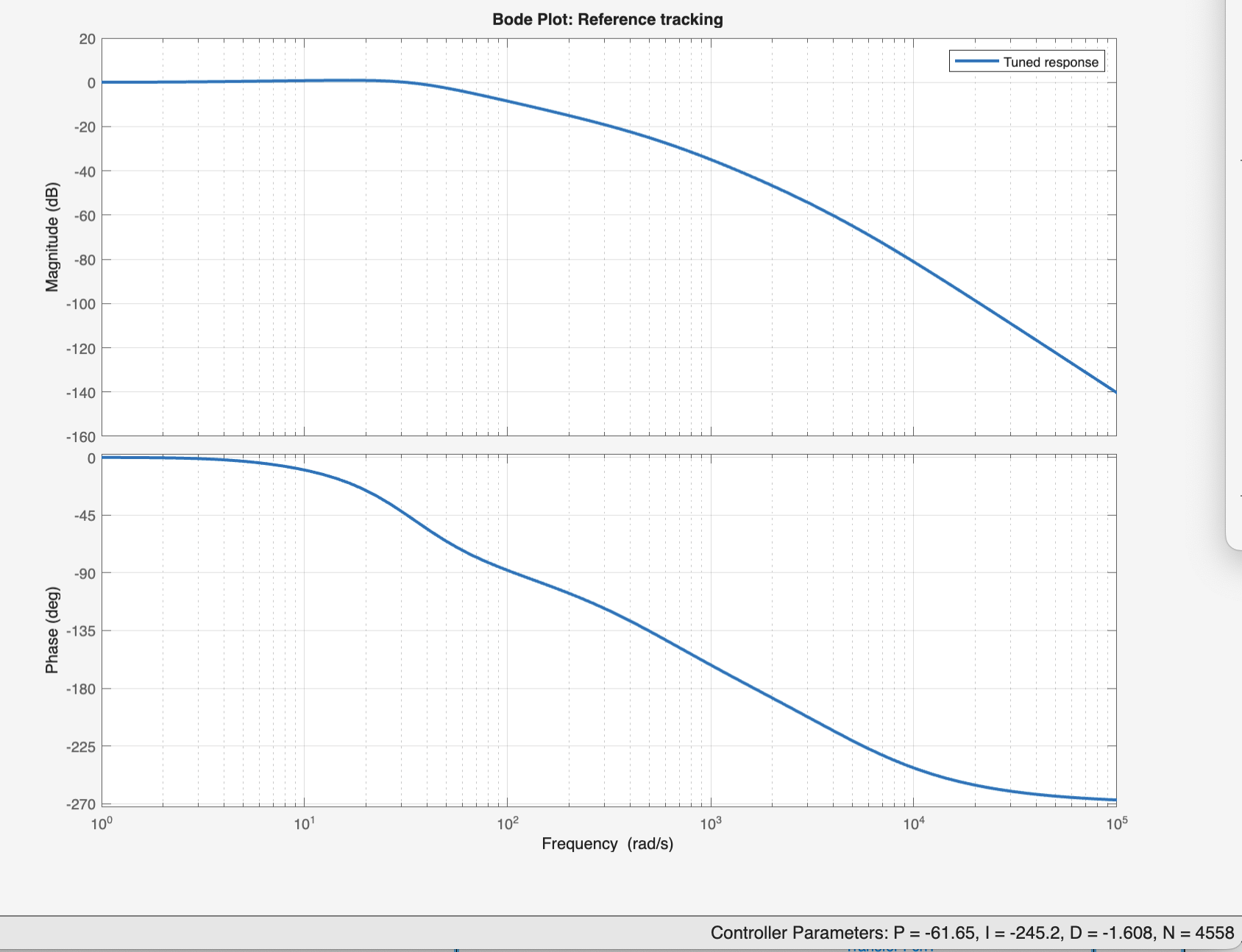} 
	\caption{PID-tuned Bode plot}
	\label{fig:bodemotor2} 
\end{figure}

    Figure \ref{fig:rlocus1} shows a root locus plot for the transfer function.  The system has 3 real zeros and 1 real pole at $p = -650$.  Since all the zeros and poles are to the left side of the imaginary axis, the system is stable.
    Figure \ref{fig:bodemotor2} shows the PID-tuned Bode plots using the same control parameters used in Figure \ref{fig:bodemotor1}.
    As Figure \ref{fig:bodemotor2} shows, the PID controller gives higher gain and phase magnitudes for the same level for frequency as those in the Bode plots in Figure \ref{fig:bode2}.

\section{System Model}
\begin{figure}[H] 
	\centering     
    \includegraphics[width=0.9\columnwidth]{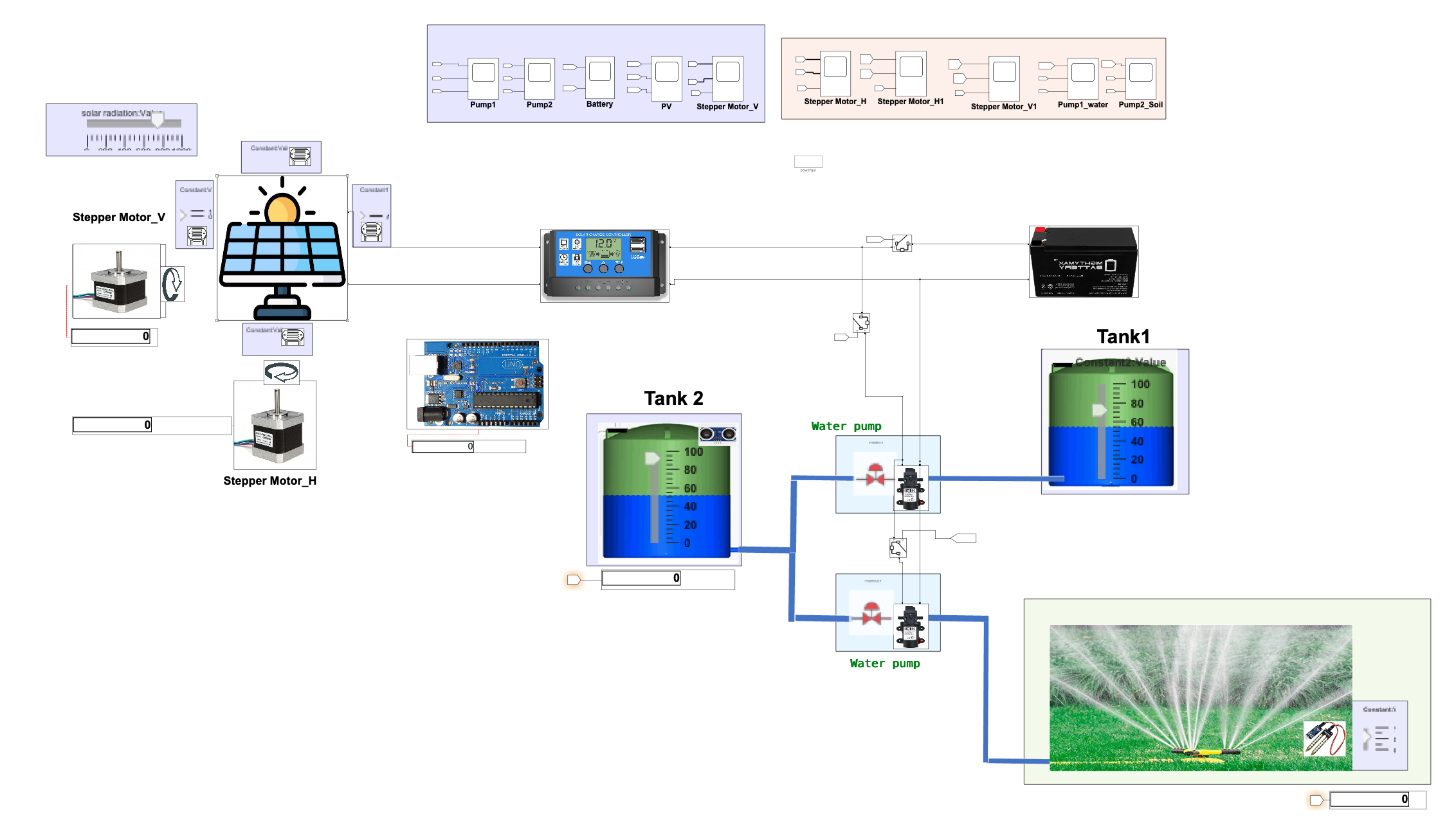} 
	\caption{Model of System}
	\label{fig:model2} 
\end{figure}

\begin{figure}[H] 
	\centering     
    \includegraphics[width=0.9\columnwidth]{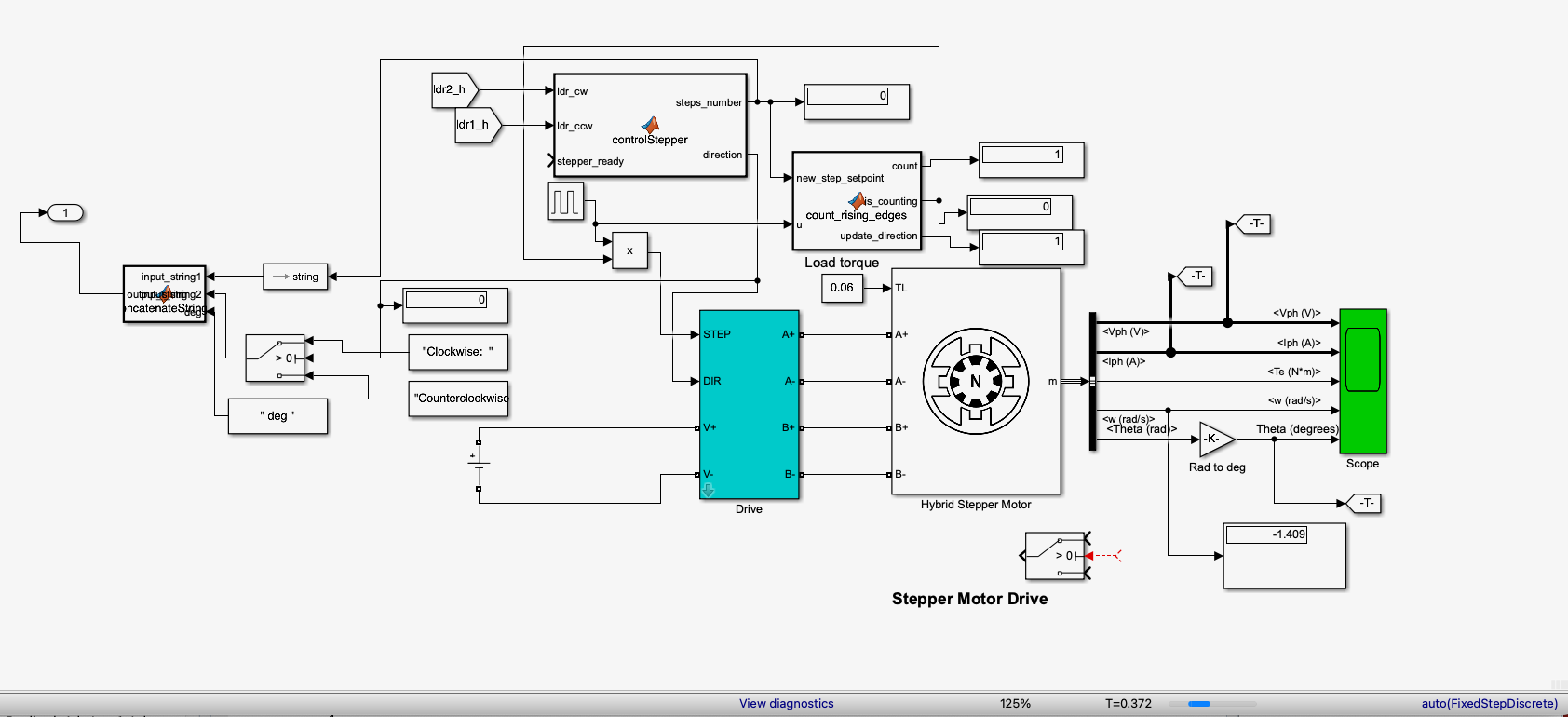} 
	\caption{Simulink Model}
	\label{fig:model1} 
\end{figure}

\begin{figure}[H] 
	\centering     
    \includegraphics[width=1\columnwidth]{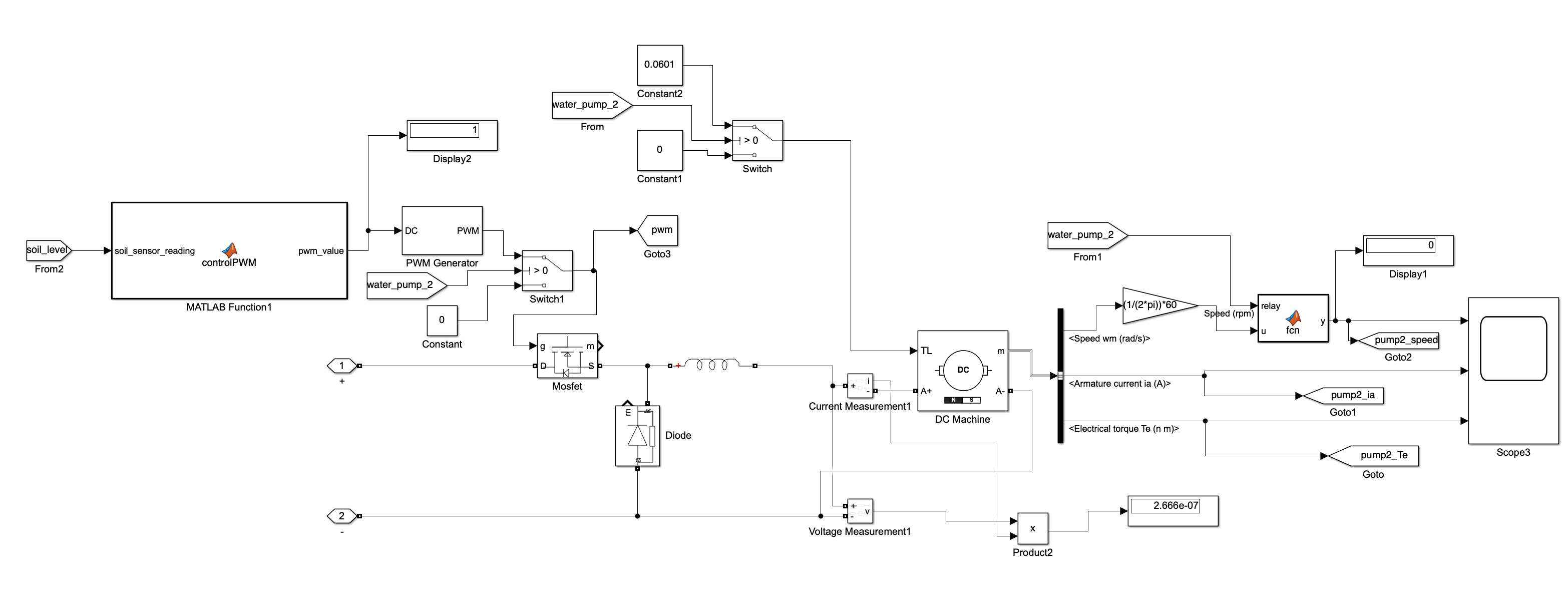} 
	\caption{Subsystem for PWM}
	\label{fig:model10} 
\end{figure}

\begin{figure}[H] 
	\centering     
    \includegraphics[width=1\columnwidth]{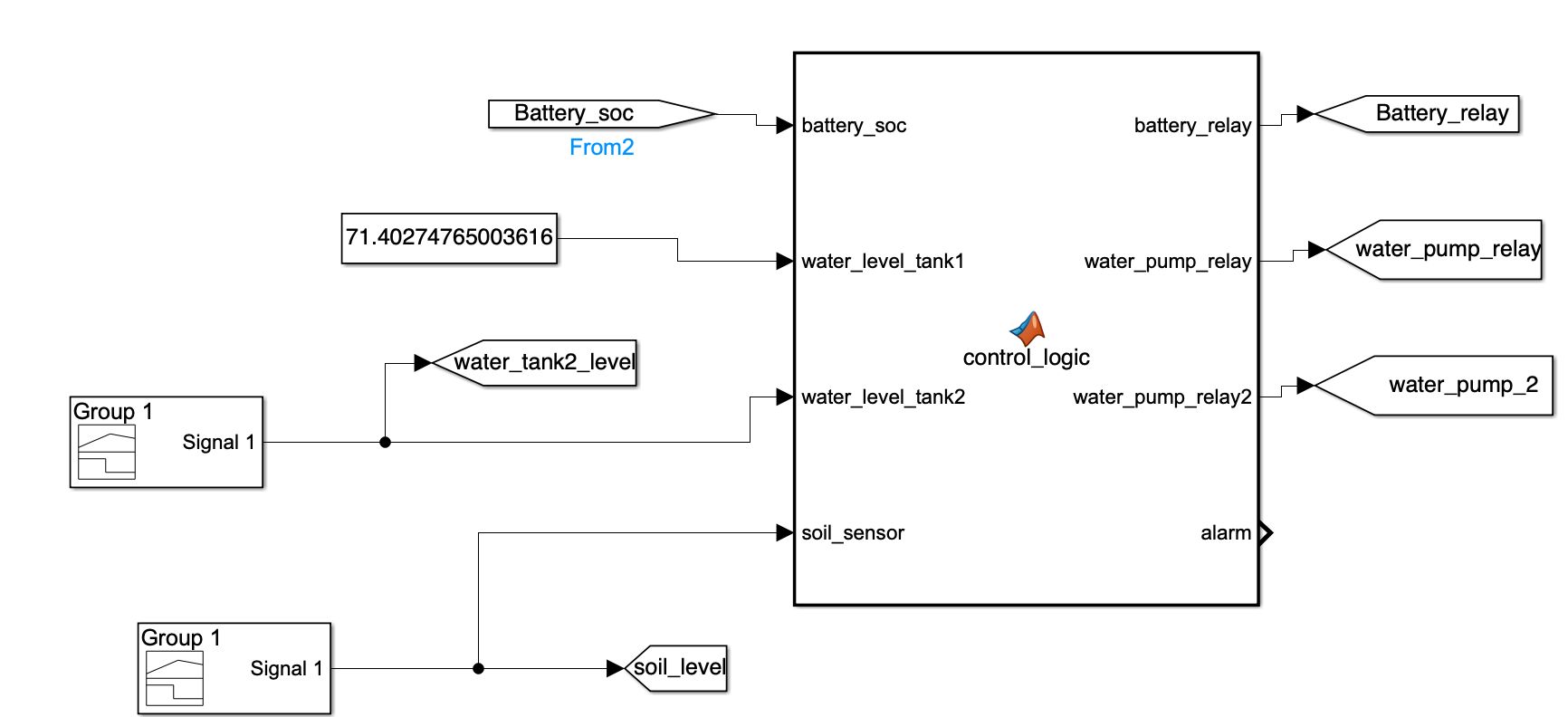} 
	\caption{Subsystem for PWM}
	\label{fig:model11} 
\end{figure}

%    Figure \ref{fig:model} shows a high level schematics of the system.
%\begin{figure}[H] 
%	\centering     
%    \includegraphics[width=0.9\columnwidth]{Images/diagram0.png} 
%	\caption{Simulink}
%	\label{fig:model} 
%\end{figure}
    Figure \ref{fig:model2} shows high level schematics of the system.  It incorporate two water pumps and two motorized ball valves that are connected to each other through a Y-connector adapter. Water is pumped to the second tank from the first. Water is then pumped from the second tank to the plant if the soil moisture falls below a certain threshold. 
    Figure \ref{fig:model1} shows the detailed Simulink system model.
    Figure \ref{fig:model10} shows a detailed Simulink subsystem for pulse width modulation (PWM) of the water pumps based on the soil sensor input level.  The subsystem uses a DC machine (motor) that implements a permanent magnet field with a torque constant of 0.0255 N m/A, a total inertia J = 0.0008 \ $\text{kg}/\text{m}^2$, armature inductance of 0.01 H, resistance of 1 Ohm, a viscous friction coefficient of 0.00001 Nms, and a Coulomb friction torque of 0.0001 Nm. 
    The subsystem includes a PWM generator that performs discrete sampling of the analog signal with a timer period of 0.01 seconds and a sample time of 0.00006.
    Figure \ref{fig:model11} shows the control logic module that includes the battery relay, water pump relays, water tank levels, soil sensor, and battery state of charge.
\subsection{Water Pump}
    Water pumps are typically operated by three-phase AC induction motors due to their high efficiency and potential to deliver the required flow rate for specific head.  Let $P$ be the power required by the motor, $H$ the power required by the motor pump, and $\epsilon$ the efficiency of the water pump.  Then 
    $P = \frac{H}{\epsilon}$.  Thus, the $\epsilon$ is the ratio of output power to input power.
    
    The current system does not require high flow rates or centrifugal pumping.  Thus, we do not need three-phase AC induction as would be the case for deep well or hydraulic pumping at a water treatment plant.    

    The current system uses two 12V miniature water diaphragm pumps and two valves used.  %One water pump diaphragm pump is 12V, with a water flow of 3L/PM 0.8GPM and a water pressure 100 Psi (7.0 bar) while the second that 
    The first water pump pumps water the first tank from the second tank based on the water level in the second tank.  The second water pump pumps water from the second water tank based on the soil moisture level in the plant through a hose connected to a spray nozzle.  Each water pump has a water flow of 5 LPM (1.35 GPM), and water pressure of 116 Psi (8.0 bar).  
    
    Each water pump is connected 3/4" motorized ball valves with open/close times of 3-5 seconds.  The valves a have a max power of 5 W with 9-24V AC/DC, max torque of 2 Nm, and max pressure of 145 PSI.  Each water pump is connected to a 5V relay switch module.      
    %second water pump is $G_{2} = \frac{5}{1 + 0.1s}$.  Both water pumps have constant gains.

    Each water tank is $17.32 \ \text{in} \times 11.81 \ \text{in.} \times 11.81 \ \text{in.}$. (2415.73 $\text{in}^{3} \approx 1.4 \ \text{ft}^3$) with a volume of 10.44 gallons or 39.5 L.  Let $q$ = flow rate (measured in liters per minute), V = empty volume of the water tank (in liters), and  $\tau$ = time in seconds.  Then $q = V/\tau$ or $5 = 39.5/\tau$ so that $\tau$ = 7.91 minutes = 475 seconds. 
    
    Therefore, the transfer function of each water pump is $G_(s) = \frac{K}{1 + \tau s} =  \frac{5}{1 + 475s}$ where $K$ is the constant gain of 5 and $\tau$ is 475. The natural frequency $\omega_{n}$ = 2.23 and damping ratio is $\zeta$ = 0.005.  The system is underdamped since $\zeta < 1$ but very close to being undamped as it is close to 0.  
    
\begin{figure}[H] 
	\centering     
    \includegraphics[width=0.9\columnwidth]{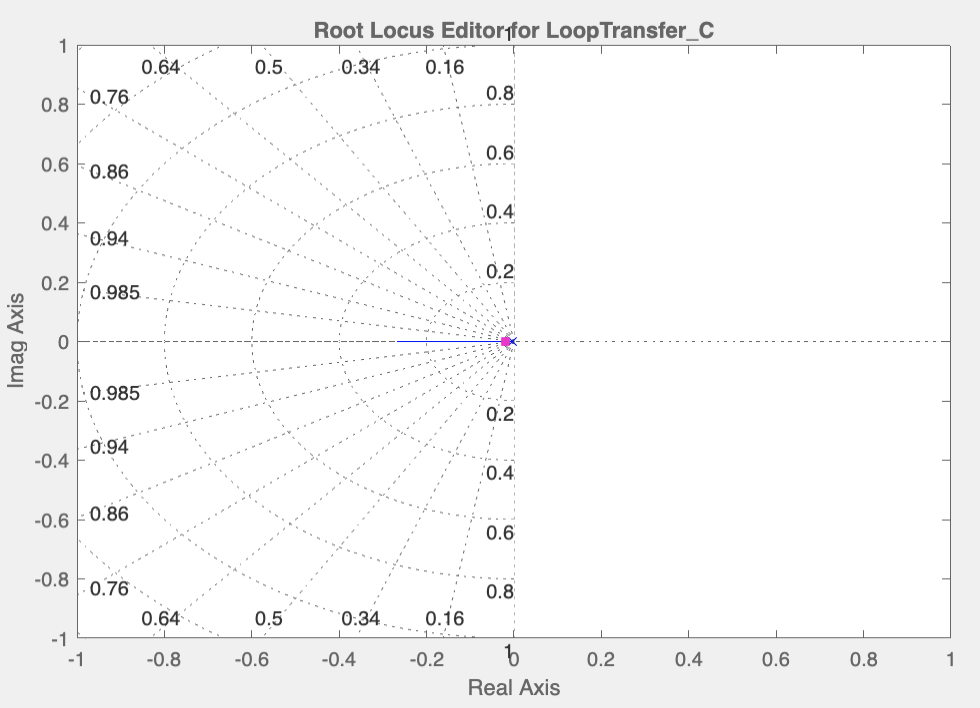} 
	\caption{Root Locus of water pump transfer function}
	\label{fig:rootlocus} 
\end{figure}

\begin{figure}[H] 
	\centering     
    \includegraphics[width=0.8\columnwidth]{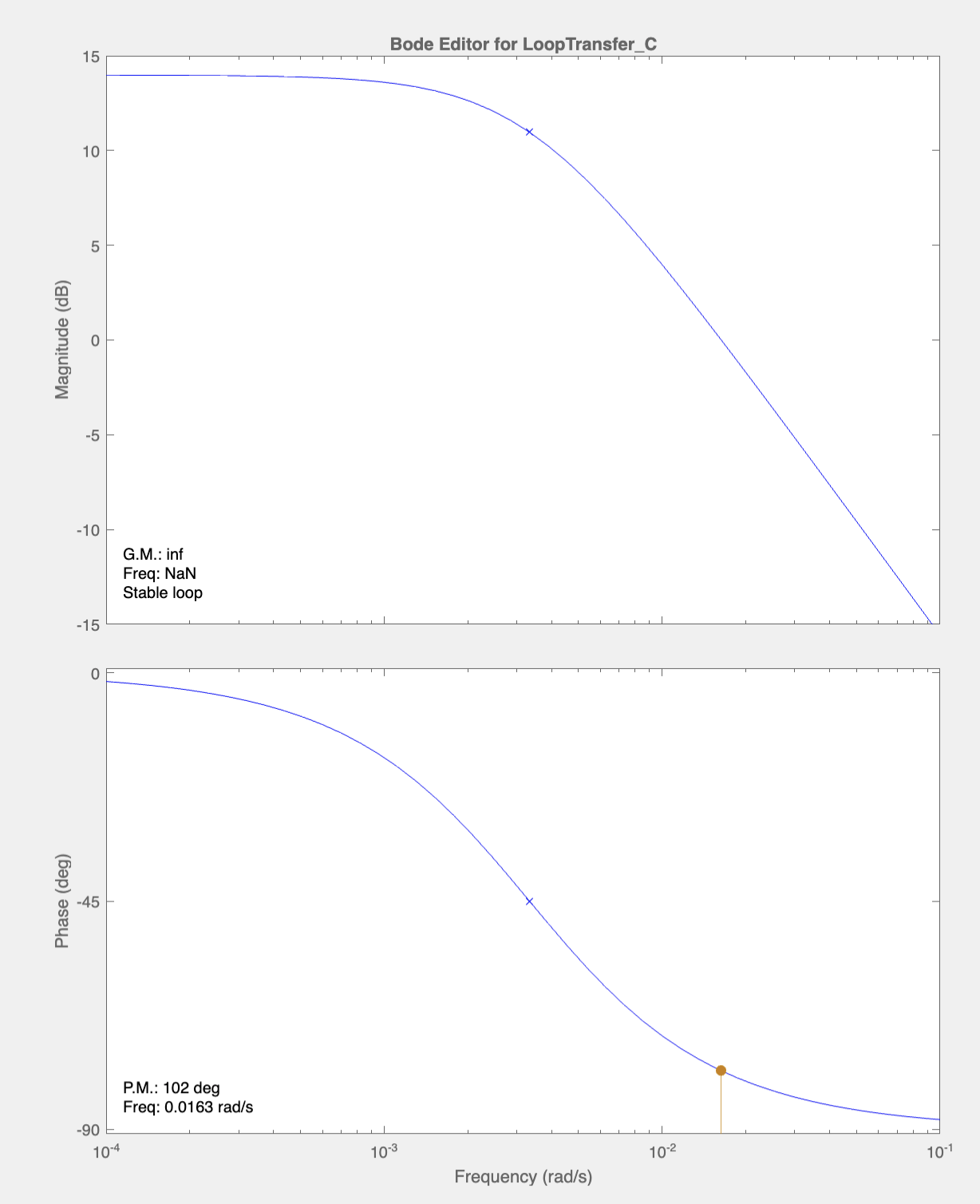} 
	\caption{Root Locus of water pump transfer function}
	\label{fig:waterpump2} 
\end{figure}

    Figure \ref{fig:water} shows the step response of the water pump assuming a unit step input.  The rise time is 174 seconds, the settling time is 310 seconds, and the steady state error $e(\infty)$ = $\frac{5}{1 + \underset{s\rightarrow 0}{\text{lim}}G(s)}$ = 0.833 so that the position constant $K_{p} = 5$. The velocity constant $K_{v} = 0$ and acceleration constant $K_{a} = 0$. The time constant is defined as the time it take for the step response to reach 63.2\% of the final value.    Looking at the step response, the time constant is 112 seconds. 
\begin{figure}[H] 
	\centering     
    \includegraphics[width=0.9\columnwidth]{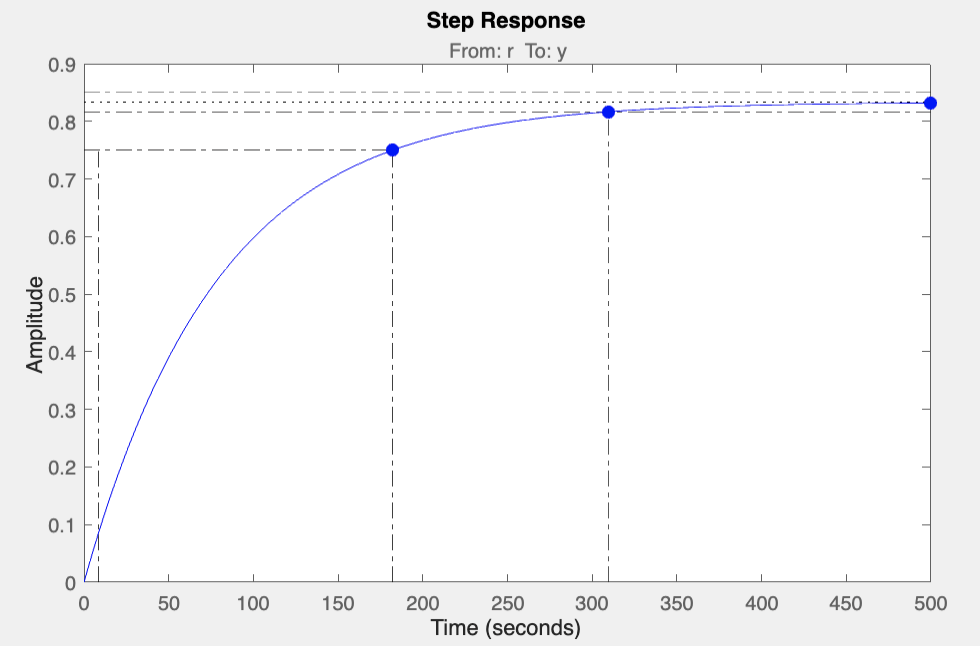} 
	\caption{Step response of water pump}
	\label{fig:water} 
\end{figure}
    Figure \ref{fig:rootlocus} shows the root locus plot.  The root locus has a zero at 0.6925 and a pole at zero so the sytem is stable. Figure \ref{fig:waterpump2} shows the Bode plots.
\subsection{Water Tank}
\begin{figure}[H] 
	       \centering     
          \includegraphics[width=0.9\columnwidth]{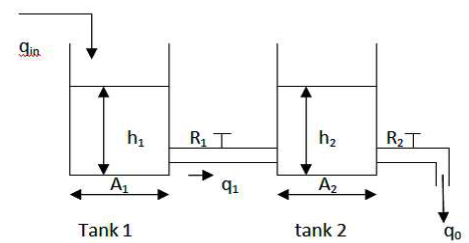} 
	       \caption{Two tank system.  Source: \cite{Singh:2015}}
	       \label{fig:tank22} 
      \end{figure}

    Let $A$ be the cross-sectional area of the water tank, $h(t)$ be the water level height at time $t$, $\frac{dh}{dt}$ the change in the water level, $g$ the gravitational constant, $q_{i}$ the water flow rate in, $q_{o}$ the water flow rate out, $\rho=$  (mass/volume) be the water (liquid) density, and $R$ be the water outflow resistance.  Let $q_{i} - q_{o}$ be the rate of change of the liquid volume so that from conservation of mass $q_{i} - q_{o}$ = $\frac{dV}{dt} = A\frac{dh(t)}{dt}$ where $V$ is the volume of liquid ($V = Ah(t)$).
    Then the ODE of the water tank level can be written as:
    \begin{equation}
        q_{i}(t) - q_{o}(t) = A\frac{dh(t)}{dt}
    \end{equation}
    Letting $q_{o}(t) = \frac{\rho}{R}h(t)$, then
    \begin{equation}
        q_{i}(t) = A\frac{dh(t)}{dt} + \frac{\rho}{R}h(t) 
    \end{equation}
    Then taking the Laplace transform, we get
    \begin{equation}
        Q_{i}(s) = AsH(s) + \frac{\rho}{R}H(s) 
    \end{equation}
    Simplifying, we get
    \begin{equation}
        Q_{i}(s) = \bigg [As + \frac{\rho}{R} \bigg ]H(s) 
    \end{equation}
    Therefore, the water tank transfer function is:
    \begin{equation}
        G(s) = \frac{H(s)}{Q_{i}(s)} = \frac{1}{As + \frac{\rho}{R}}
    \end{equation}
    Since the density of water is approximately 1g/$\text{cm}^{3}$ and multiplying by $R$, we get the transfer function
    \begin{equation}
        G(s) = \frac{R}{1 + RAs}
        \label{eq:tanktf}
    \end{equation}
      We do not multiply $\rho$ by the gravitational constant $g$ = 9.81 $m/s^{2}$ since we assume in our model setup that water is not flowing down from tank 1 into the top of tank 2 but at the same level as the other tank through a connecting hose as illustrated in Figure \ref{fig:tank22}. Setting $\tau = RA$, then $G(s)$ = $\frac{R}{\tau s + 1}$.   
     Various liquid control level models assume that the tank outflow rate $q_{o}$ is proportional to the square root of the height level (nonlinear), e.g. $q_{o} = \frac{\rho}{R}\sqrt{h}$, but the transfer function will still be the same first order \cite{Singh:2015}, \cite{Mahmood:2020}, \cite{Janevska:2013}.
      
      \subsection{Water Tank Coupled to Valve}
      Consider a valve coupled to the motor pump that opens and closes so that in effect, the flow of water into and out of the tank is controlled by  the valve coupled to the tank.  Then, the rate of change of the water level is
      \begin{equation}
        q_{i}(t) - q_{v}(t) = \frac{dV}{dt} = A\frac{dh(t)}{dt}
      \end{equation}
      We assume the valve functions like a standard sharp edged orfice so that the flow through the valve is related to the water (fluid) level in the tank, $h \equiv h(t)$, by the expression \cite{Marizan:2001}:
      \begin{equation}
            q_{v}(t) = c_{v}a_{v}\sqrt{h}
            \label{eq:valve}
      \end{equation}
      where 
      \begin{itemize}
      \item $q_{v}$ is the flow rate out of the valve.
      \item $a_{v}$ is the cross sectional area of the orfice and represents the dimensions of the valve and flow channel in which it is founded.  Since this dimension can change along the length of the channel. $a_{v}$ would be the mean value; 
      \item $c_{v}$ is the discharge coefficient of the valve and takes into account all liquid characteristics, losses, and irregularities in the system so that the two sides of equation \ref{eq:valve} balance.
      \end{itemize}
      If the water if flowing in from the top into the tank, the gravitation constant $g$ is included in the square root in equation \ref{eq:valve}, but it does not in our model so we do not include it.
      We assume $c_{b}$ is constant and that $q_{v}$ is proportional to the square root of the level $h$.   In practice, the valve flow rate $q_{v}$ is some general non-linear function of the level $h$, $q_{v} = f(h)$ so that 
      \begin{equation}
      q_{i}(t) = A\frac{dh}{dt} + f(h)
      \label{eq:valve2}
      \end{equation}
      which is a first order ODE.  To make it useful for control systems, we can linearize it.  Let the output water level $h = h_{o} + \delta h$ where $h_{o}$ is the normal operating level (a constant) and $\delta h$ be a small variation around $h_{o}$. 
    
      Let the valve flow rate $q_{v} = \bar{q}_{v} + \delta q_{v}$ and inflow rate be $q_{i} = \bar{q}_{i} + \delta q_{i}$ where
      $\bar{q}_{i}$ is the steady inflow level component of $q_{i}$ and $\delta q_{i}$ is a small change around $\bar{q}_{i}$  We can write equation \ref{eq:valve2} as:
      \begin{equation}
            \bar{q}_{v} + \delta q_{v} + A \frac{dh}{dt} = \bar{q}_{i} + \delta q_{i} 
      \label{eq:valve4}
      \end{equation}
       which can be written as:
       \begin{equation}
            A\frac{dh}{dt} + f(h_{0}) + D\delta h = \bar{q}_{i} + \delta q_{i}
       \label{eq:flow2}
       \end{equation}
      where coefficient $D$ is the slope of the valve characteristics at the level $h_{0}$:
      \begin{equation}
                D = \frac{\delta f(h_{o})}{\delta h}
      \end{equation}
      For instance, applying a first order Taylor expansion to eq. \ref{eq:valve} about the steady state height $h_{0}$, yields 
      \begin{equation}
        f(h) = c_{v}a_{v} \bigg (\sqrt{h_{0}} + \frac{h - h_{0}}{2\sqrt{h}} \bigg ) 
        \label{eq:state}
      \end{equation}
      where $D = 1/(2\sqrt{h})$ and $\delta h = h - h_{0}$.  If the water level is constant such that $\delta q_{i} = 0$, then eq. \ref{eq:state} yields the steady state relation for flow and water height level $h_{0}$,  
      \begin{equation}
      f(h_{0}) = \bar{q}_{i} = c_{v}a_{v}\sqrt{h_{0}}
      \label{eq:flow}
      \end{equation}
      Subtracting equation \ref{eq:flow} from equation \ref{eq:flow2} and then rearranging provides the linear first order differential equation for a single tank system:
      \begin{equation}
            A\frac{dh}{dt} + D \delta h = \delta q_{i}
      \end{equation}
      Let $k_{v} = D^{-1} = 1/D$, then $D = 1/k_{v}$.  Define the time constant $\tau_{v} = A/D$ so that $A = \tau_{v}D$.  Therefore,
      \begin{align}
            \tau_{v}D\frac{dh}{dt} + \delta h\frac{1}{k_{v}} &= \delta q_{i} \\
            \tau_{v}\frac{1}{k_{v}}\frac{dh}{dt} + \delta h\frac{1}{k_{v}} &= \delta q_{i} \\
            \frac{1}{k_{v}}\big (\tau_{v}\frac{dh}{dt} + \delta h \big ) &= \delta q_{i} \\
            \tau_{v}\frac{dh}{dt} + \delta h &= k_{v} \delta q_{i}
      \end{align}
      Taking the Laplace transform gives the single tank transfer function:
      \begin{align}
            \tau_{v}H(s)s + H(s) &= k_{v}Q_{i}(s)  \\
            H(s)(\tau_{v}s + 1) &= k_{v}Q_{i}(s)
      \end{align}  
      or
      \begin{equation}
            \frac{H(s)}{Q_{i}(s)} = \frac{k_{v}}{\tau_{v}s + 1}
            \label{eq:tftank}
      \end{equation}
       where $H(s)$ and $Q_{i}(s)$ are the Laplace transforms of $\delta h$ and $\delta q_{i}$, respectively.
       Let $k_{i}$ be the motor pump gain constant and let $k_{s}$ be the sensor gain constant.  Let $v_{i}$ be the applied voltage to the pump motor amplifier and let $v_{o}$ be the output voltage.   Then 
       \begin{equation}
       q_{i} = k_{i}v_{i}
       \label{eq:sensor2}
       \end{equation}
       \begin{equation}
       v_{o} = k_{s}h
        \label{eq:sensor1}
       \end{equation}
       Combining eqs. \ref{eq:sensor2} and \ref{eq:sensor1} with the the tank transfer function in equation \ref{eq:tftank}, yields a first order system transfer function
       \begin{equation}
            \frac{V_{o}}{V_{i}} = \frac{G}{\tau_{v}s + 1}
       \end{equation}
        where $G = k_{i}k_{s}k_{v}$ is the system gain \cite{Marizan:2001}.

\section{Simulation Analysis}
\begin{figure}[H] 
	\centering     
    \includegraphics[width=0.9\columnwidth]{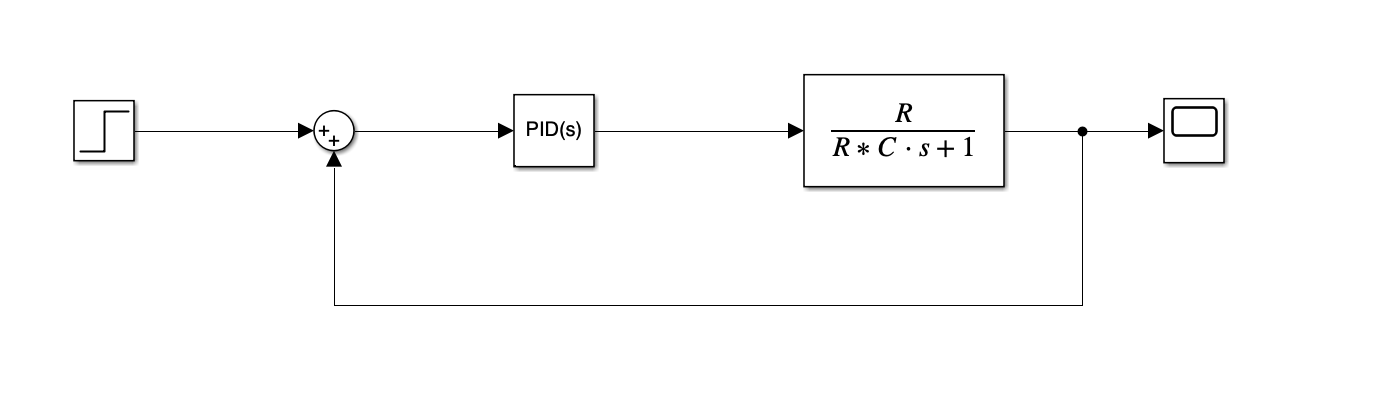} 
	\caption{Closed-loop feedback system}
	\label{fig:water4} 
\end{figure}

\begin{figure}[H] 
	\centering     
     \includegraphics[width=0.9\columnwidth]{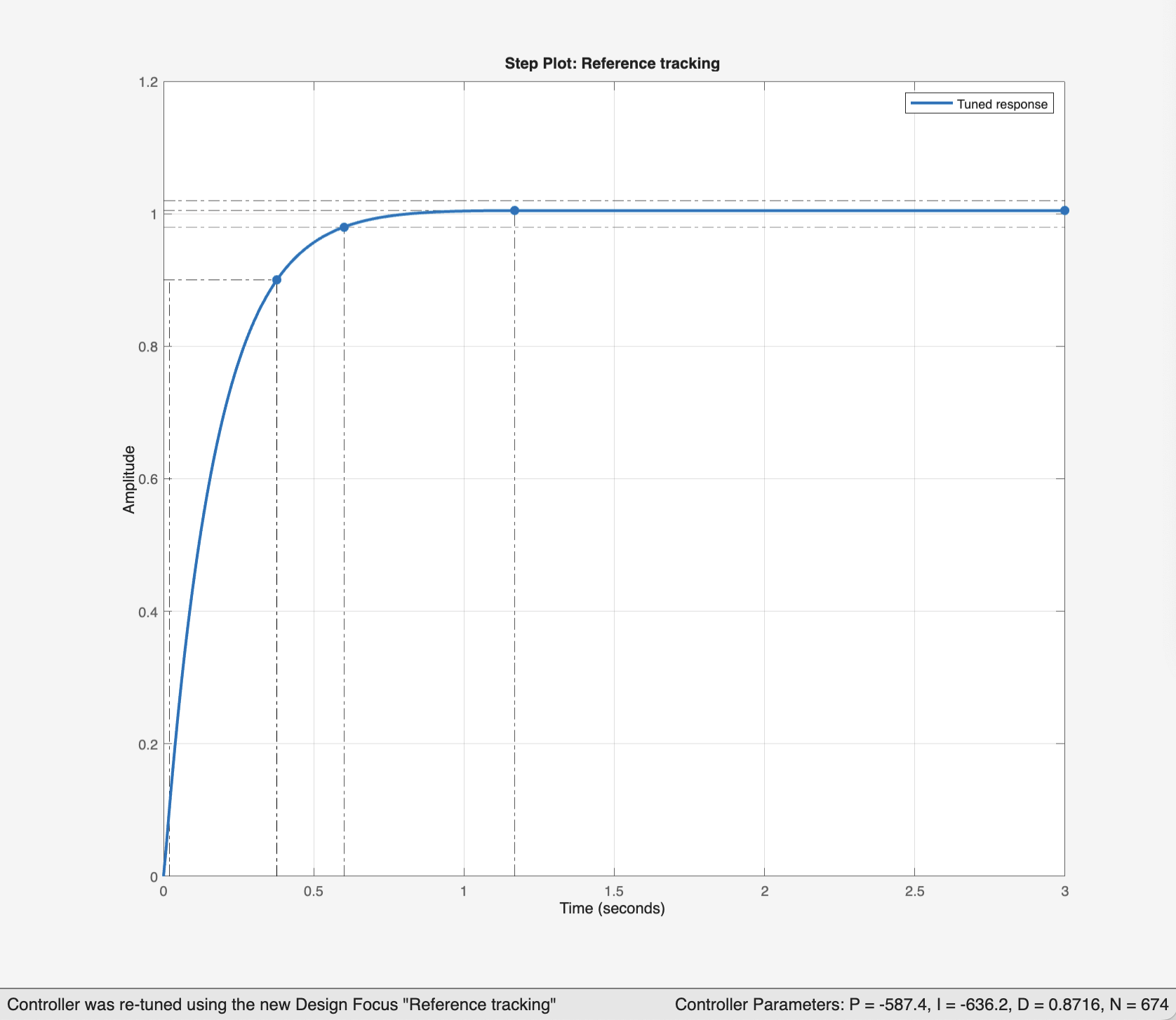} 
	\caption{Step Response}
	\label{fig:pid1} 
\end{figure}

\begin{figure}[H] 
	\centering     
    \includegraphics[width=0.9\columnwidth]{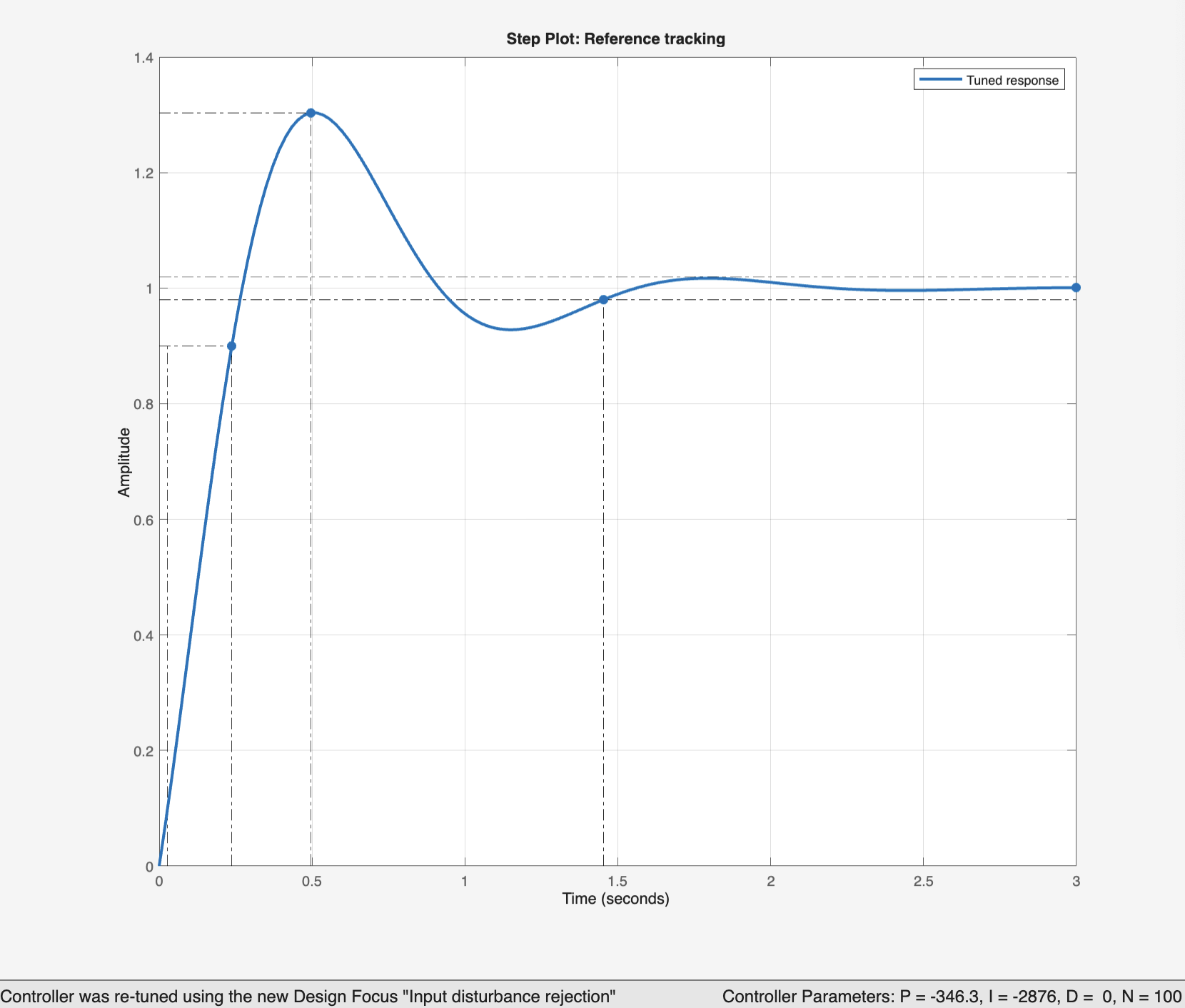} 
	\caption{Step-response}
	\label{fig:pid2} 
\end{figure}

\begin{figure}[H] 
	\centering     \includegraphics[width=0.9\columnwidth]{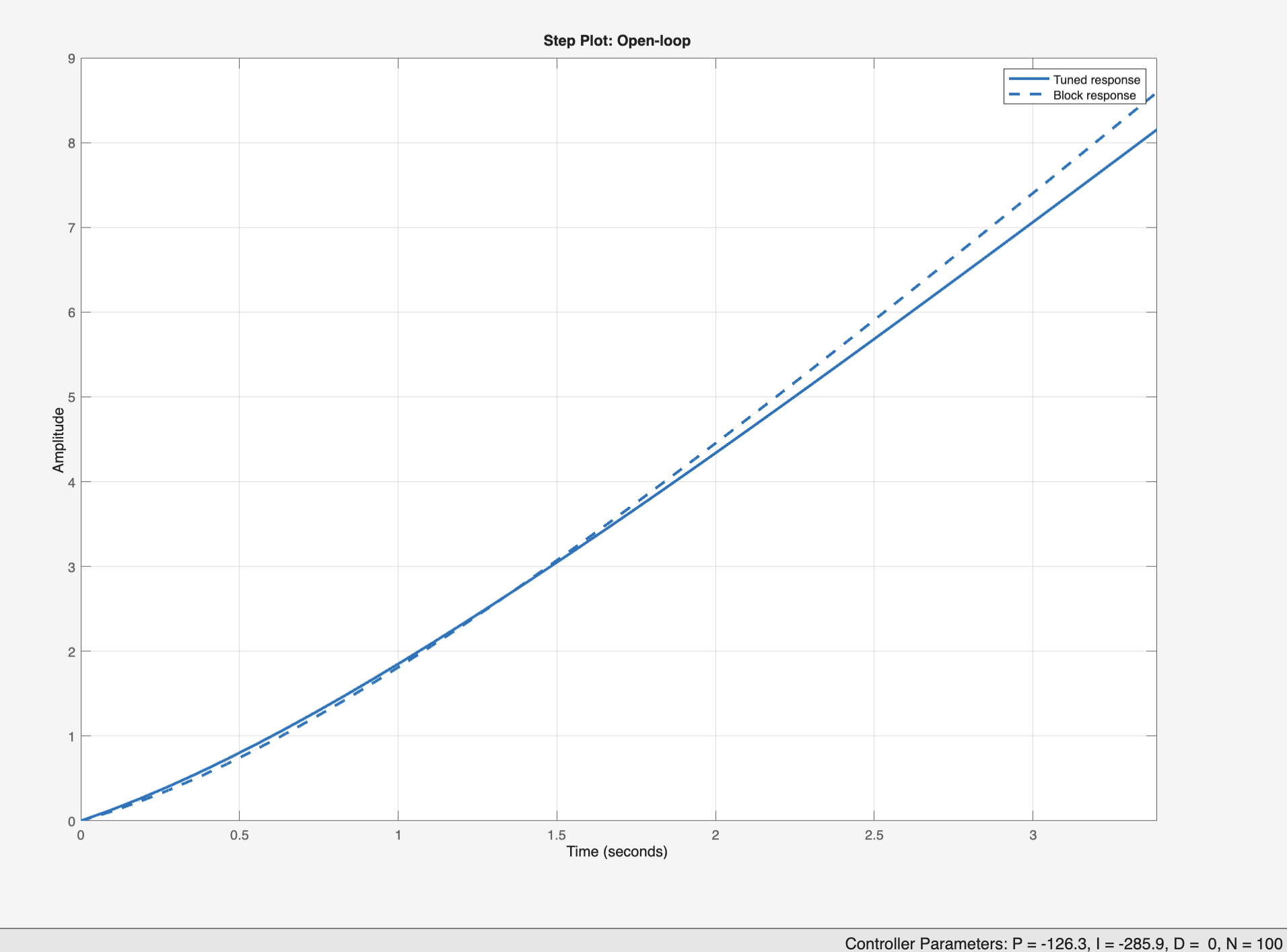} 
	\caption{Open-loop feedback system}
	\label{fig:pid3} 
\end{figure}

\begin{figure}[H] 
	\centering     
    \includegraphics[width=0.9\columnwidth]{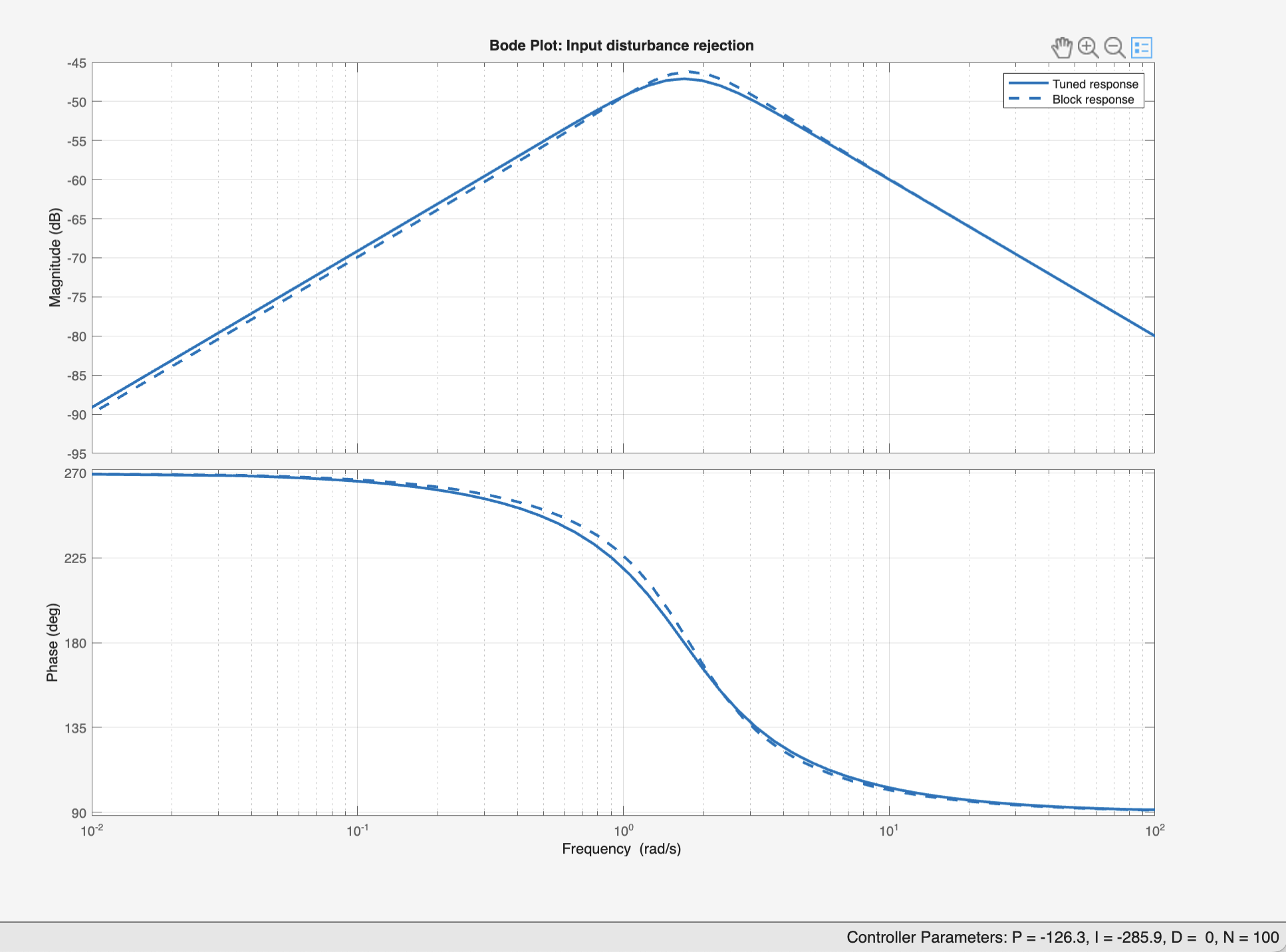} 
	\caption{Bode-plot with input disturbance rejection}
	\label{fig:pid4} 
\end{figure}

\begin{figure}[H] 
	\centering     
    \includegraphics[width=0.9\columnwidth]{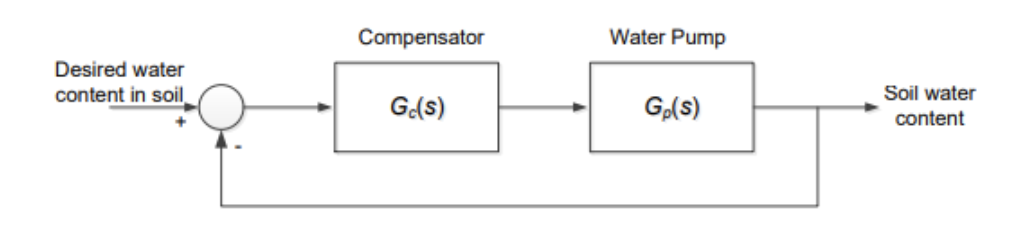} 
	\caption{Block diagram for soil watering}
	\label{fig:soil} 
\end{figure}

\begin{figure}[H] 
	\centering     
    \includegraphics[width=0.9\columnwidth]{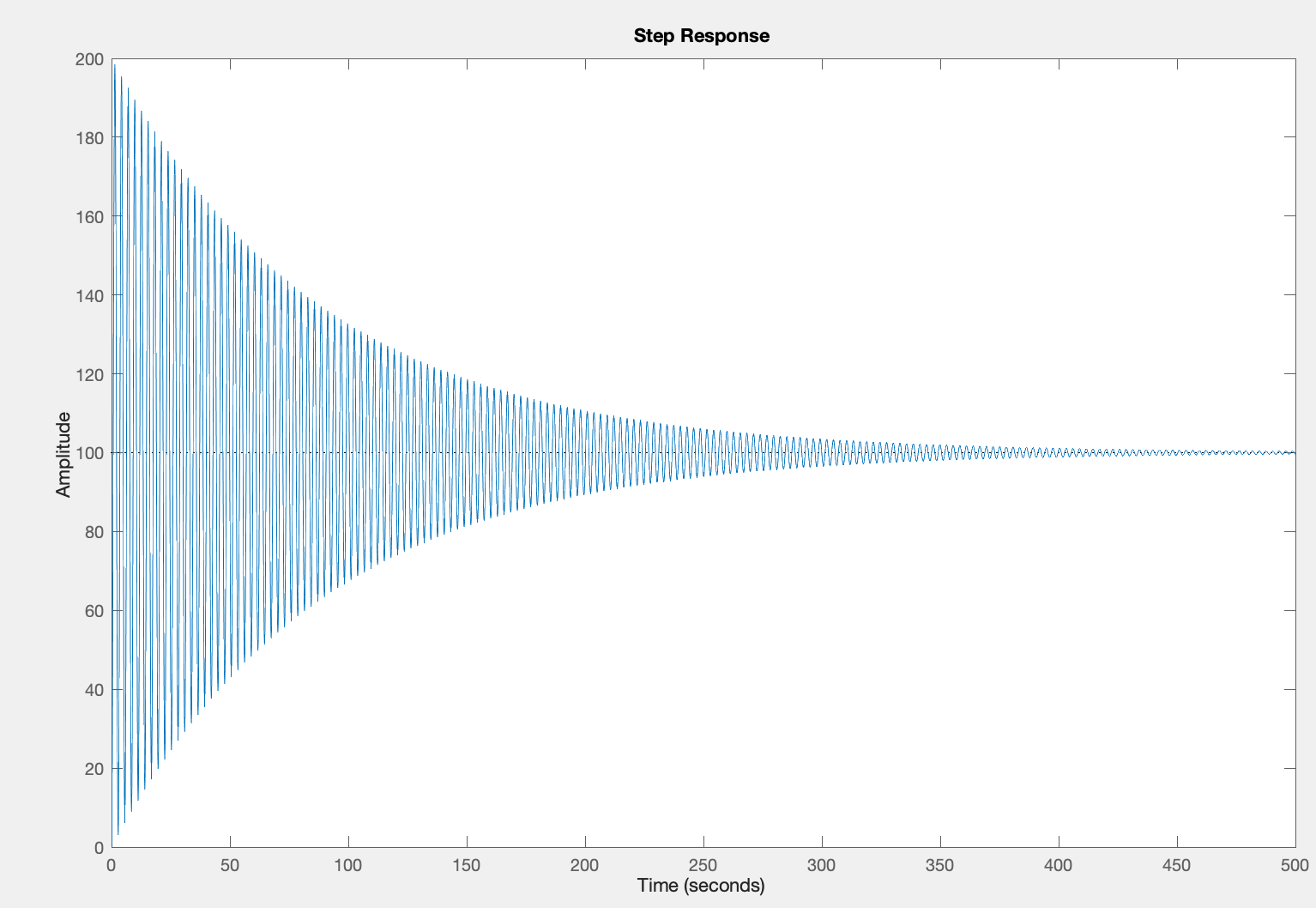} 
	\caption{Step Response for transfer function in \ref{eq:osc}}
	\label{fig:oscillations} 
\end{figure}

\begin{figure}[H] 
	\centering         \includegraphics[width=0.9\columnwidth]{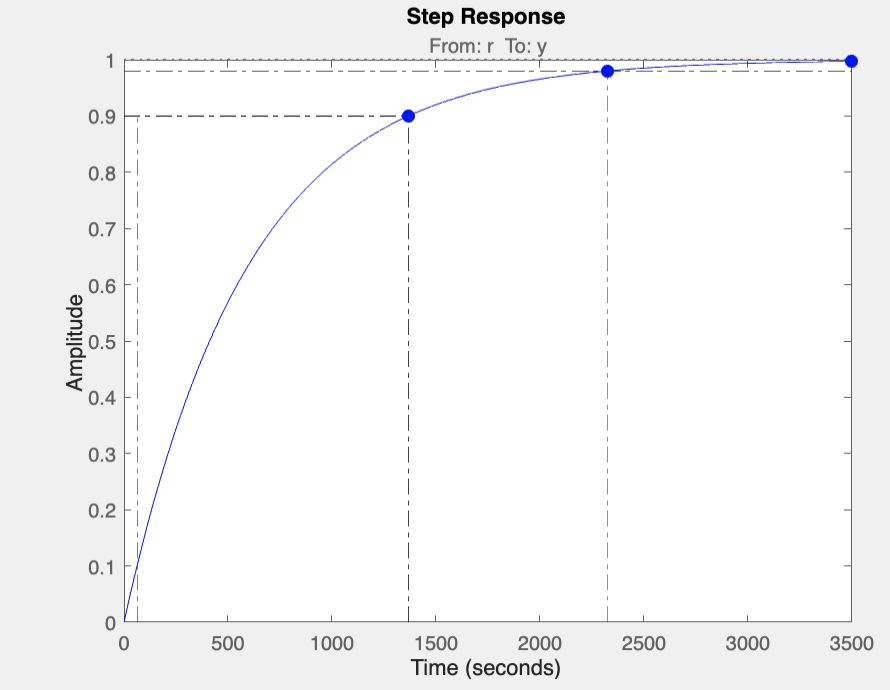} 
	\caption{Step Response with PID} 
	\label{fig:100} 
\end{figure}

\begin{figure}[H]
	\centering  
     \includegraphics[width=0.9\columnwidth]{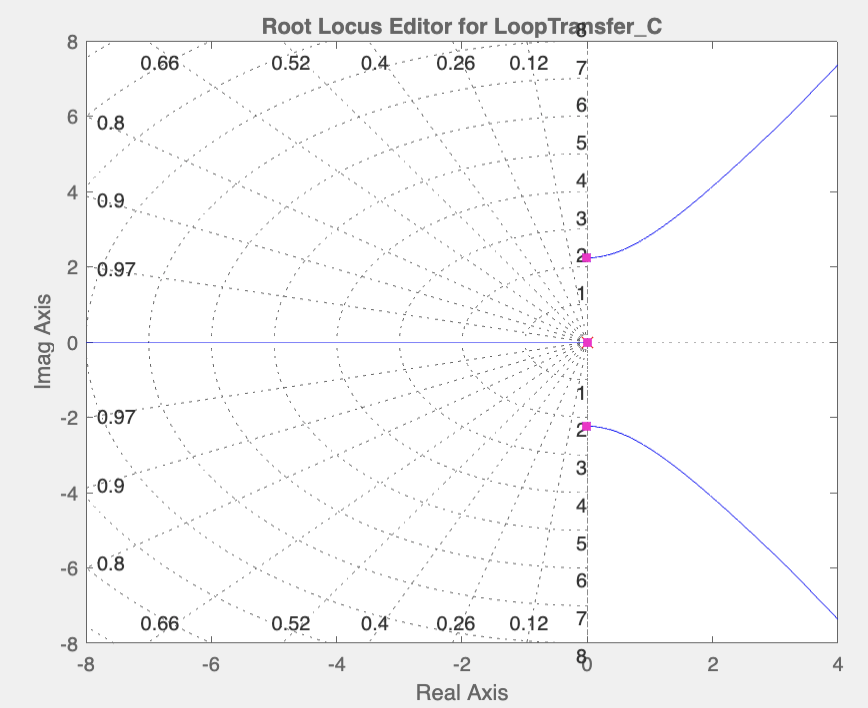} 
	\caption{Root Locus with PID}
	\label{fig:101} 
\end{figure}

\begin{figure}[H]
	\centering  \includegraphics[width=0.75\columnwidth]{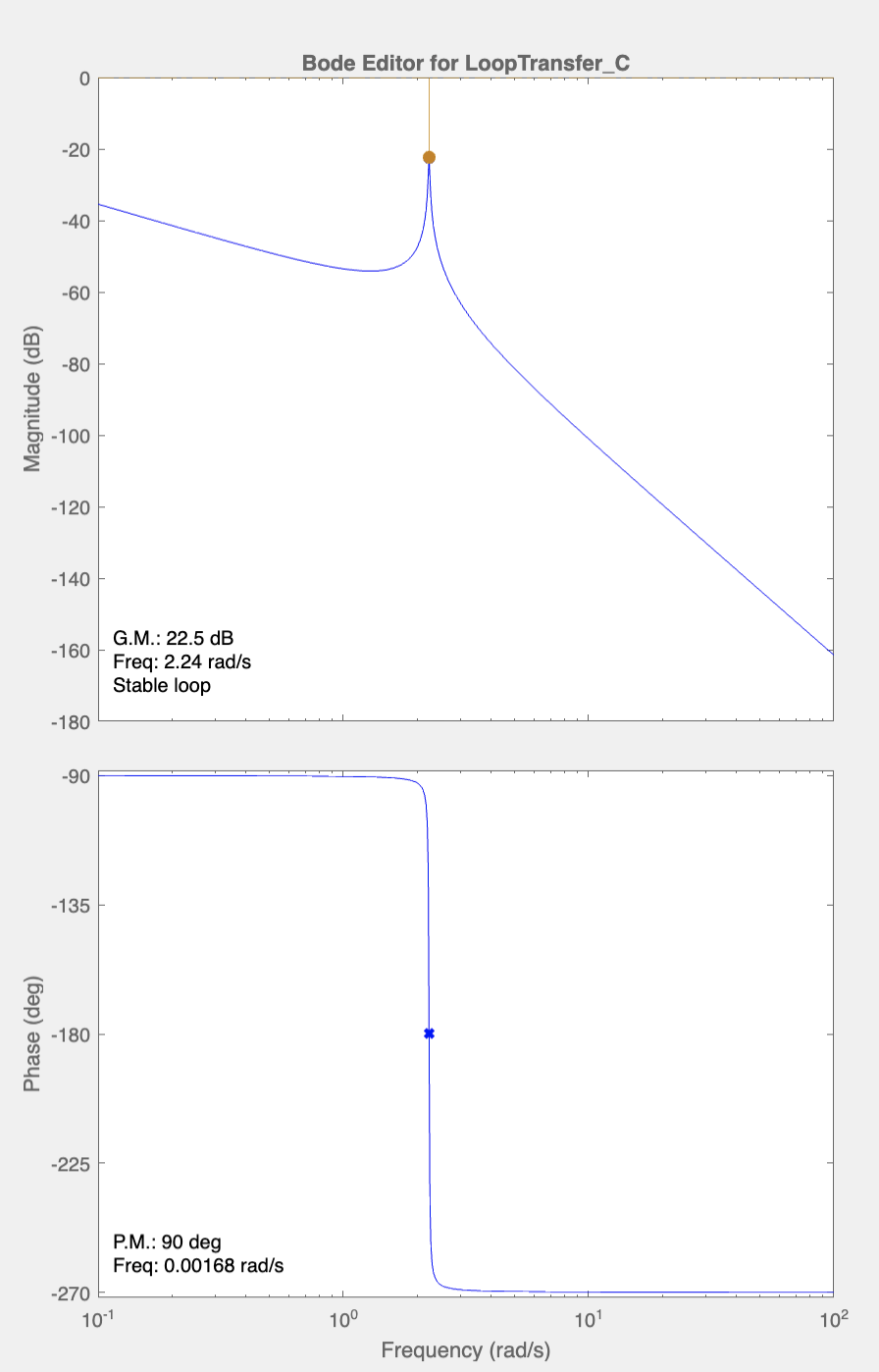} 
	\caption{Bode Plot with PID}
	\label{fig:102} 
\end{figure}

\begin{figure}[H] 
	\centering         
 \includegraphics[width=0.9\columnwidth]{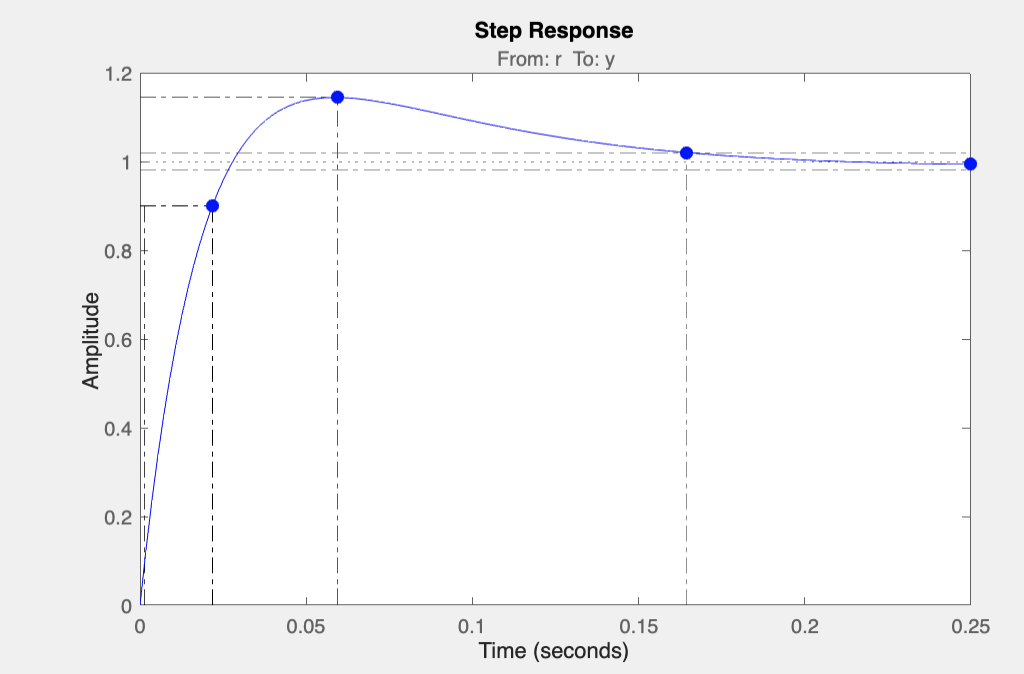} 
	\caption{Step Response with PID 2} 
	\label{fig:103} 
\end{figure}

\begin{figure}[H]
	\centering  \includegraphics[width=0.9\columnwidth]{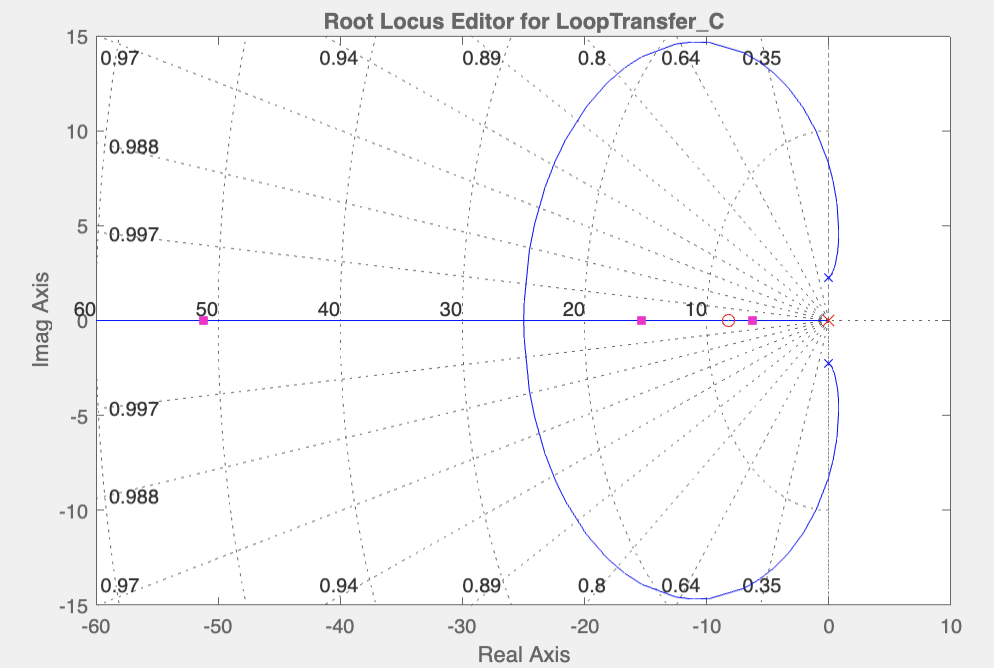} 
	\caption{Root Locus with PID 2}
	\label{fig:104} 
\end{figure}

\begin{figure}[H]
	\centering  \includegraphics[width=0.75\columnwidth]{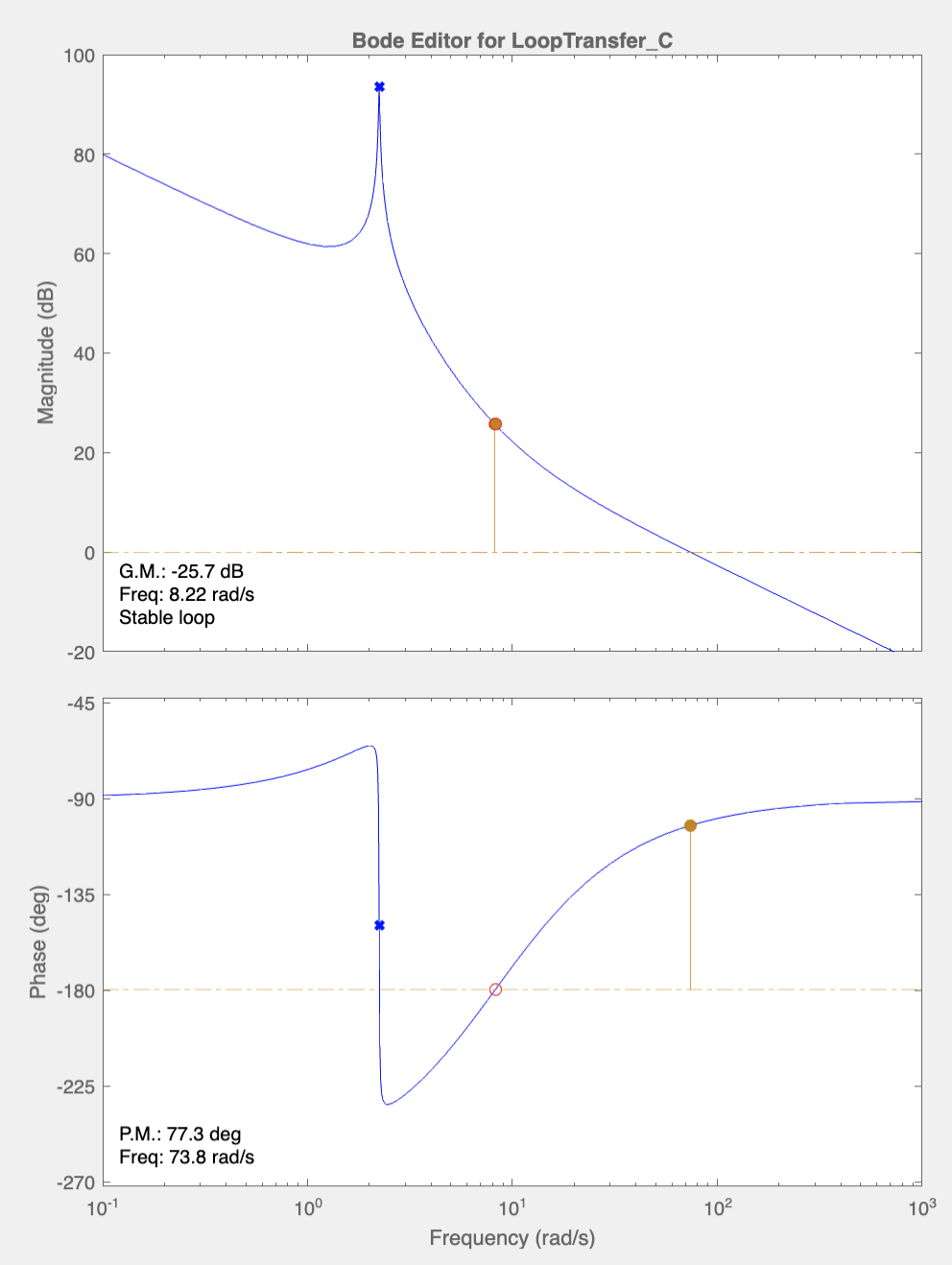} 
	\caption{Bode Plot with PID 2}
	\label{fig:105} 
\end{figure}

    A Simulink diagram of a closed-loop feedback PID controller system for the water tank transfer function in eq. \ref{eq:tanktf} is shown in Figure \ref{fig:water4}.  We use $R = 0.01$ L$\Omega$ (liter ohms) and $A = 100$, so that $G(s) = \frac{0.01}{s + 1}$.
    Figure \ref{fig:pid1} shows a PID-tuned step response function for reference tracking using $K_{p} = -587.4$, $K_{i} = -636.2$, $K_{d} = 0.8716$, and $N=674$. 
    Figure \ref{fig:pid2} shows a PI-tuned step response plot with input disturbance rejection using $K_{p} = -346.3$, $K_{i} = -2876$, $K_{d} = 0$, and $N = 100$.
    Figure \ref{fig:pid3} shows a PID-tuned open-loop plot of the system using $K_{p} = -126.3$, $K_{i} = -285.9$, $K_{d}$ = 0, and $N=100$.
    Figure \ref{fig:pid4} shows a PID-tuned Bode plot of the closed loop input disturbance rejection plot using $K_{p}$ = -126.3, $K_{i}$ = -285.9, and $K_{d}$ = 0, and $N=100.$
\begin{figure}[H]
	\centering     \includegraphics[width=0.9\columnwidth]{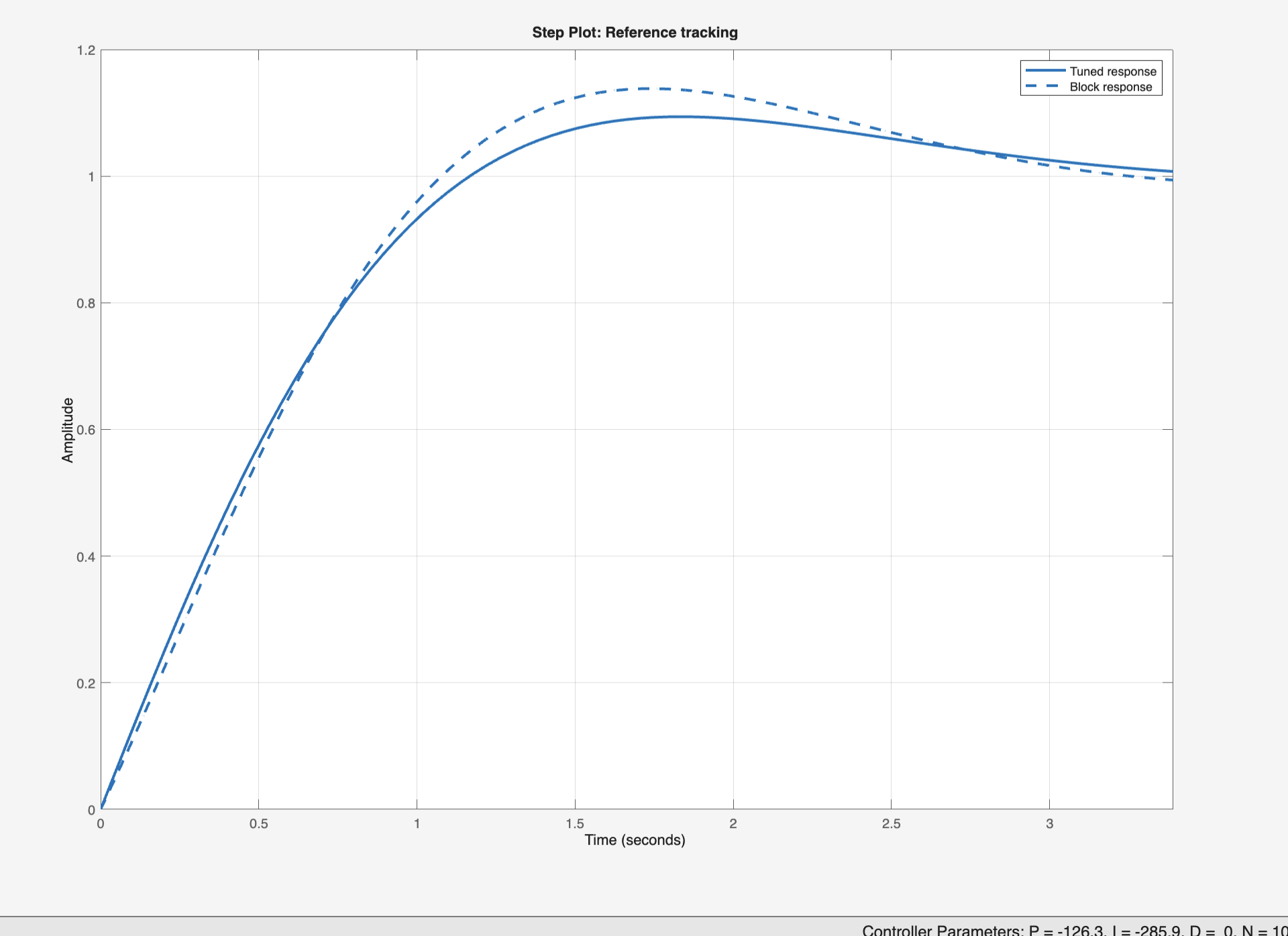} 
	\caption{Closed-loop feedback system}
	\label{fig:pid5} 
\end{figure} 
  In our cascaded system, one water pump supplies water to a water tank so that a another water pump can supply water from the tank through a hose to a plant(s) if the soil moisture level falls below a certain threshold.  
  The block diagram is shown in Figure \ref{fig:soil}
  The water tank transfer function using feedback from a sensor can be shown to be a second order \cite{Craig:2002}: 
  \begin{equation}
        \frac{H(s)}{V(s)} = \frac{K \omega_{n}^{2}}{s^{2} + 2 \zeta \omega_{n} s + \omega_{n}^{2}}
  \end{equation}
  where $H(s)$ is the height in the water tank, $V(s)$ is input voltage, and $K$ is the gain.  For instance, using the calculated values of $\omega_{n}$ and $\zeta$ above, then the transfer function becomes
  \begin{equation}
    \frac{H(s)}{V(s)} = \frac{5K}{s^{2} + 0.02241s + 5} 
    \label{eq:osc}
  \end{equation}
  but the system has two imaginary poles at $-0.0112 \pm 2.236i$ and the step response has high frequency oscillations as shown in Figure \ref{fig:oscillations}. 
  A PID controller needs to be used to provide a stable unit response. For instance, a compensator of $C(s) = \frac{0.00010028}{s}$
  will generate the step response, root locus, and bode plots in Figures \ref{fig:100}, \ref{fig:101}, and \ref{fig:102}, respectively.
  From the Bode plot, we see that there is a fundamental frequency at 2.5 rad/sec and from the step response is overdamped and takes a long time to reach a steady state which is not acceptable performance.  But if we adjust the PID controller to  $C(s) = 747.5\frac{(1 + .12s)(1+.12s)}{s}$, then we get the step, root locus, and Bode plots in Figures \ref{fig:103}, \ref{fig:104}, and \ref{fig:105}, respectively.
    With this PID, The rise time is 0.0204, overshoots at 14.4\% at  0.059 sec with a peak of 1.14, and settling time is 0.165 seconds.  
    There are two real zeros $z = -8.24$ and a pole at zero making the system stable.

  A metering pump which is capable of delivering a precise flow rate of fluid can provide better control and stability. Most metering pumps \enquote{consist of an electric motor that varies the strike length of a shaft, allowing more or less fluid to pass through its body} \cite{Nise:2015}.  

  The open-loop transfer function of the pump has been found to be \cite{Yu:2011}
  \begin{equation}
    \frac{Y(s)}{U(s)} = \frac{1.869}{s^2 + 12.32s + 0.4582}
  \end{equation}
  where the output of the system, Y(s), represents the water flow and input, U(s), is the command signal to the motor that varies the pump's plunger strike length.

\subsection{Stability and Error Analysis}
\begin{figure}[H]
	\centering  \includegraphics[width=0.75\columnwidth]{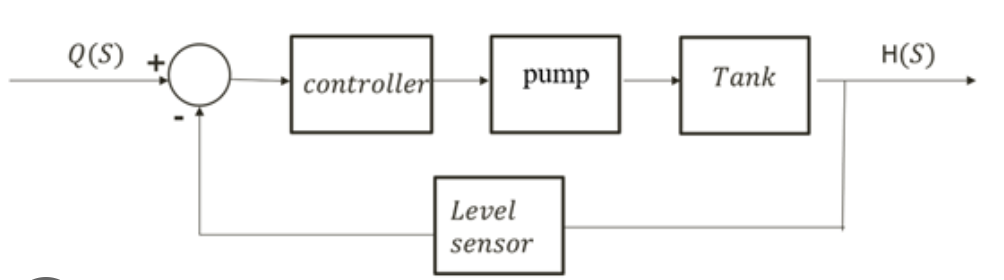} 
	\caption{Feedback loop with PID controller, plant (water pump and tank), and sensor}
	\label{fig:loop1} 
\end{figure}

    We consider the closed feedback loop block diagram in Figure \ref{fig:loop1} 
    The PID is the controller, the plant $G(s)$ is the multiplication of the feed forward transfer functions of the water pump and the water tank, namely $G(s)= \frac{0.05}{0.1s^2 + 1.1s + 1}$, and the sensor level $H(s) = K_{s}$ is the gain of the sensor. To achieve the desired dynamic step response, $K_{s}$ is set to 50.  We also use a transfer function of $\frac{5}{.1s + 1}$ for the water pump.  Using $\tau$ as a time constant (475 s) in the transfer function as previously calculated does not provide acceptable step response performance and so we treat $\tau$ as a flow in rate (0.1) instead which does yield a desired step response.
    
    $H(s)$ is modeled as a derivative transfer function because the \textit{change} in the soil moisture level feeds back into the system.   $G(s)$ is a second order transfer function and has a natural frequency $\omega_{n}$ = 1 rad/s at real pole $p_{1}=-1$ and 
    $\omega_{n} = 10$ rad/s at real pole $p_{2} = -10$ with a damping ratio $\zeta = 1$ so the system stable, but critically damped.
    
\begin{figure}[H]
	\centering  \includegraphics[width=0.75\columnwidth]{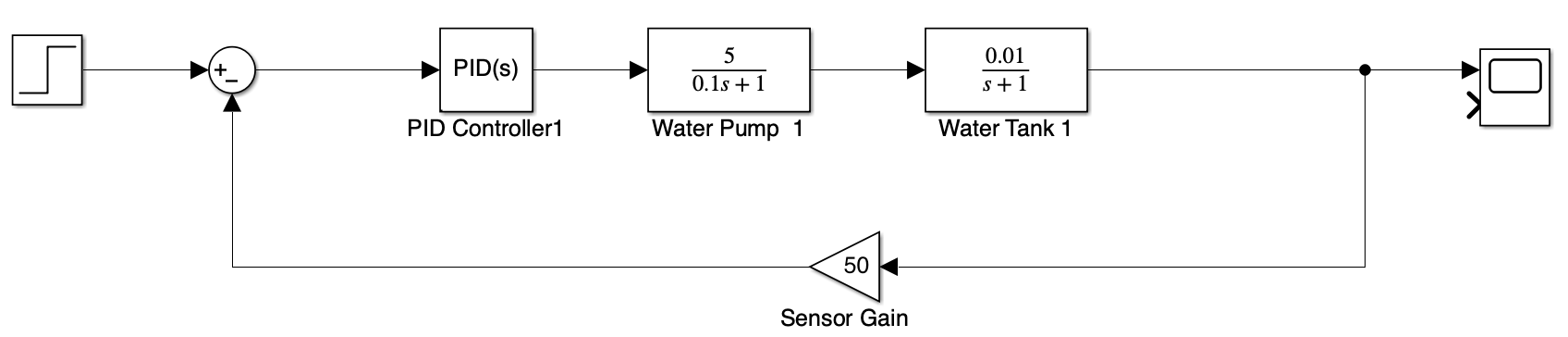} 
	\caption{Block diagram of closed feedback loop}
	\label{fig:loop2} 
\end{figure}

    Figure \ref{fig:loop2} shows the specific transfer functions used in the system.  For instance, $H(s)$ is the soil sensor.  It receives a +5V input and outputs a value, namely, the soil moisture value.  
    If we define $K$ as the gain of $G(s)$, the Routh-Hurwitz stability Table \ref{table:table3} can be shown to be
\begin{table}[H] 
\centering
\begin{tabular}{c | c  c }
\hline
    coeff. &    & \\
\hline
   $a_{2}$ & $\frac{1}{10}$ & $\frac{K}{20}$+1 \\
   $a_{1}$ & $\frac{11}{10}$ &  0 \\
   $a_{0}$ & $\frac{K}{20}$+1 &  0  \\
   \hline 
\end{tabular}
\caption{Routh-Hurwitz table} 
\label{table:table3}
\end{table}
       
\begin{figure}[H]
	\centering  
  \includegraphics[width=0.9\columnwidth]{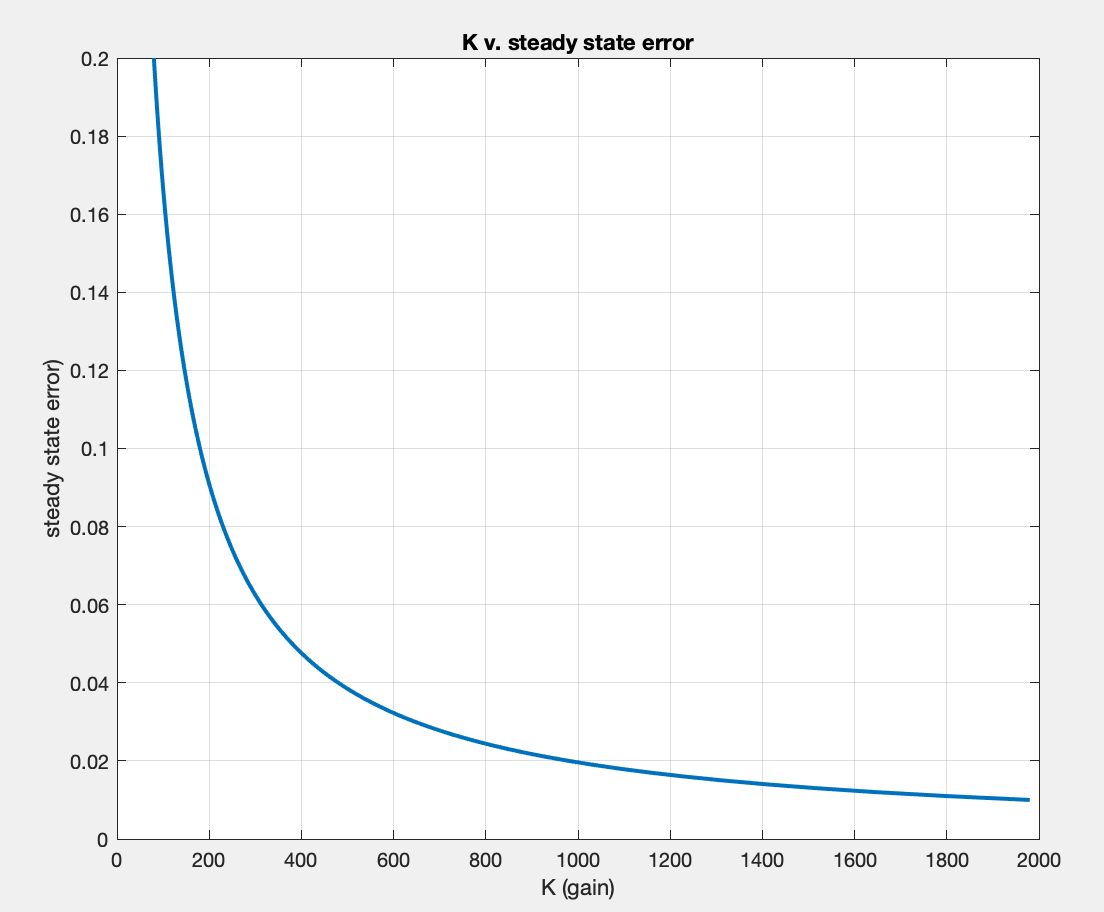} 
	\caption{Steady-state error v. system gain}
	\label{fig:error2} 
\end{figure}

Therefore, the system is stable for all $K > 0$.  A plot of the steady-state error vs. gain is shown in Figure \ref{fig:error2}.
    For a steady state error of 0.1, the system gain is 0.05 and the gain parameter is required to be $K=180$ while for a steady state error of 0.01, the gain parameter is required to be $K=1980$.  For a unit step input, $R(s) = \frac{1}{s}$, $e(\infty) =  \frac{sR(s)}{1 + G(s)} = \frac{0.05}{1 + \underset{s\rightarrow 0}{\text{lim}}G(s)} = \frac{0.05}{1.05} = 0.048$. 

\begin{figure}[H]
	\centering  
\includegraphics[width=0.9\columnwidth]{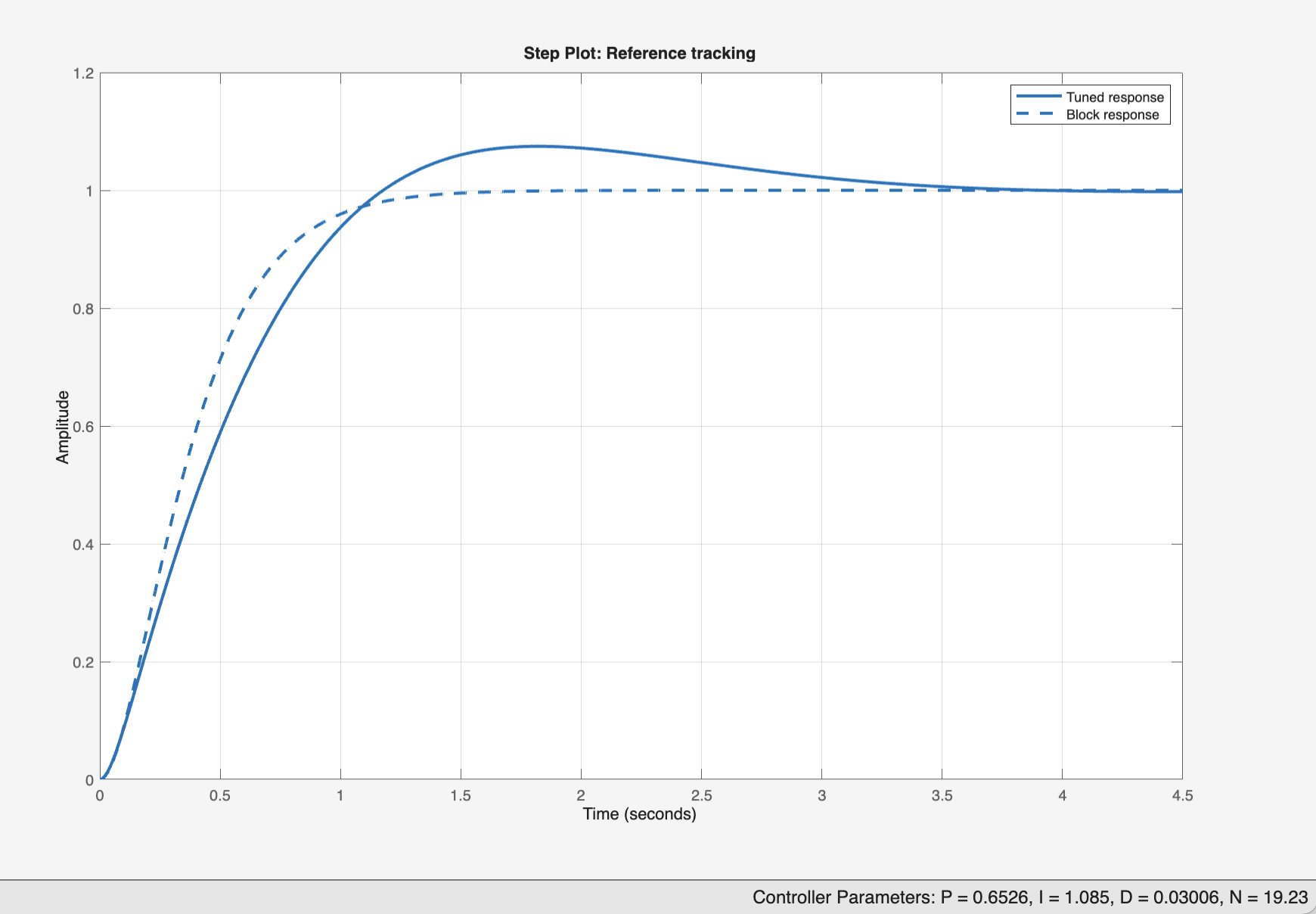} 
	\caption{Step plot of PID}
	\label{fig:tunedPID2} 
\end{figure}

    Since $G(s)$ is critically damped, we use a tuned PID controller.  Using tuned PID parameters $K_{p} = 0.653$, $K_{d} = 0.03$, $K_{i} = 1.085$ and $N=19.23$, we generate a step plot with the reference block response is given in Figure \ref{fig:tunedPID2}.
    The rise time is 0.812 seconds, the settling time is 3.04 seconds, the overshoot is $7.47\%$, and the peak is 1.07.  
    
\begin{figure}[H]
	\centering  
  \includegraphics[width=0.9\columnwidth]{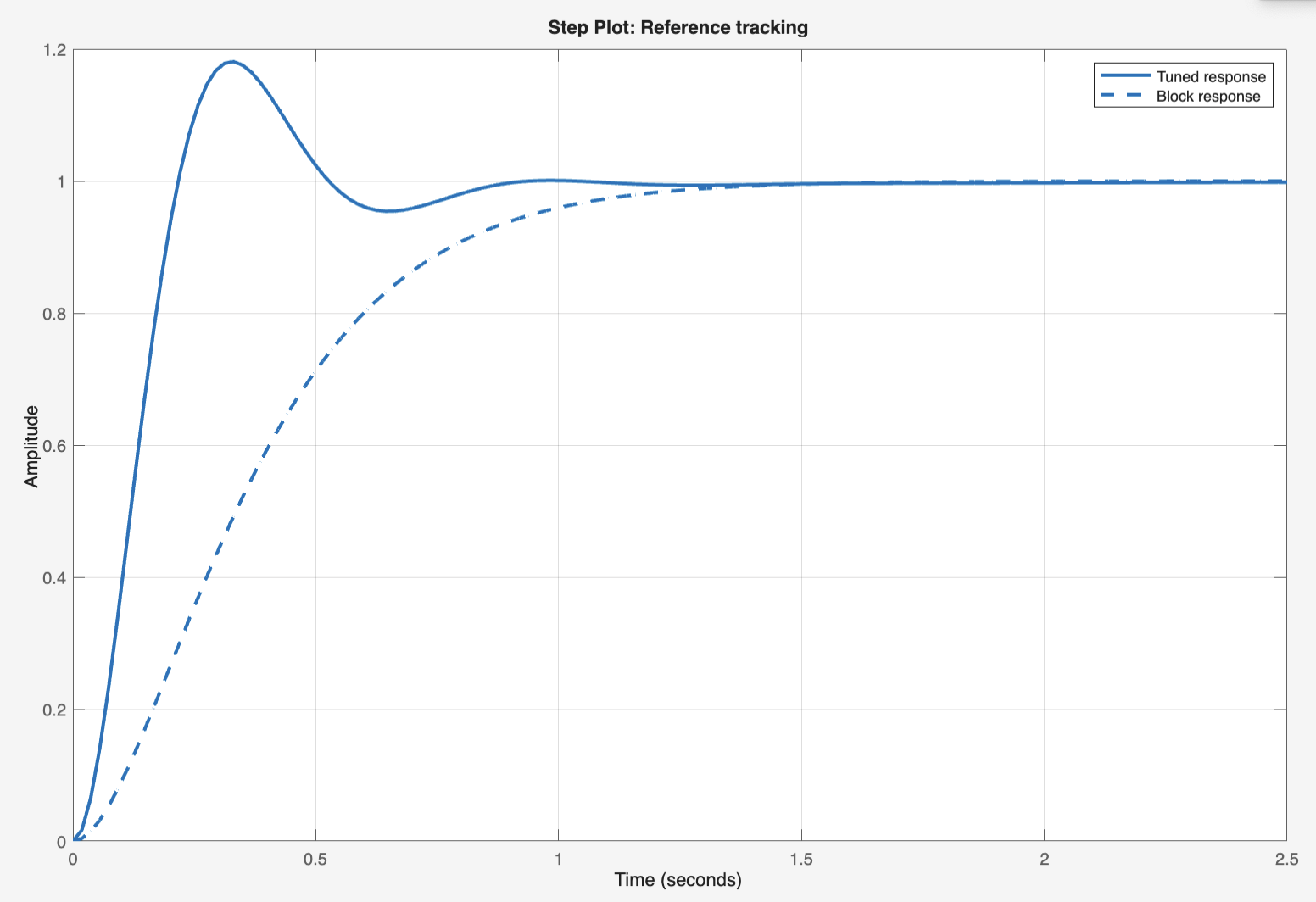} 
	\caption{Step plot of modified tuned PID}
	\label{fig:tuned4} 
\end{figure}

    As shown in Figure \ref{fig:tuned4} we can improve the performance by using a modified tuned PID controller with $K_{p} = 4.67$, $K_{i}= 3.91$, and $K_{d} = -0.0047$, and $N=1002.69$.  The rise time is now 0.146 seconds, the settling time is 0.796 seconds, the overshoot is $18.1\%$, the peak is 1.18, the gain margin is 38.9 dB @ 101 rad/s, and the phase margin is $49.3^{\circ}$ @ 8.77 rad/s.

  %Two Nema 17 1.5A stepper motors are used for the rotation and tilting of the PV panel that in turn powers a
\section{Results}
%\subsection{Three Phase Motor Induction}
\subsection{Simulation}
\begin{figure}[H] 
	\centering  \includegraphics[width=0.9\columnwidth]{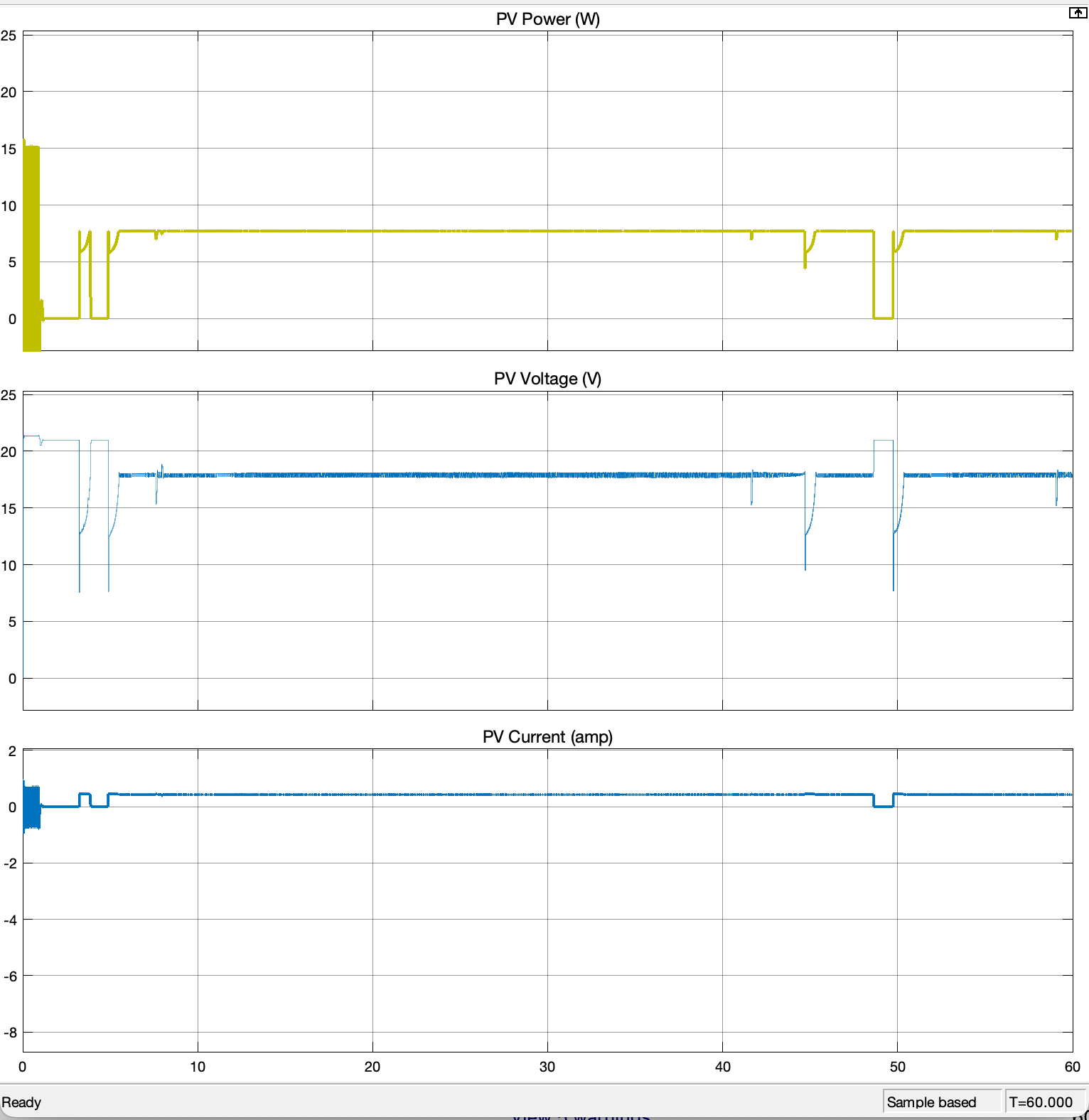} 
	\caption{Simulated PV voltage, current, and power}
	\label{fig:v2} 
    \end{figure}

\begin{figure}[H] 
	\centering  \includegraphics[width=0.9\columnwidth]{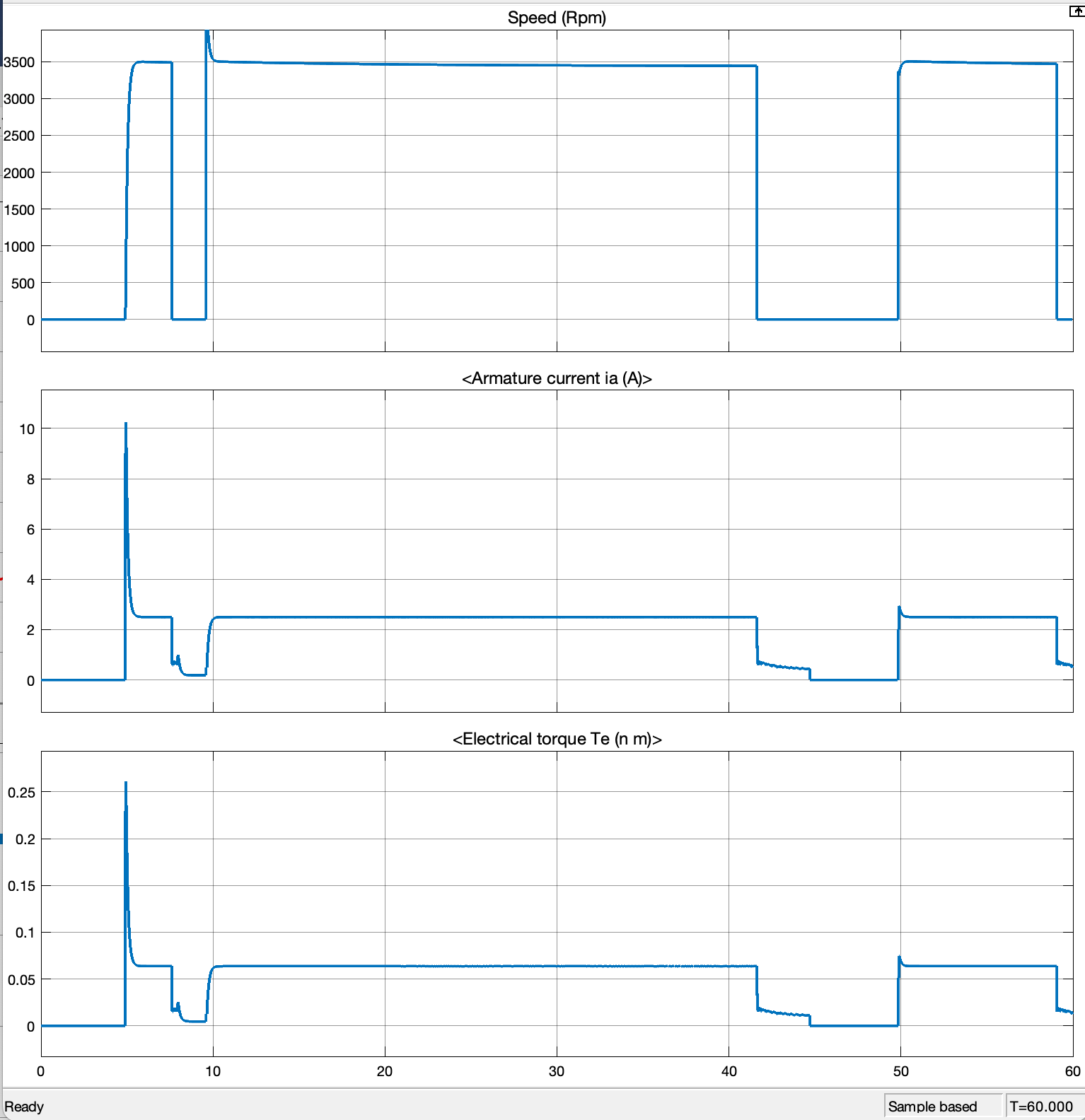} 
	\caption{Simulated motor speed, current, and torque}
	\label{fig:v4} 
    \end{figure}

    %Figure \ref{fig:model3} shows the Simulink diagram to generate (current-voltage) IV and PV (power-voltage) characteristics of a PV array @ $1000 W/m^{2}$.  
%\begin{figure}[H] 
%	\centering  \includegraphics[width=0.9\columnwidth]{Images/model3.png} 
%	\caption{Simulink Model to Generate IV-PV}
%	\label{fig:model3} 
%\end{figure} 
    Figure \ref{fig:v2} shows the simulated voltage, current, and power generated by the PV panel.  In general, these values remain roughly constant over short periods of time since the MPPT algorithm matches the voltage of the PV to the  voltage of the battery that stores the converted solar energy as charge.
    Figure \ref{fig:v4} shows the stepper motor speed (Rpm), armature current (A), and electric torque (Nm) as the solar irradiance and light photocell resistor level changes. 

\begin{figure}[H] 
	\centering  \includegraphics[width=0.9\columnwidth]{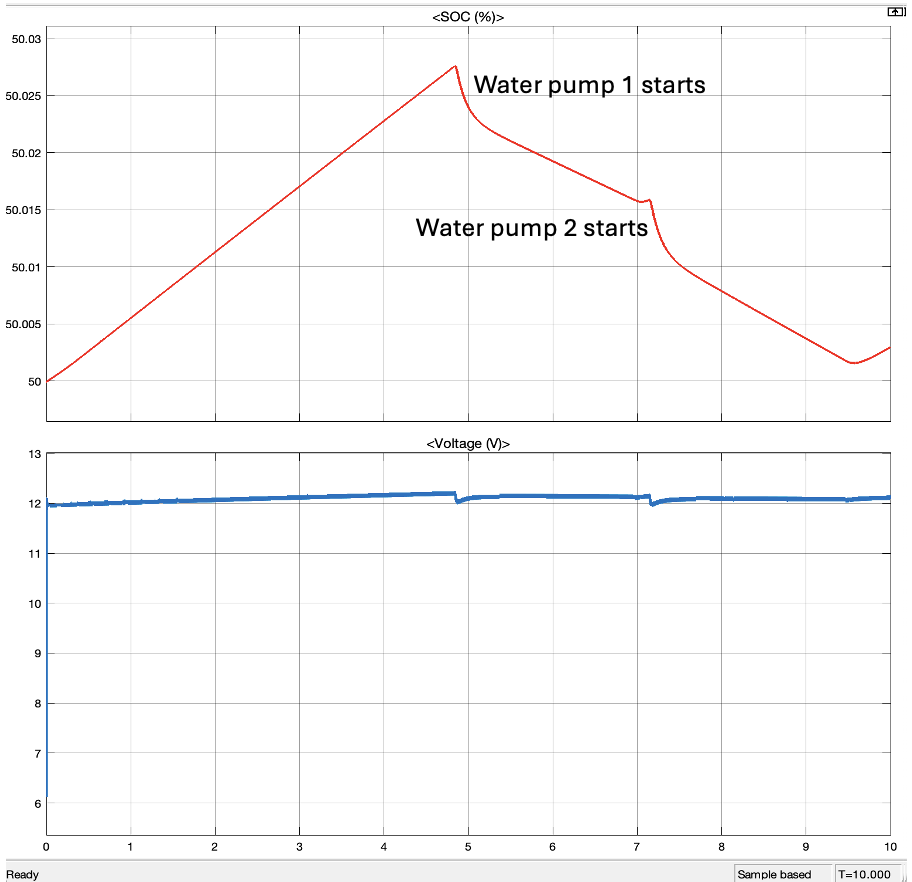} 
	\caption{SOC\% and voltage}
	\label{fig:current10} 
    \end{figure}

    Figure \ref{fig:current10} shows simulated battery state of charge (SOC) percentage of a cell.  The value at a given point in time denotes the capacity that is currently available as a function of the rated capacity.  
    As shown, initially the SOC\% increases as the battery is charged by the solar PV.  However, in the first SOC dissipation phase, water pump 1 start.  In the second dissipation phase, water pump 2 starts.  The voltage remains roughly constant except for small drops when the water pumps start as they serve as loads and draw power from the battery.
    
\begin{figure}[H] 
	\centering  \includegraphics[width=0.9\columnwidth]{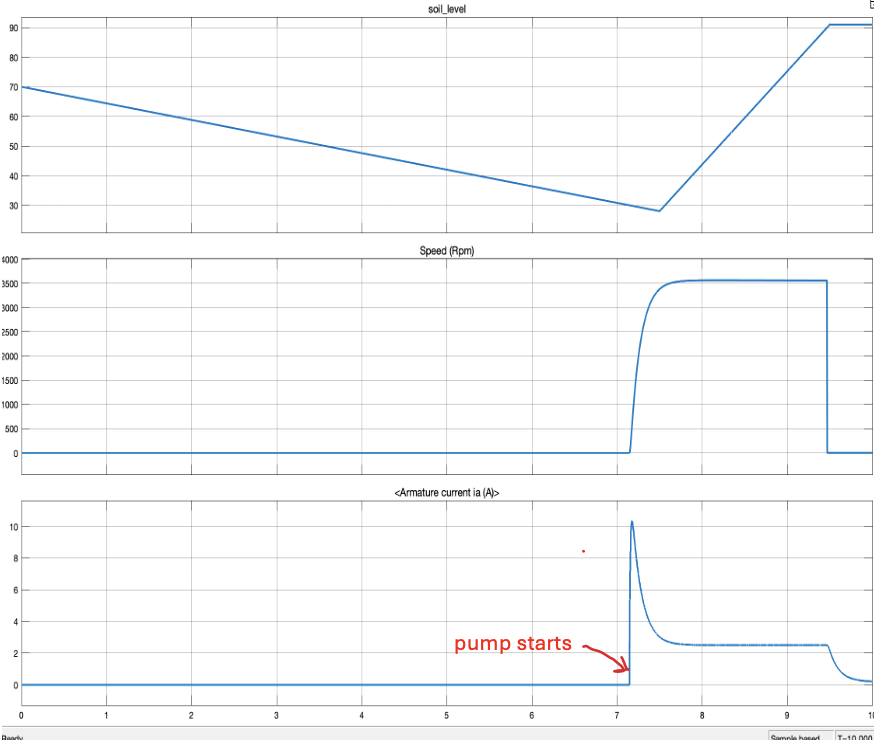} 
	\caption{Simulated soil moisture level}
	\label{fig:water2} 
    \end{figure}

\begin{figure}[H] 
	\centering  \includegraphics[width=0.9\columnwidth]{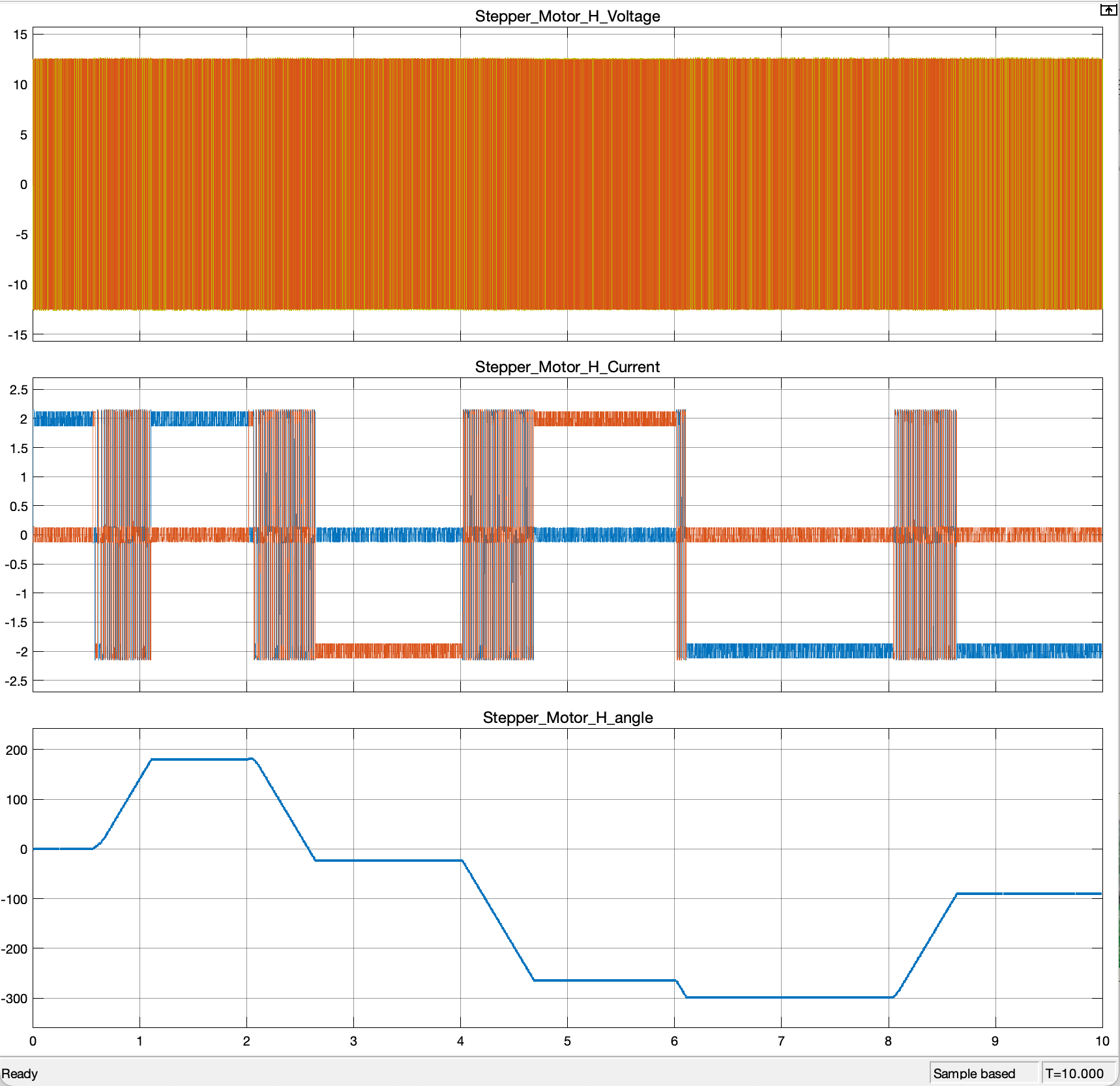} 
	\caption{Stepper motor H voltage, current, and angle}
	\label{fig:current} 
    \end{figure}

\begin{figure}[H] 
	\centering  \includegraphics[width=0.9\columnwidth]{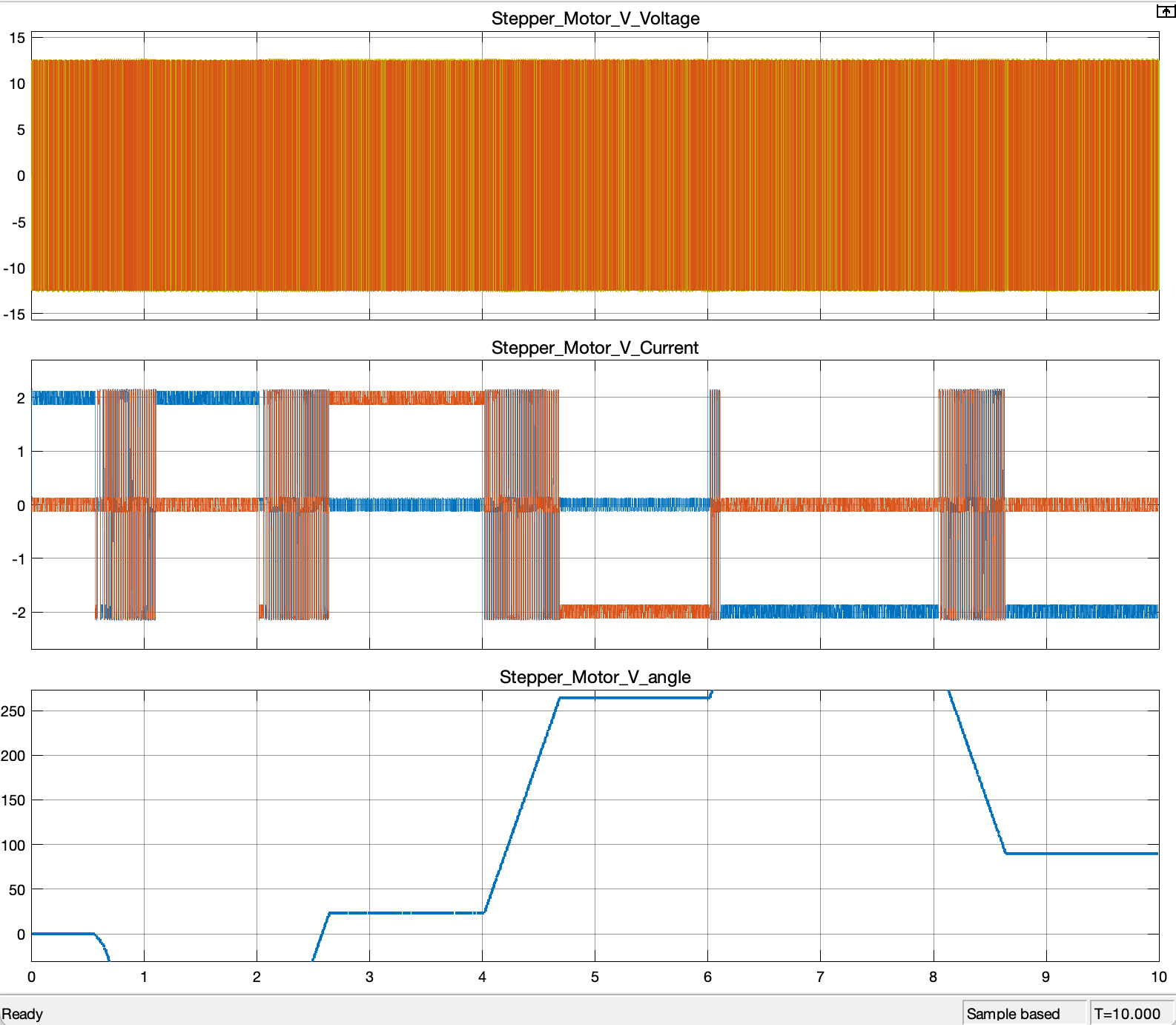} 
	\caption{Stepper motor H voltage, current, and angle}
	\label{fig:current2} 
    \end{figure}

    Figure \ref{fig:water} shows the water tank level in tank 2.  When the water level fall below 20\%, water pump 1 pumps water to tank 2 which until it is full.  Both the speed and armature current of water pump 1 increase until the relay switch turns it off once the water level sensed by the ultrasonic sensor sense the tank is full at 90\%.
Figure \ref{fig:water} shows the stepper motor speed (Rpm), armature current (A), and electric torque (Nm) as the solar irradiance and light photocell resistor level changes. 
    \begin{figure}[H] 
	\centering  \includegraphics[width=0.9\columnwidth]{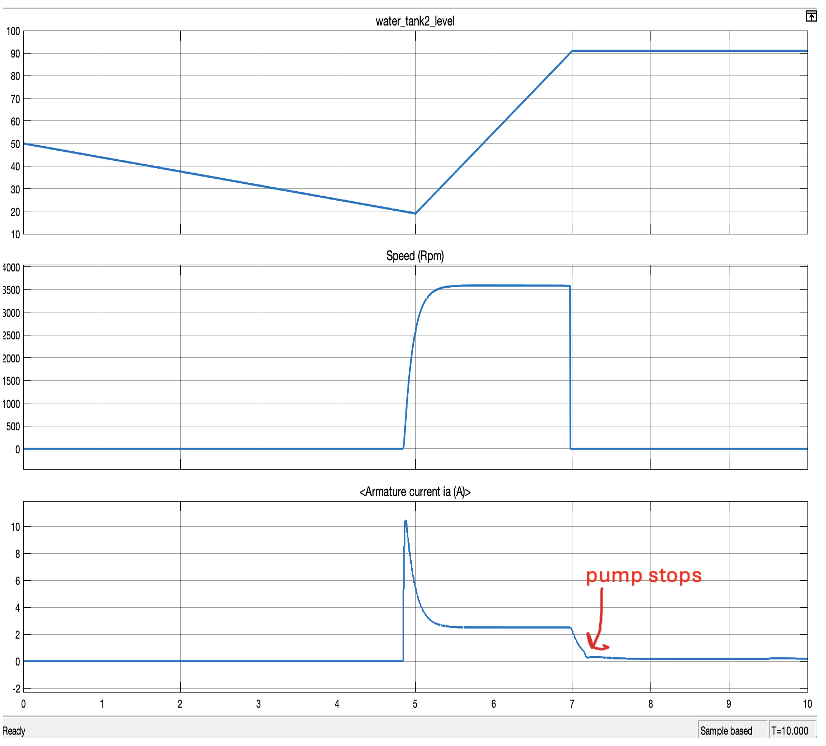} 
	\caption{Simulated water level of tank 2 and water pump 1 speed and armature}
	\label{fig:water} 
    \end{figure} 
When the soil moisture level falls below a humid threshold, e.g., at a dry level, then water pump 2 turns on and pumps water through the hose and nozzle spray to the plant until the soil moisture level is wet.  Figure \ref{fig:water2} simulates the soil moisture level, water pump 2 speed  (rpm), and the pump's armature
Figure \ref{fig:current} and Figure \ref{fig:current2} show simulated horizontal stepper motor and vertical stepper motor voltage, current, and angle, respectively.
\subsection{Experimental}
\begin{figure}[H] 
	\centering  
     \includegraphics[width=0.9\columnwidth]{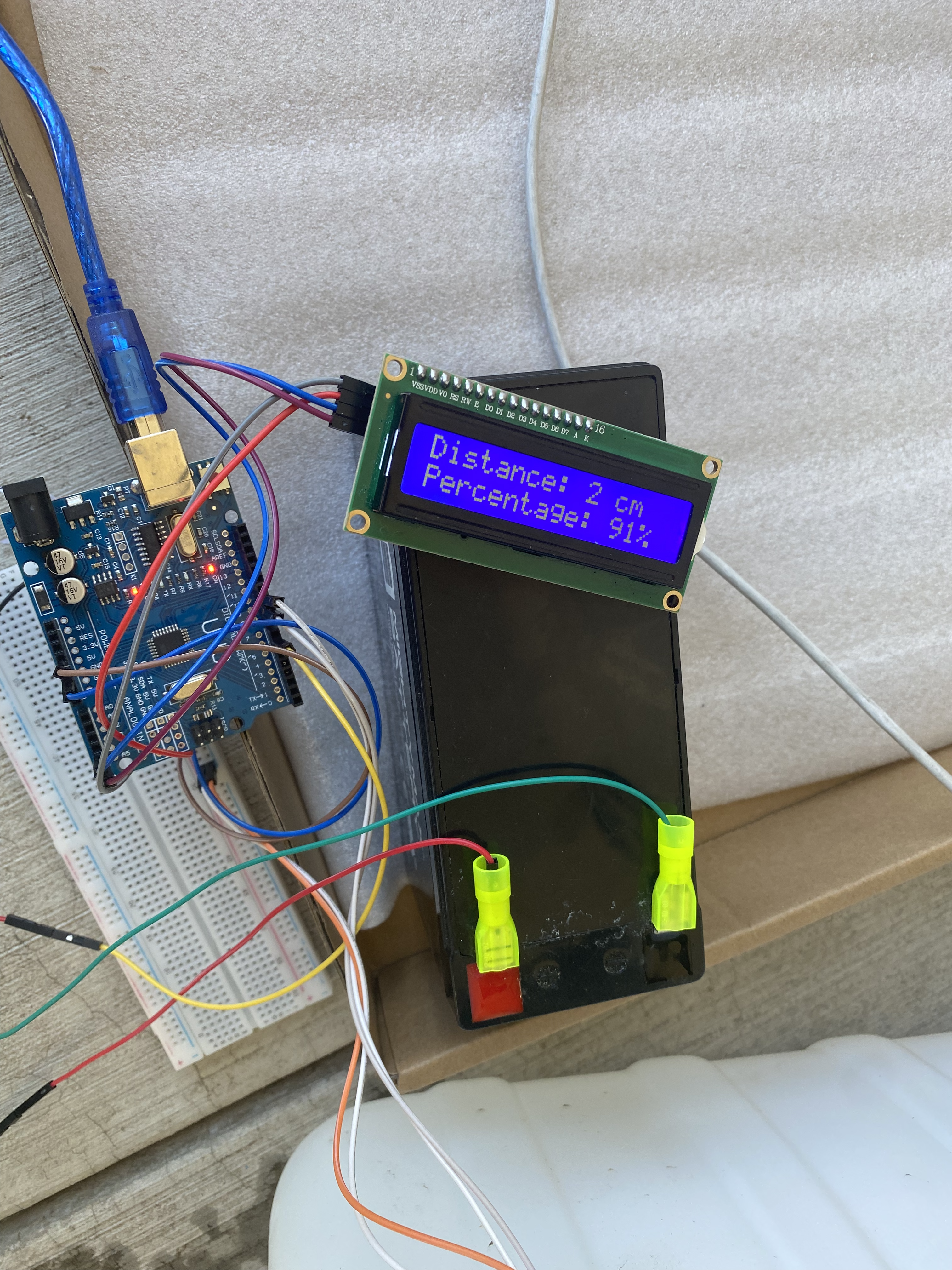} 
	\caption{}
	\label{fig:bread} 
\end{figure}

\begin{figure}[H] 
	\centering  
     \includegraphics[width=1\columnwidth]{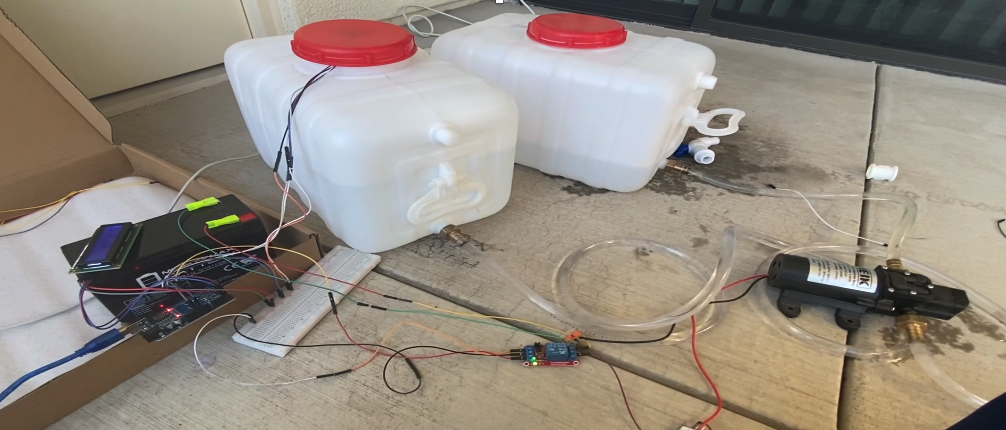} 
	\caption{Arduino and Breadboard}
	\label{fig:setup1} 
\end{figure}

\begin{figure}[H] 
	\centering  
     \includegraphics[width=0.9\columnwidth]{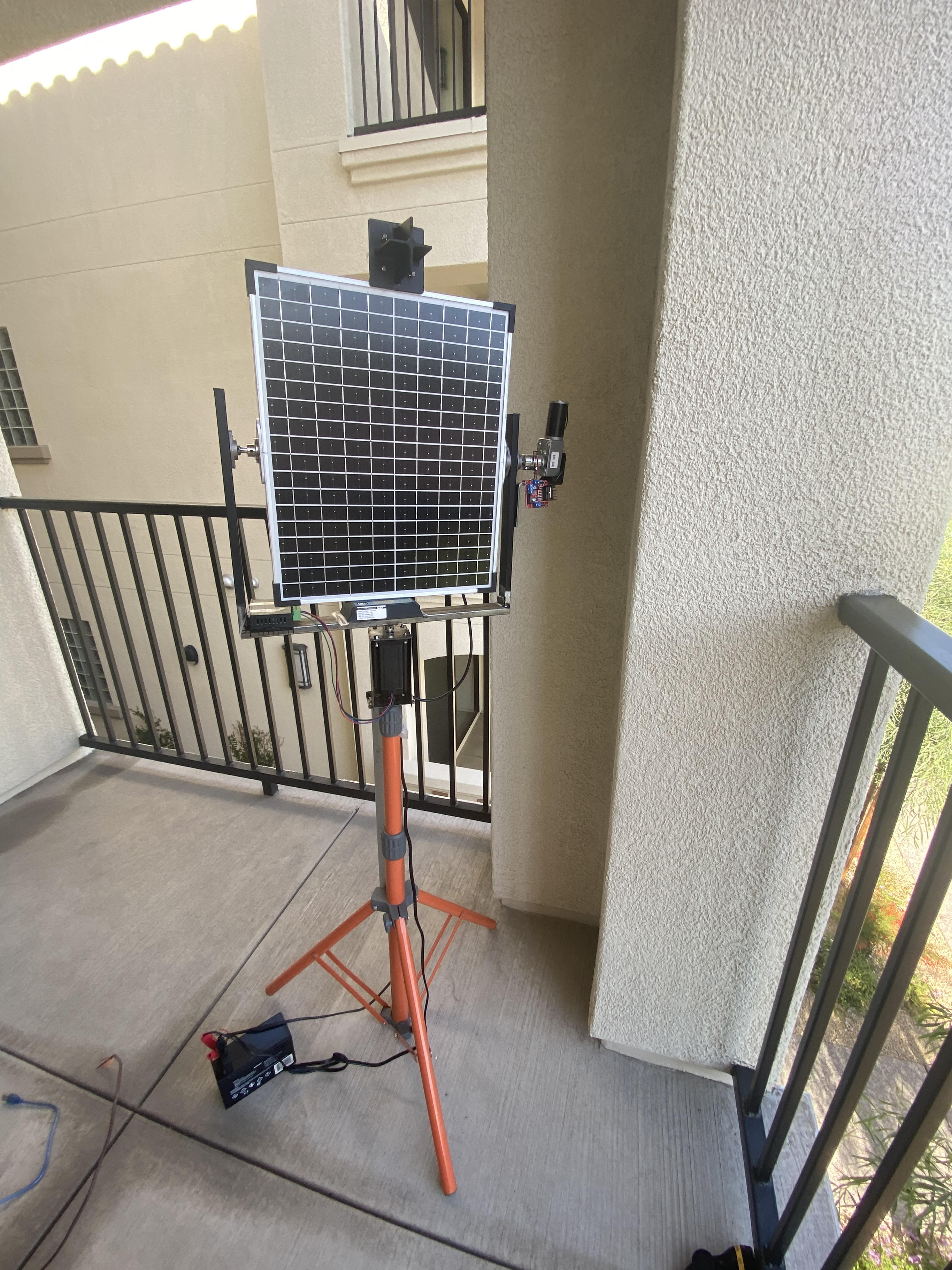} 
	\caption{Solar Tracker on tripod}
	\label{fig:setup2} 
\end{figure}

\begin{figure}[H] 
	\centering  
     \includegraphics[width=0.9\columnwidth]{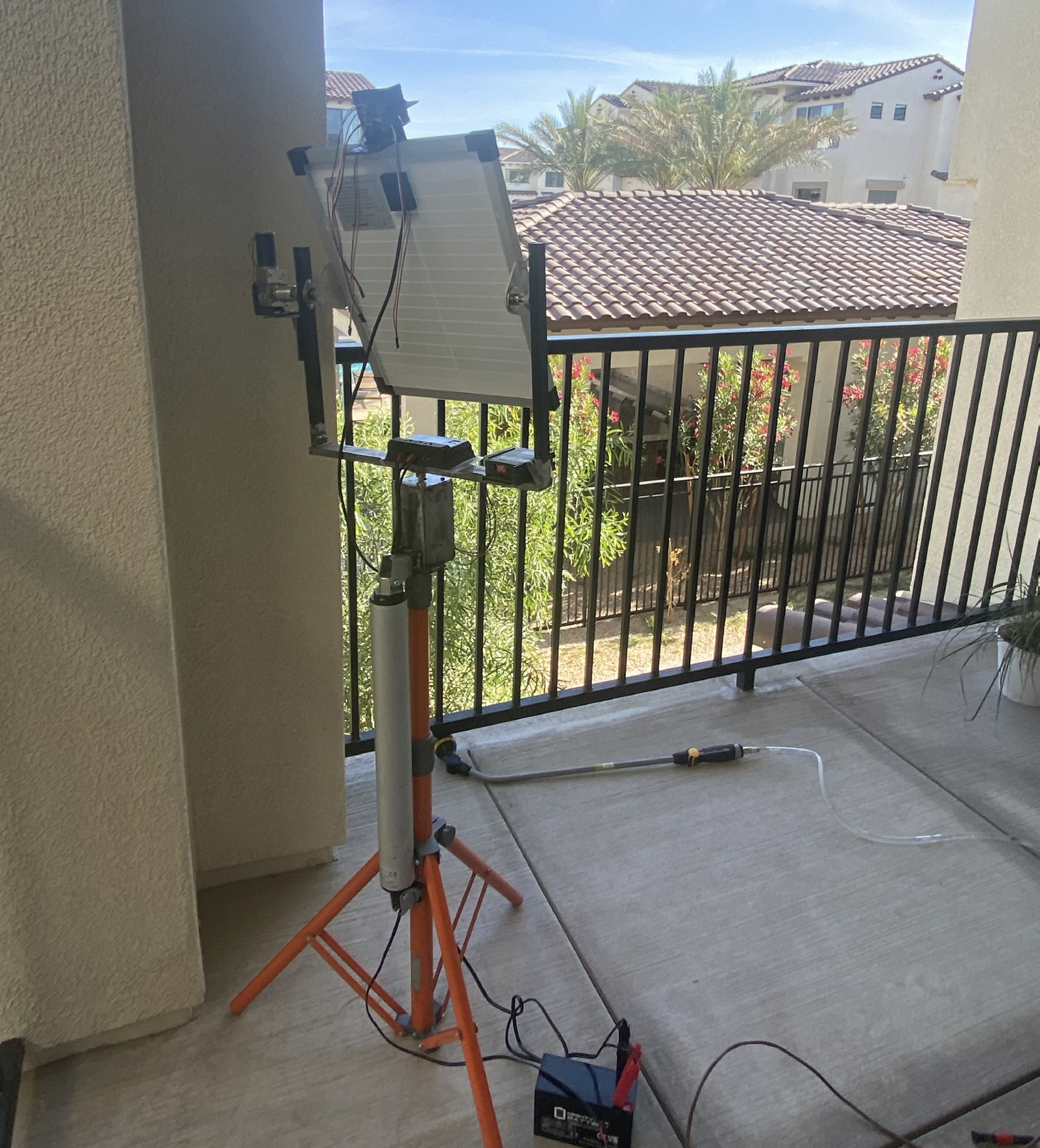} 
	\caption{Solar Tracker facing sunlight}
	\label{fig:solar7} 
\end{figure}

\begin{figure}[H] 
	\centering  
     \includegraphics[width=0.9\columnwidth]{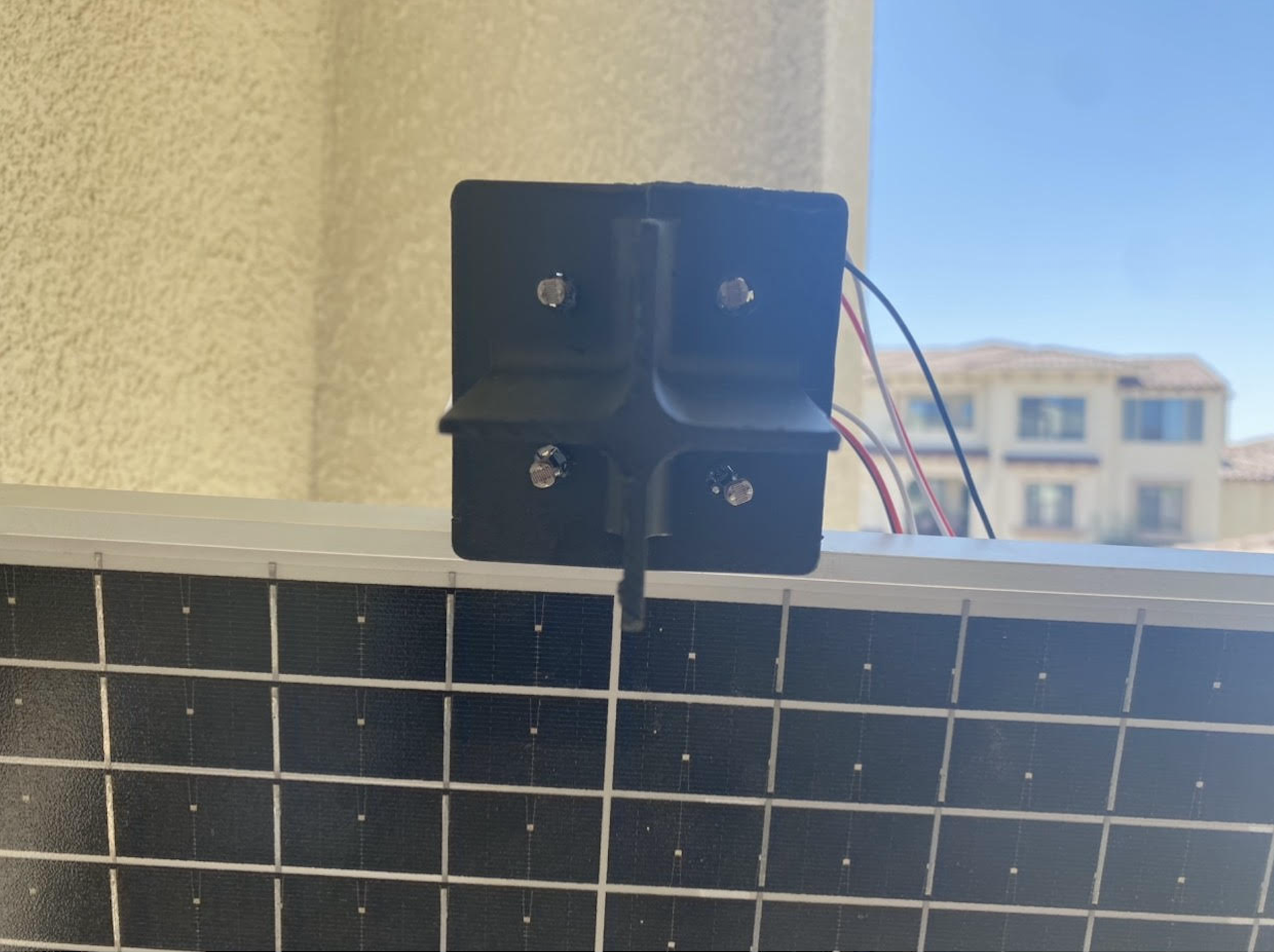} 
	\caption{LPR in 4 quadrants}
	\label{fig:light} 
\end{figure}

\begin{figure}[H] 
	\centering  
     \includegraphics[width=0.9\columnwidth]{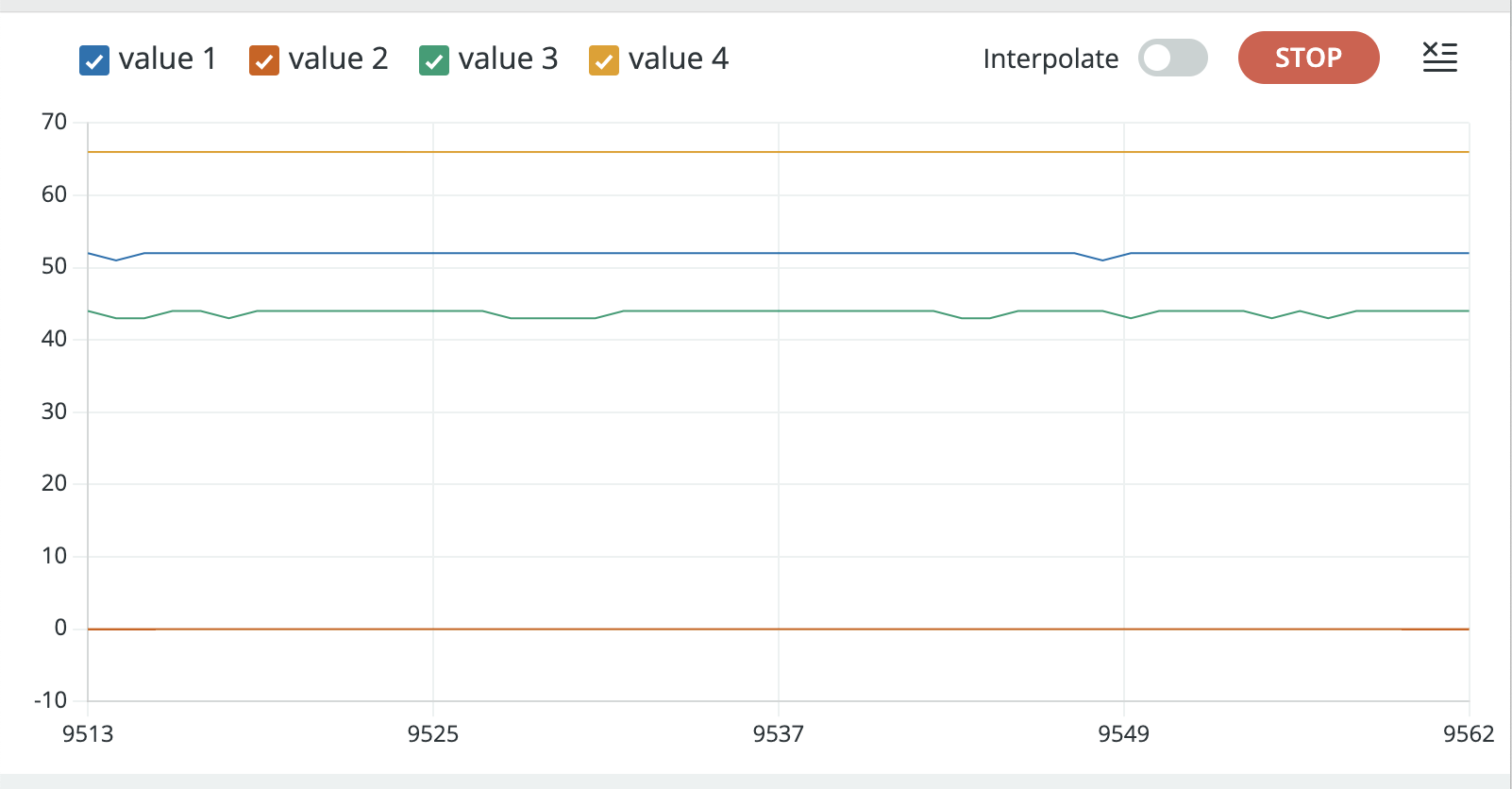} 
	\caption{LPR Values}
	\label{fig:LPR} 
\end{figure}

\ \ \ Figure \ref{fig:bread} shows a display of the water level measured by the ultrasonic sensor and shown on the LCD display along with wiring to the Arduino and breadboard.  The ultrasonic sensor measures the distance from the top of the tank to the top of the water.  As the water level rises in tank 2 from pumped water from tank 1, the total water percentage in the tank increases and the ultrasonic sensor measured distance decreases.
Figure \ref{fig:setup1} shows the two water tanks with the ultrasonic sensor wires coming out from underneath the lid of the second water tank along with wiring to the Arduino and breadboard.
Figure \ref{fig:setup2} shows the DC motor (on right arm) and stepper motor (at base) connected to the PV panel on the tripod.
Figure \ref{fig:solar7} shows a different angle of solar tracker with the front side of the panel facing the sunlight.
Figure \ref{fig:light} shows the four LPRs cells in 4 quadrants.  The tracking algorithm determines the optimal rotation and tilt adjustment of the  PV panel based on the relative intensities between the LRPs.  For instance, if the intensity is greatest in the northeast (first quadrant), the stepper and DC motors will move the panel to optimize the sunlight in this direction. 
Figure \ref{fig:LPR} shows actual LPR values generated by each of the LPRs in Scottsdale, Arizona when the temperature outside was roughly $88^{\circ}$C with sunny skies. As shown, most values stay relatively constant until the solar irradiance changes based on the position of the sun and/or the temperature changes.

Figure \ref{fig:setup4} shows the solar charge controller measuring the charge to the 12 V lead acid battery and the TB6600 stepper motor driver.
\begin{figure}[H]
	\centering  
     \includegraphics[width=0.9\columnwidth]{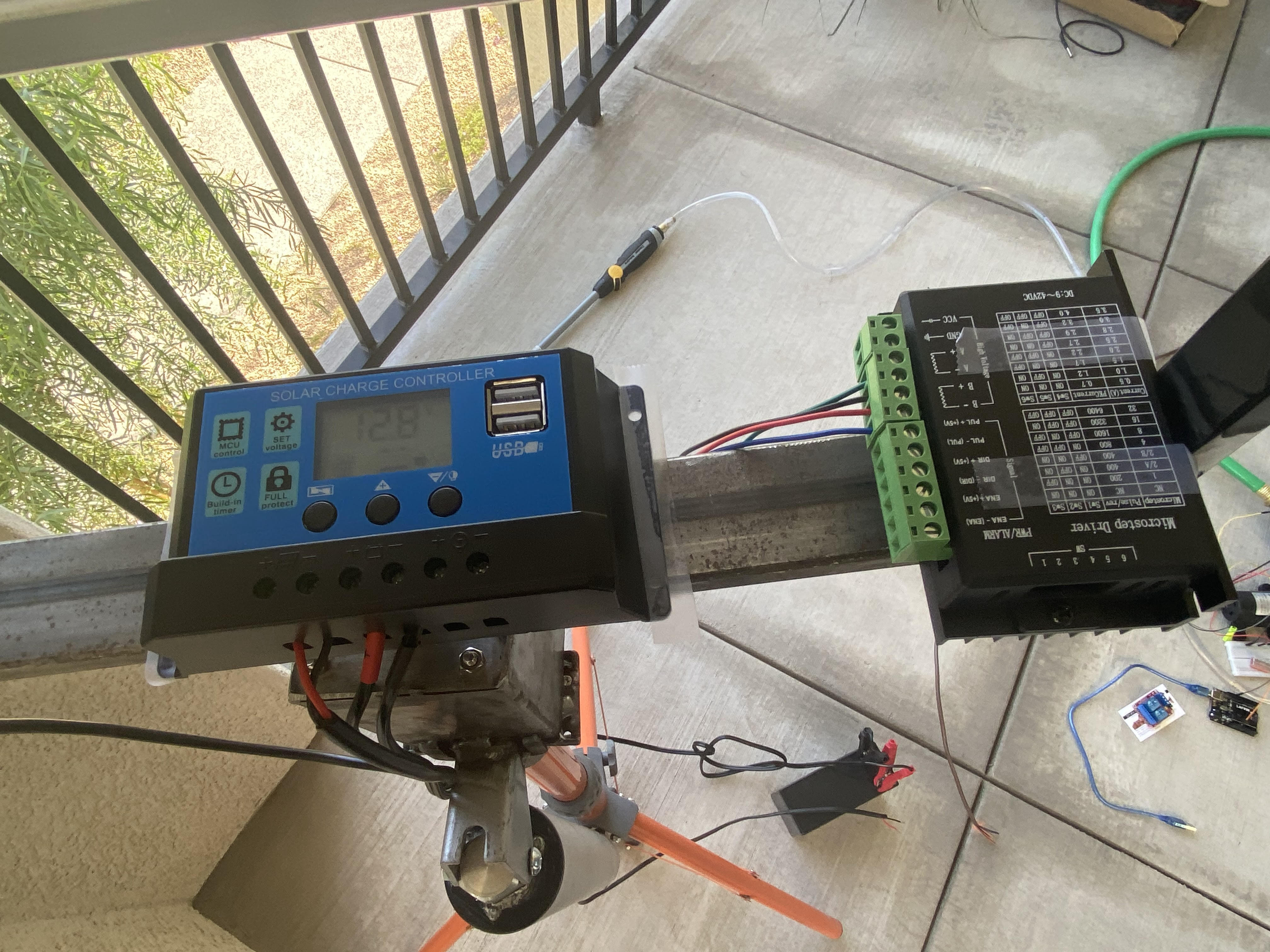} 
	\caption{Solar charge control}
	\label{fig:setup4} 
\end{figure}
    Figure \ref{fig:setup4} shows the solar tracker raised up by the linear actuator.
\begin{figure}[H]
	\centering  
     \includegraphics[width=0.75\columnwidth]{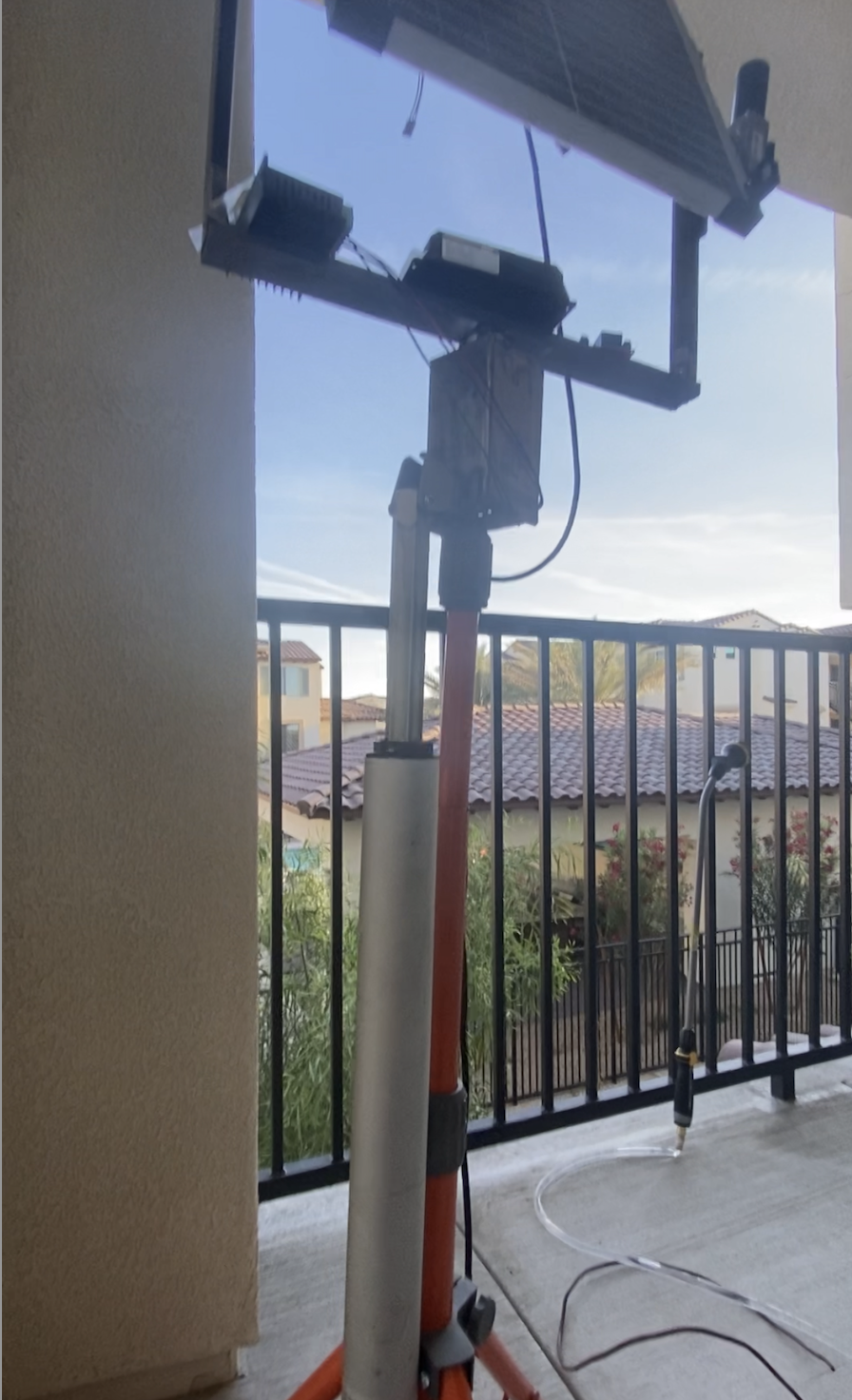} 
	\caption{Linear actuator}
	\label{fig:setup4} 
\end{figure}

\section{Summary}
    A novel 3-axis solar tracker was introduced that is energy and cost-efficient.  The system was analyzed from a control systems analysis perspective.  The transfer function for the DC motor was used to generate step response plots.  The armature-controlled stepper motor parameters such as the moment of inertia, inductance, and back emf torque determine the coefficients of the transfer function characteristic equation.  In order to reduce steady-state errors and enhance system stability, a PID controller was used in the simulations.  Since the poles of the characteristic equation shown in the root loci plots are on the left side of the imaginary axis (the real parts are negative), the proposed system is stable.
    
    The simulations indicate that tuning of the gain parameter $K$ in the motor transfer function plays is a key factor in the step response function and the type of system damping.  The use of various sensors to measure the level of solar irradiance, water level, and soil moisture were used to provide feedback control to the system and worked as simulated.  The LPR cells measure the level of solar irradiance, a key input into the system.  The solar tracking algorithm optimizes the the angle and positioning of the PV panel based on the LPR values. The MPPT perturb and observe algorithm is embedded software in the solar charge controller and serves as an effective DC-DC step-down converter between the panel and DC lead acid battery.

    However, there are various challenges that need to be resolved for the prototype system to be modified for use in a commercial application.  First, the various number of crossing jumper wires, motor drivers, relay switches, breadboards, and wago connectors made diagnosing issues difficult.  Second, the stepper motor requires a lot of experimentation and tuning, e.g. changing the DIP switches, to get the proper microstepping.  Third, the weight of the materials used in the solar tracker frame impacts the microstepping performance because the shaft requires higher torque. 
    
    Thus, future work will involve designing a PCB with integrated circuits and frame fabrication using lighter weight  materials such as aluminum instead of steel to reduce the torque needed for microstepping. Finally, more real-time data needs to be extracted from the system for analysis and comparison to simulated results using Matlab hardware support packages for Arduino.

\newpage
\printbibliography

@misc{Sharma:2019,
title = {Solar Water Pumping System Real-Time Testing and MATLAB Simulink Validation},
author = {Sharma, A. and Parmar, R. and Kumar, S.},
year = {2019}
}

@article{Chin:2011,
author = {Chin, C. and Babu, A. and McBride, W.},
title = {Design, modeling and testing of a standalone single axis active solar tracker using MATLAB/Simulink},
journal = {Renewable Energy},
volume = {36},
year = {2011},
pages = {3075-3090}
}

@article{Habib:2023,
author = {Habib, S. and Liu, H. and Tamoor, M. and Zaka, M. and Jia, Y. and Hussien, A. and Zawbaa, H. and Kamel, S.},
title = {Technical modelling of solar photovoltaic water pumping system and evaluation of system performance and their socio-economic impact},
journal = {Heliyon},
volume = {9},
number = {5},
pages = {e16105},
year = {2023},
}

@article{Riley:2014,
author = {Riley, D. and Hansen, C.},
title = {Sun-Relative Pointing for Dual-Axis Solar Trackers Employing Azimuth and Elevation Rotations},
journal = {Sandia National Laboratories Report},
year = {2014}
}

@article{Moron:2017,
author = {Moron, C. and Ferrandez, D. and Saiz, P. and Vega, G. and Diaz, P.},
title = {New Prototype of Photovoltaic Solar Tracker Based on Arduino},
journal = {Energies},
year = {2017}
}

@article{Musa:2023,
author = {Musa, A. and Alozie, E. and Suleiman, S. and Ojo, J. and Imoize, A.},
title = {A Review of Time-Based Solar Photovoltaic Tracking Systems},
journal = {Information},
volume = {14},
doi = {https://doi.org/10.3390/info1404211},
year = {2023}
}

@article{Hanwate:2018,
author = {Hanwate, S. and Hote, Y.},
journal = {International Journal of Computational Intelligence Systems},
title = {Design of PID controller for sun tracker system using QRAWCP approach},
volume = {11},
year = {2018},
pages = {133-145},
}

@article{Amar:2022,
author = {Amar, A. and Hamraoui, K. and Belguellaoui, M. and Kheldoun, A.},
title = {Performance Enhancement of Photovoltaic Water Pumping System Based on BLDC Motor under Partial Shading Condition},
journal = {Engineering Proceedings},
volume = {14},
year = {2022},
number = {1},
url = {https://www.mdpi.com/2673-4591/14/1/22},
}

@book{Khatib:2021,
author = {Khatib, T and Muhsen, D.},
title = {Photovoltaic Water Pumping Sytems: Concept, Deesign, and Methods of Optimization},
publisher = {Academic Press: Elsevier},
isbn = {978-0-12-821231-8},
year = {2021}
}

@book{Khatib:2016,
author = {Khatib, T. and Elmenreich, W.},
title = {Modeling of Photovoltatic Systems Using Matlab},
publisher = {Wiley},
year = {2016}
}

@book{Nise:2015,
author = {Nise, N.},
title = {Control Systems Engineering, 7th Ed.},
publisher = {Wiley},
year = {2015}
}

@article{Yu:2011,
author = {Yu, S. and Wu, J. and Zhang, D.},
title = {Robust Non-fragile PID Controller Design for the Stroke Regulation of Metering Pumps},
journal = {Chinese Journal of Chemical Engineering},
volume = {19},
number = {1},
year = {2011},
pages = {83-88}
}

@misc{Craig:2002,
author = {Craig, K.},
title = {Sensors \& Actuators in Mechatronics : Liquid Level Control},
url = {http://engineering.nyu.edu/mechatronics/Control_Lab/Criag/Craig_RPI/SenActinMecha/Liquid_Level_Control_Example.pdf},
year = {2002}
}

@article{Mahmood:2020,
author = {Mahmood, Q. and Mawaf, A. and Mohamedali, S.},
title = {Simulation and Performance of Liquid Level Controllers for Linear Tank},
journal = {Jurnal Teknologi},
volume = {82},
number = {3},
pages = {75-82},
year = {2020},
url = {https://journals.utm.my/jurnalteknologi/article/view/14245/6656},
}

@inproceedings{Janevska:2013,
author = {Janevska, G.},
title = {Mathematical Modeling of Pump System},
booktitle = {Electronic International Interdisciplinary Conference},
year = {2013},
url = {https://www.researchgate.net/publication/283348821_Mathematical_Modeling_of_Pump_System}
}

@misc{Marizan:2001,
author = {Marizan, Y.},
title = {Modelling and Control of Liquid Level System},
url = {https://core.ac.uk/download/pdf/12006856.pdf},
year = {2001}
}

@misc{Singh:2015,
author = {Singh, R.},
title = {Design of Internal Model Control and Internal Model Feed-Forward Control for Liquid Level System},
year = {2015},
url = {https://core.ac.uk/download/80148226.pdf},
}
%\end{flushleft}
%\caption{Performance and Robustness}
%\label{tab:table1}
%\end{table}
\end{document}